\documentclass[twocolumn]{aastex631}
\usepackage{natbib}
\usepackage{amsmath}
\usepackage{blindtext, rotating}

\newcommand{\Msun}{$\mathrm{M}_\odot$}

\graphicspath{{./}{figures/}}
\usepackage{array,xcolor,colortbl}
\usepackage{multirow}
\renewcommand{\arraystretch}{1.2}
\usepackage[normalem]{ulem}

\usepackage{xcolor}
\usepackage{xspace}
\newcommand{\redmagic}{redMaGiC\xspace}
\newcommand{\redmapper}{redMaPPer\xspace}

\received{2021 July 19}
\revised{2022 May 5}
\accepted{2022 May 15}
\published{2022 July 8}
\shortauthors{Lokken et al.}

\begin{document}

\title{\large Superclustering with the Atacama Cosmology Telescope and Dark Energy Survey \\ \normalsize I. Evidence for thermal energy anisotropy using oriented stacking}
%@arxiver{unoriented_stacking.png,oriented_stacking.png,act_buzzard_comparison_meq4_nocuts_cuts_1032_2632_18}

\correspondingauthor{Martine Lokken}
\email{m.lokken@mail.utoronto.ca}

\author[0000-0001-5917-955X]{M.~Lokken}
\affil{David A. Dunlap Department of Astronomy \& Astrophysics, University of Toronto, 50 St. George St., Toronto, ON M5S 3H4, Canada}
\affil{Canadian Institute for Theoretical Astrophysics, University of Toronto, 60 St. George St., Toronto, ON M5S 3H4, Canada}
\affil{Dunlap Institute of Astronomy \& Astrophysics, 50 St. George St., Toronto, ON M5S 3H4, Canada}

\author[0000-0002-0965-7864]{R.~Hlo\v zek}
\affil{David A. Dunlap Department of Astronomy \& Astrophysics, University of Toronto, 50 St. George St., Toronto, ON M5S 3H4, Canada}
\affil{Dunlap Institute of Astronomy \& Astrophysics, 50 St. George St., Toronto, ON M5S 3H4, Canada}

\author[0000-0002-3495-158X]{A.~van~Engelen}
\affil{School of Earth and Space Exploration, Arizona State University, Tempe, AZ 85287, USA}

\author[0000-0001-6740-5350]{M.~Madhavacheril}
\affil{Perimeter Institute for Theoretical Physics, 31 Caroline Street N, Waterloo ON N2L 2Y5 Canada}
\affil{Department of Physics and Astronomy, University of Southern California, Los Angeles, CA, 90007, USA}

% DES and ACT data contributors, minus tenured professors

\author{E.~Baxter}
\affiliation{Institute for Astronomy, University of Hawai'i, 2680 Woodlawn Drive, Honolulu, HI 96822, USA}

\author[0000-0002-0728-0960]{J.~DeRose}
\affiliation{Lawrence Berkeley National Laboratory, 1 Cyclotron Road, Berkeley, CA 94720, USA}

\author{C.~Doux}
\affiliation{Department of Physics and Astronomy, University of Pennsylvania, Philadelphia, PA 19104, USA}

\author{S.~Pandey}
\affiliation{Department of Physics and Astronomy, University of Pennsylvania, Philadelphia, PA 19104, USA}

\author[0000-0001-9376-3135]{E.~S.~Rykoff}
\affiliation{Kavli Institute for Particle Astrophysics \& Cosmology, P. O. Box 2450, Stanford University, Stanford, CA 94305, USA}

\author{G.~Stein}
\affil{Berkeley Center for Cosmological Physics,
341 Campbell Hall, University of California, Berkeley, CA 94720, USA}
\affil{Lawrence Berkeley National Laboratory, 1 Cyclotron Road, Berkeley, CA 94720, USA}

% ...

\begin{abstract}
The cosmic web contains filamentary structure on a wide range of scales. On the largest scales, \textit{superclustering} aligns multiple galaxy clusters along inter-cluster bridges, visible through their thermal Sunyaev-Zel'dovich signal in the Cosmic Microwave Background. We demonstrate a new, flexible method to analyze the hot  gas signal from multi-scale extended structures. We use a Compton-$y$ map from the Atacama Cosmology Telescope (ACT) stacked on \redmapper cluster positions from the optical Dark Energy Survey (DES). Cutout images from the $y$ map are oriented with large-scale structure information from DES galaxy data such that the superclustering signal is aligned before being overlaid. We find evidence for an extended quadrupole moment of the stacked $y$ signal at the 3.5$\sigma$ level, demonstrating that the large-scale thermal energy surrounding galaxy clusters is anisotropically distributed. We compare our ACT$\times$DES results with the Buzzard simulations, finding broad agreement. Using simulations, we highlight the promise of this novel technique for constraining the evolution of anisotropic, non-Gaussian structure using future combinations of microwave and optical surveys.
\\
\end{abstract}


\section{Introduction}\label{Sec:Intro} 
The anisotropic clustering of galaxies, galaxy clusters, and intergalactic matter provides a unique insight into the development of large-scale structure (LSS) in our universe. Superclusters and filaments, often referred to as discrete objects, are in reality part of a continuous network of matter. Novel statistical methods focused on the anisotropic scale-dependent aligning tendencies of clusters, galaxies, gas, and dark matter in this network are needed to assess the relative amplitudes that the various species contribute to the overall `superclustering.' We use this term to refer to elongated nonlinear overdense structures that span a wide range of scales (typically tens of Mpc, but also including near-cluster-core sizes of a few Mpc as well as some correlated structures which extend beyond 100 Mpc). The formation and evolution of superclustering is highly dependent on the cosmological model, and thus studying large populations of filaments and superclusters may provide key constraining power to discriminate between different cosmologies \citep{ Cen1994, Frisch1995, Basilakos2001, Kolokotronis2002, Bharadwaj2004, Hopkins2005, Bagchi2017, Ho2018}.

In this paper, we present a novel way of using the \textit{oriented stacking} method with a combination of multi-wavelength data. As a proof-of-concept, we use the method to assess the anisotropy of the thermal energy distribution surrounding galaxy clusters. We begin with an overview of the theoretical and observational landscape of superclustering.

\subsection{An Overview of Superclustering}
Observations, simulations, and analytic theory have converged upon a common model of LSS formation. The early universe was a near-uniform field of Gaussian random density fluctuations which evolved through gravitational instabilities into the web-like structure that exists today. This continuous network of matter, dubbed the `cosmic web' by \citet{BKP} (hereafter BKP), has complex features across a wide range of scales. Its large-scale pattern is predictable from features in the early universe density field, namely the locations of rare mass-peaks and the surrounding large-scale tidal fields \citep{BBKS1986, BKP, BondMyers1996}. Meanwhile, the small-scale details are products of complex local gravitational interactions in the late-time universe.

% Modern cosmological simulations show the web of dark matter---and sometimes gas---in exquisite detail \citep[e.g.,][]{Springel2005, Klypin2011, Dubois2011, Skillman2014, Vogelsberger2014,McAlpine2016}. The algorithms typically track dark matter particles as they move and interact under the current standard model of cosmology, $\Lambda$CDM. Rare small-scale peaks in the dark matter field gravitationally collapse into clumps called halos. Large-scale overdensities ellipsoidally collapse, beginning with the shortest and intermediate principal axes, forming filaments. Filaments sometimes become joined by intervening matter, creating walls. As matter overdensities compactify, underdense regions expand and become increasingly emptier voids. Dark matter halos tend to flow along filaments towards the dense nodes in the web, hierarchically merging along the way; the coalescence of many massive galaxy-hosting halos at the nodes creates galaxy clusters. Hydrodynamic simulations model how gas traces the web of dark matter at a range of temperatures and densities, reaching the hottest and densest levels within galaxy clusters. %original submitted text went to here; the following paragraph was not included
% Thus, as simulations evolve in time, density variations grow and the cosmic web comes into sharper relief. In addition to the dark matter, hydrodynamic simulations model the way gas follows dark matter and heats as the universe evolves. 

From the very earliest cosmological simulations, filaments emerged as the dominant characteristics of the cosmic web. Two distinct theories of their formation (the pancake picture of \citealt{zeldovich:1970} and hierarchical clustering) were synthesized into the BKP model, as reviewed in \citet{vdWetal:2008a,vdWetal:2008b}. This model demonstrates how close pairs of clusters are bridged by filaments with a strength determined by the proximity and alignment of the cluster neighbors. However, the alignment of structure is not limited to the cluster-bridge scale. Clusters themselves can align and cluster, forming \textit{superclusters}. This superclustering of clusters and galaxies is ubiquitous at lower redshifts, and is also played out at higher redshifts in protoclusters.

% Superclusters played a central role in observations of LSS during the late 70s and into the 80s, and those easiest to identify were named \citep{Oort1983}. Recent research has continued to search for the superclusters in available galaxy data using various methods \citep{Tully2014, Chow-Mart2014, Chon2015, Bagchi2017}. However, treating superclusters as distinct objects has mostly grown out of fashion due to the inter-connective nature of the cosmic web.

Supercluster regions consist not only of clusters and filaments; their 3D complexity includes membranes joining filaments and large voids. Despite being the largest nonlinear structures in the universe, they are far from dynamical equilibrium. Hence there is no simple (e.g., spherical) shape readily usable for analysis. Though supercluster regions always have some alignment, a simple straight filament picture does not adequately capture the way in which filaments arc between clusters as the orientations of the clusters pass from perfect alignment to partial alignment. 
% Alignments of clusters and of galaxies in and around clusters were important topics for observers in the 80s \citep{Binggeli1982, Argyres1986} and theorists (Bond 1987). That tide was the key tensor rather than the Hessian of density was emphasize by BM and BKP, but was fundamental to the Zeldovich school in the 70s, which, through Einasto, inspired the quest for auperclustering in the east. In the west, the COMA supercluster, not to mention the Local Supercluster, the Great Attractor, the Bootes void, all brought the rare nonlinear but nonequilibrium structures on the largest scales to the fore of cosmological thinking. Paradoxically, the advent of larger and larger surveys led to a concentration on power spectra of various sorts and the cosmic parameters that could be estimated from the likelihood functions derived from them.

In recent years, the advent of larger and larger surveys has driven the cosmology field away from localized measures of LSS and towards global statistical descriptors. The isotropic two-point correlation function and its Fourier transform, the power spectrum, are frequently used to describe the clustering of matter \citep[a review is presented in][]{Coil2013} and have provided strong constraints on cosmological parameters \citep[see][for the cosmological results from DR12 of BOSS]{boss_dr12}. However, these methods are only sensitive to isotropic clustering, as they are functions of a directionless distance (for the correlation function) or wavenumber magnitude $\vert k\vert$ (for the power spectrum). The lowest-order statistic beyond power is the bispectrum \citep[a review is provided in Section 4 of][]{Desjacques2018}. This is the Fourier transform of the 3-point correlation function \citep{Peebles1980}. The bispectrum encodes information about non-Gaussianity in the late universe, measuring significant cosmological information beyond the power spectrum which can be used to probe dark energy \citep{Takada2004, Sefusatti2006}. However, it is expensive to measure in its full glory with all possible wavenumber triangle configurations, and reductions of the wavenumber possibilities to specific choices restricts the measurable information. \citep{Desjacques2018}. In the era of large surveys, localized measurements of superclustering are still needed, especially those which consider local alignments and can be applied to a wide range of scales.

\subsection{Motivation}
% The formation and evolution of superclustering is highly dependent on the cosmological model, so studying large populations of superclusters may provide key constraining power to discriminate between cosmological models. 
% In recent years, an intense focus on power spectra has dominated LSS data analyses. 
\begin{deluxetable*}{cccc}[htbp!]
\tablewidth{0pt}
\tablecaption{Motivation for making localized measurements of gas anisotropy at various scales.}\label{tab:science_cases}
\tablehead{
\colhead{Object class(es)} & Approx. long-axis length $L$ &  \colhead{Science case(s)}}
\startdata
Galaxy clusters and local surroundings & $L \lesssim 3h^{-1}$ Mpc & Filaments feeding clusters,\\
& & cluster assembly processes,\\
& & baryonic feedback\\
& & & \\
Inter-cluster filaments, & $3 h^{-1}$ Mpc $\lesssim L \lesssim 40~h^{-1}$Mpc &  Cosmological model (dark energy, dark matter),\\
small to mid-sized superclusters & & gas dynamics from filament compression,\\
& & baryonic feedback\\
& & & \\
Largest superclusters & $L\gtrsim 40~h^{-1}$Mpc &  Potential tests of\\ & & primordial non-Gaussianity\\
\enddata

\end{deluxetable*}

With the goal of pushing beyond the limitations of correlation functions, which are ensemble averages of localized measurements centered at \textit{every point} in the universe, this paper presents a new way to explore anisotropic structure through localized measurements at \textit{selected centers} in real space. We emphasize the distinction between \textit{statistical} and \textit{localized} anisotropy; in this paper we refer only to local anisotropies, i.e., the local variations as a field is rotated around a selected point. While the superclustering of matter can be examined through multiple probes, this work focuses on signals from hot gas for a proof-of-concept. Measurements of the gas anisotropy surrounding galaxy clusters have the potential to address cosmological and astrophysical questions across a wide range of scales; Table~\ref{tab:science_cases} provides a brief overview. We focus the discussion on low redshifts ($z<1$), where galaxy survey data (a key component of our methods) is most abundant.

For a homogeneous and statistically isotropic Gaussian random field in the linear phase, all information, including local anisotropic aspects, is encoded in the 3D power spectrum. If there is primordial non-Gaussianity, the power spectrum is insufficient to capture the full information of the LSS fields even on linear scales. Thus higher-order statistics like the bi- and tri-spectra are often used to search for primordial non-Gaussianity, in the hopes of distinguishing between different models of inflation \citep{Giannantonio2012}. However, some types of primordial non-Gaussianity may also be detectable through methods (such as those presented in this work) which measure, in real space, any excess in the anisotropy of different tracers around selected points in the cosmic web. These searches would be best applied at large scales which are less affected by the non-Gaussianity induced by late-time gravitational evolution.

On scales for which the universe has evolved beyond the linear regime, gravitational collapse drives runaway local anisotropy and non-Gaussianity of the dark matter (and baryons). At late times, dark energy changes the superclustering pattern. Hence, localized probes of extended structure ranging from scales of a single inter-cluster bridge to a many-cluster superstructure encode information on the nature of the dark components that may be obscured in the global $k$-space compression of the information onto isotropic power. Understanding the gas and galaxy content of filaments is also important for galaxy evolution studies. For example, studies have shown that the position of a galaxy relative to filaments correlates with its spin and spin alignment \citep{CodisJindal2018, Krolewski2019, Welker2020} as well as its mass, morphology, star formation rate, nuclear activity, and feedback mechanisms \citep{DarraghFord2019, Kraljic2020, SantiagoBautista2020}. Overall, the relationship between the dark matter, galaxies, and gas in filamentary structures contains a wealth of cosmological and astrophysical information. Our succeeding paper will focus on measuring the relationship between the galaxy number density and gas thermal energy using the methods presented herein.

On smaller scales, the orientation of the gas and galaxies within and on the outskirts of clusters is determined by highly nonlinear dynamics. While the anisotropy is certainly influenced by the properties of dark energy and dark matter, the many unknowns in complex dynamical processes such as mergers and splashback make cosmological information difficult to disentangle. However, the dynamics and assembly history of such objects is interesting in its own right. Studying how the cluster gas orientation relates to its surrounding filament(s) could provide insight into the process by which the cosmic web feeds cold gas and galaxies into clusters \citep[e.g., as recently studied in The ThreeHundred project in][]{Kuchner2022, Kotecha2022}.

At all scales, baryonic feedback processes are a confounding factor in studying anisotropies in the gas thermal energy content for cosmological purposes. The complex small-scale processes which move baryons out of dark matter halos and change gas temperatures and pressures are not well understood, and must be constrained in order to glean cosmological information from thermal energy measurements. Understanding the effects of feedback on clusters, groups, and filaments will in turn improve our understanding of the feedback sources, namely black holes and massive stars.

Thus, localized measurements of thermal energy anisotropy are motivated by various science cases across a wide range of scales. For any scale, the questions of interest can be addressed by comparing the characteristics of the observational signal with simulations run under varying sets of cosmological and astrophysical assumptions.

For the purposes of a proof-of-concept we focus this work on the scales of inter-cluster filaments and cluster pairs ($\sim4-12 h^{-1}$ comoving Mpc, further motivated in Sec~\ref{subsec:smoothing}). For this scale, we focus on the redshift range where the available galaxy and cluster data are most abundant (described in Sec~\ref{sec:Data}), $0.2<z<0.7$.

% A subsequent paper will analyse how the signals found herein can be used to constrain the thermal energy content of gas in and beyond filament halos. 

% Thus measuring anisotropic structure is useful and possibly essential to distinguishing between cosmological models with high precision. 

% Generally, global correlation functions--whether two-point or three-point or higher--are ensemble averages of localized measurements centered at each point in the universe. These statistics lack any association between the position of the real-space vectors with the cosmic field on which the measurement is being made. More information can be gleaned by using selected centers as opposed to random ones. This is possible with real-space stacking (a technique described in \ref{subsec:intro_to_methods}), by limiting the stacked objects to, e.g., clusters or galaxies above a certain mass.

% Even within the standard $\Lambda$CDM model, the evolving properties of extended large scale structure over cosmic time are relevant to many sub-fields of astronomy. 
% The manner in which filaments form and feed into nodes in the cosmic web affects the distribution of stars, dust, and gas in galaxies and clusters at varying redshifts. 
%  Thus, our methods enable both cosmological and astrophysical applications.

\subsection{Introduction to Methods} \label{subsec:intro_to_methods}
% Signals from hot gas hold information about the state of baryons in groups, clusters, and filaments. 
We measure the superclustering of thermal energy through the signals imprinted on cosmic microwave background (CMB) data from hot gas in galaxy clusters, groups, and filaments. We also incorporate data from a large-sky galaxy survey into the method to provide necessary LSS information. The galaxy number density, a biased tracer of the total matter, is also of interest as a signal and not only an intermediate step; our succeeding paper will address this. The mass field can be more directly probed with weak lensing maps (as done for filaments in \citealt{Yang2020}); we leave this for future exploration. 

Hot gas in the universe is visible through the thermal Sunyaev-Zel'dovich (tSZ) effect. The tSZ effect is observed along lines of sight which pass through hot gas, arising because a small percentage of incoming cold photons from the primary CMB scatter off of hot free electrons in the intervening gas \citep{SZ1969,SZ1970,SZ1972}. In this inverse Compton scattering process, the photons gain energy and the observed CMB spectrum shifts towards higher frequencies. This causes a shortage of CMB photons at frequencies lower than $\sim$218 GHz and an excess of photons at higher frequencies as compared to the usual CMB spectrum. Thus galaxy clusters appear as decrements in CMB maps with frequency $<218$GHz, and increments in higher frequency maps. The strength of the effect is parameterized by the dimensionless Compton-$y$ parameter,
\begin{equation}\label{eq:tSZ}
    y =  \int d\ell n_\mathrm{e} \frac{k_\mathrm{B} T_\mathrm{e}}{m_\mathrm{e} c^2} \sigma_\mathrm{T},
\end{equation}
a line-of-sight integral of the number density of electrons $n_\mathrm{e}$ and the electron temperature $T_\mathrm{e}$. $\sigma_\mathrm{T}$ is the Thomson cross section and $m_\mathrm{e}$ is the electron mass. Maps of $y$ are typically constructed with component separation techniques from linear combinations of CMB maps at multiple frequencies both higher and lower than 218 GHz \citep{Remazeilles2011}.

% The fractional energy gain per scatter is $4T_e /m_e c^2$, and the fractional diffusive variance is $2T_e /m_e c^2$, with the system obeying the Kompaneets equation, a Bose-Einstein extension of the Fokker-Planck equation for scattering with small energy transfers per scatter (e.g., \citet{Bondleshouches1996}). 

Recent improvements in ground-based tSZ surveys have yielded catalogues of galaxy clusters readily apparent in Compton-$y$ maps due to their high temperatures and densities. Details of the interior cluster gas are revealed through $y$ maps, as the $y$ profile of a cluster is a 2D (cylindrical) projection of the cluster ionized gas 3D pressure profile \citep{carlstrom/etal:2002,Mroczkowski/etal:2019}. The relationship between the cluster halo mass $M_c$ and angular-integrated Compton-$y$ parameter, $Y$, which is proportional to $\langle P_e \rangle V_c$ (the electron pressure times cluster volume), is fairly consistent with $Y\propto M^{5/3}$, as occurs in a gas with adiabatic index $\gamma =5/3$ in equilibrium \citep[see][and references therein.]{BondMyers1996,mccarthy/etal:2003,giodini/etal:2013}

However, for lower-mass halos ($M\leq10^{14}$ \Msun)---such as those that host galaxy groups along inter-cluster filaments---this relationship no longer holds. \citet{Lim2018, Hill2018} and others have shown that these smaller halos have a steeper $Y$--$M$ relation, primarily because gas is blown out of halos by feedback from active galactic nuclei (AGN), changing the relation from that which is expected by gravitational arguments alone. The extent to which feedback mechanisms from AGN, supernovae, magnetic fields and other sources determine the $\langle P_e\rangle V_c$ of lower mass clusters and groups and redistribute gas beyond halo boundaries is still not fully understood. Even for massive clusters, AGN feedback has been a needed addition to relate cluster observations to theory \citep{Sijacki2007, Puchwein2008, Sijacki2008, Battaglia2010, Gitti2012}.

Therefore, it is difficult to probe superclustering from the tSZ effect due to the wide variation in gas pressure in different parts of the cosmic web. The weak tSZ signals from lower-mass halos are drowned out by noise in individual images, and the signals from gas outside of halos are even weaker. Researchers frequently employ \textit{stacking} to extract information from these low-signal regions. In stacking, multiple similar images are averaged. The signal is overlaid throughout the stack and the random noise averages down with more images. If all of the $N$ component images have roughly the same signal, the final stack has a higher signal-to-noise ratio (SNR) than any of individual images by a factor of $\sqrt{N}$. Stacking has been used to study the relationship between the tSZ signal and other properties of clusters using data from both space- and ground-based instruments \citep[see e.g.][]{Plagge2010,Hand2011,Sehgal2011, PlanckStackedSZ2013, Sehgal2013,Planck2016tSZ}. It has also been used to study small halos through the thermal and kinetic Sunyaev-Zel'dovich effects in recent work by \citet{Schaan2021, Amodeo2021}, who found that the gas profile extends well beyond the virial radius of the dark matter halo.

In recent years, various teams employed stacking to study inter-cluster filaments and attempt to detect the warm-hot intergalactic medium (WHIM), low-density gas thought to have been blown out from dark matter halos by feedback processes. This gas at $10^5-10^7$K may make up $\sim40-50\%$ of baryons \citep{CenOstriker2006, Shull2012}. Thus, despite its lower tSZ signal compared to collapsed objects, it is an important contributor to the census of cosmological baryons \citep{CenOstriker1999}. By stacking $\sim$1 million pairs of Luminous Red Galaxies (typically found at the center of clusters), \citet{deGraaff2019} and \citet{Tanimura2019} detected a filament tSZ signal. The signal emerged from a combination of galaxies, groups, and the WHIM. The former group also claimed detection of the WHIM gas itself by accounting for the massive galaxies in each filament. In addition to tSZ evidence, recent studies have found tentative signals from the WHIM through absorption lines from filaments intersecting the lines-of-sight of quasars  \citep{Tejos2016, Pessa2018, Bouma2021}. Altogether, these works have been important confirmations of the existence of a large amount of baryons in low-mass halos and intergalactic gas, accounting for most or all of the so-called `missing baryons' \citep[as reviewed in][]{Tumlinson2017}.

Despite these successes, thus far, most studies of filamentary structure have been limited in scale. Cluster-pair stacking is limited by the distance between pairs: in the aforementioned studies, the filaments ranged from 6--10$\,h^{-1}$Mpc. Therefore, this method cannot be used to study very small-scale alignments of galaxies nearby a cluster, nor superclusters tens or even hundreds of Mpc long. Some recent work has used alternative methods to probe a wider range of scales \cite[e.g.,][]{Tanimura2020}, but more work is needed to fully explore the multi-scale information content of tSZ anisotropy.

%  with the DisPerSE algorithm \citep{Disperse2013} to study the baryon and dark matter content with data from the \textit{Planck} satellite \citep{Planck2016tSZ}

\subsection{Oriented stacking}

\begin{figure}
\centering
\includegraphics[width=0.45\textwidth]{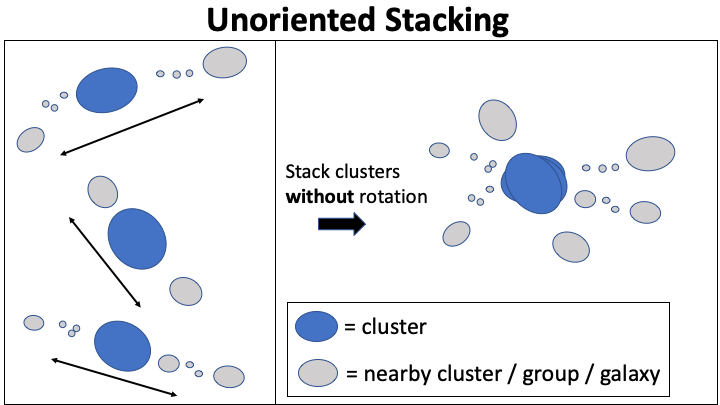}

\includegraphics[width=0.45\textwidth]{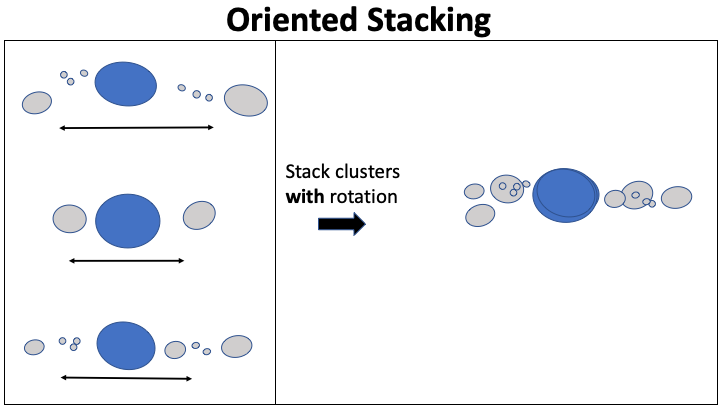}

\caption{A visualization of unoriented (top) versus oriented (bottom) stacking. When stacking cluster cutout images from the Compton-$y$ map with random orientation, the signal from external structure is averaged out such that the final image, as the number of cutouts increases, approaches perfect isotropy. In oriented stacking, we determine the direction of strongest alignment around each cluster using the curvature of galaxy overdensity maps. We rotate each cutout image before stacking such that the excess $y$ signal from anisotropic superclustering is overlaid.}\label{fig:oriented_stacking_vis}
\end{figure}

Oriented stacking is a more general approach to stacking elongated structures which does not necessitate the identification of cluster pairs nor distinct filaments. This method seeks to quantify anisotropic superclustering over a range of scales. Oriented stacking was used in \citet{Battaglia2012a} to examine the anisotropy of gas around clusters on relatively small ($\sim5$ Mpc) scales in simulations; it was also applied to studies of CMB polarization in \citet{Planck2016_Isotropy_Statistics}. Our paper presents its first application to a combination of CMB and galaxy data. This work complements a recent theoretical exploration of anisotropic superclustering in full 3D and in 2D projections given in \cite{Regaldo2021}.

In Figure~\ref{fig:oriented_stacking_vis} we illustrate how oriented stacking is applied to a combination of a CMB $y$-map and galaxy and cluster surveys to tease out the faint tSZ signals from superclustering. In this method, we stack the tSZ signal from both clusters and the surrounding gas by aligning along a measured axis of large-scale structure. We go beyond previous filament studies by applying characteristics of the galaxy field to select on areas of high superclustering, which we control, augmenting the signal relative to the fluctuation noise. We demonstrate how multi-scale selections on superclustering features can probe how the gas signal changes with scale. After performing oriented stacks, we quantify our results using a multipolar decomposition of the results. We compare the observational results with simulations to search for any discrepancies indicating inaccuracies in the cosmology, galaxy and cluster sample selection, and/or gas prescription in the simulations.

Because the number of stacked objects in this work is 2--3 orders of magnitude smaller than the cluster-pair stacking studies, we do not expect to significantly detect the WHIM signal. Rather, the dominant signals come from clusters and galaxy groups, and thus our methods generally probe the anisotropic clustering of thermal energy. In future work, we will determine whether the methods described herein can be used to find explicit evidence for hot gas outside of halos. Future work will also explore the potential of oriented stacking to address the various science questions in Table~\ref{tab:science_cases}, including how to disentangle the small-scale baryonic physics from cosmological effects and ultimately search for signs of physics beyond the standard model.

% \dick{this needs to be expanded on.  for us every object has a pair, one of larger scale. ie we are using one point stats plus constraint. they are using one point stacks plus constraint of nearby cluster. counting pair products overcounts. however connor and i did stack with planck on many more clusters because it was all sky y.} 

This paper is organized as follows. In Section \ref{sec:Data}, we describe each data product. Section \ref{sec:galaxy_field_characteristics} describes the properties of the galaxy density field we use to find regions of high superclustering. Section \ref{Sec:Methods} describes the stacking methods used. Section \ref{sec:expectations_from_theory} presents the expected signal from the Websky \citep{Stein2020} and Buzzard \citep{DeRose2019} simulations with various cluster populations. In Section \ref{Sec:Results} we compare the ACT$\times$DES results with the Buzzard simulations and forecast for future data. Finally, in Section \ref{sec:conclusion} we discuss the prospects of applying these novel methods to future superclustering analysis.

Throughout the paper, when a cosmological model is assumed for conversions from redshift to comoving distance and from angular size to transverse comoving distance, we use the \texttt{Planck15} cosmology from the \texttt{astropy.cosmology}\footnote{\url{https://docs.astropy.org/en/stable/cosmology/index.html\#module-astropy.cosmology}} package which implements a Flat $\Lambda$CDM model with parameters from the \citet{Planck2016}. The model has $\Omega_M=0.308$, $\Omega_k=0$, $\Omega_\Lambda=0.691$, a single species of massive neutrinos with mass 0.06 eV, and  $H_0=67.7\,\mathrm{km}\, \mathrm{s^{-1}}\, \mathrm{Mpc^{-1}}$. All quoted distances are in comoving units.

\section{Data} \label{sec:Data}

In this work we combine the tSZ Compton-$y$ signal maps derived from high-resolution ACT measurements with DES cluster and galaxy catalogs, and then compare our observational results to simulations of large scale structure. 
% This section describes the observational data we use as well as our selection process to match the simulation outputs to data.

\subsection{ACT Compton-y map}
ACT is a 6-meter off-axis Gregorian telescope, located in the Atacama Desert of northern Chile, at an elevation of 5190~m on the Cerro Toco stratovolcano \citep{Fowler2007, Swetz2011}. The telescope has been operating since 2008, first measuring only temperature fluctuations in the microwave regime. In 2013 the ACTPol receiver was deployed, enabling ACT to observe both temperature and polarization data at 98 GHz and 150 GHz \citep{thornton/etal:2016}. The receiver was subsequently upgraded to Advanced ACTPol, adding three more frequencies \citep{Henderson2016, Shuay2017, Choi2018, Li2018}. ACT has recently produced maps covering 18,000 square degrees of the sky \citep{Naess2020, Aiola2020}. This paper focuses only on the region for which a tSZ map has been made which also overlaps with part of the DES footprint. This 456 square degree region is called `D56' and was first presented in \cite{Louis2017}.

% This contained 391 hours of observational data after cuts, obtained from July to December 2014 and Oct 2015 to Jan 2016. The D56 data was first presented in \cite{li/etal:2017}.
%%%%%
We use the Compton-$y$ map in D56 first presented in \citet{Madhavacheril2020}. This map was reconstructed from a combination of 2015--2016 night-time data from the ACTPol receiver at 98 and 150 GHz as well as multifrequency data from the \textit{Planck} satellite \citep{Planck2016tSZ}. Data from \textit{Planck} were isolated to include only modes between $20<\ell<300$ for the \textit{Planck} low-frequency instruments (LFI) at 30 GHz and 44 GHz respectively, $20<\ell<2000$ for the 70 GHz LFI and modes between $20<\ell<5800$ for the high frequency instruments (HFI) at 100, 143, 217, 353 and 545 GHz. Data from ACT include modes $500<\ell<24000$. The maps are combined via an internal linear combination (ILC) algorithm to isolate the $y$-component in each Fourier
pixel using the estimated covariance between the map arrays at different frequencies. The resulting $y$ map has an effective 1.6 arcminute beam. 

The $y$ map may contain residual contribution from the primary CMB and astrophysical foregrounds. For the purposes of this work, the residual of greatest concern is the cosmic infrared background (CIB): this emission from dusty galaxies is highly correlated with the tSZ effect. Radio point sources are a far sub-dominant source of contamination because they are much less correlated with tSZ sources \citep{Sehgal2010}, and the primary CMB is not expected to bias our results as it is uncorrelated. To test the impact of the CIB contamination, we make use of an additional map described in \citet{Madhavacheril2020}, made with a constrained ILC algorithm in which the frequency-combined map is required to have a null response to the CIB in addition to having unit response to Compton-y. For our key results in this paper (Sec. \ref{Sec:Results}) we run our pipeline with this CIB-deprojected map and find that it biases the signal slightly lower compared to the results with the original map. The signal is reduced by $\sim6\%$, which corresponds to only $\sim12\%$ of the original 1-sigma errorbars. Because it is a negligible effect compared to the errors, we choose to present results using the less noisy map without CIB deprojection.

In addition to the residual contamination from non-$y$ components, the $y$ map contains stripey instrument noise which is oriented along the ACT scan direction and uncorrelated with any structure in the Galactic or extragalactic sky. We emphasize that this does not affect the oriented stacking procedure, as orientations are entirely determined by the galaxy data described in Sec.~\ref{sec:field_galaxy_data} and collected by a different survey (DES).

A complete description of the $y$ map procedure is found in \citet{Madhavacheril2020} and the maps are publicly available on the NASA Legacy Archive Microwave Background Data Analysis \cite[LAMBDA,][]{lambda}.\footnote{\url{https://lambda.gsfc.nasa.gov/product/act/act_dr4_derived_maps_info.cfm\#compsep}} Figure \ref{fig:desdata_plus_actoutline} shows the sky area of the map (outlined in black) overlaid on a map of the positions of galaxies (red points) and galaxy clusters (black points) from the DES. We discuss these cluster samples below.

\subsection{Galaxy cluster data}\label{subsec:clusterlocs}

\begin{figure*}[t]
    \includegraphics[width=1\textwidth,trim={1cm 6cm 1cm 6cm}, clip]{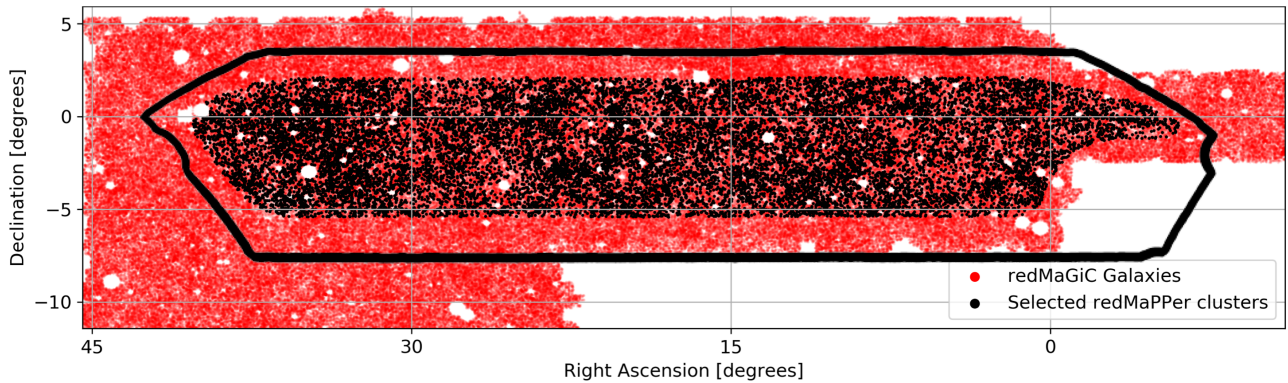}
    \caption{Selected \redmapper clusters (black points), \redmagic galaxies (red points), and the outline of the ACT Compton-$y$ map (black line). Clusters are cut from the edge of the $y$ map and galaxy footprint to avoid edge effects from the $y$ map when stacking, and to ensure that there is ample galaxy information surrounding each cluster.}
    \label{fig:desdata_plus_actoutline}
\end{figure*}

We stack cutout images from the ACT Compton-$y$ map on locations of galaxy clusters identified in the Dark Energy Survey (DES) Y3 data. DES \citep{DES2005} recently completed a six year survey (2013--2019) of 5,000 square degrees of the southern sky in five optical filters (\textit{grizY}). The survey was conducted by the 4-meter Blanco Telescope, fitted with the Dark Energy Camera \citep{Flaugher2015}, at the Cerro Tololo Inter-American Observatory (CTIO) in Chile. The \redmapper algorithm, originally introduced in  \citet{Rykoff2014}, identifies galaxy clusters by searching for overdensities of red galaxies. The first \redmapper catalog for DES was published using the Science Verification Data \citep{Rykoff2016}. The algorithm determines a value for richness, $\lambda$, for each cluster by summing the membership probability of each galaxy which has some likelihood of belonging to that cluster, within a defined radius. $\lambda$ is therefore related to the mass; a detailed study of the mass-richness relation for DES was done in \citet{McClintock2019}. Our study uses a catalog generated from Y3 Gold data from the first three years of the survey \citep{DESY32021}. This \redmapper catalog (titled \texttt{v6.4.22+2 Full}) extends out to $z\sim1$, includes all $\lambda>5$ clusters, and provides a photometric redshift (photo-$z$) for each cluster with uncertainties $\sigma_z/(1+z)\sim0.01--0.02$ \citep{McClintock2019}. 

From this catalog, we select only clusters which overlap with the ACT D56 region. We further limit the cluster area by enforcing that clusters must be over 2 degrees inside the edge of D56, as shown in Figure \ref{fig:desdata_plus_actoutline} (black points versus black D56 outline). This ensures that no edge effects are present in any of the $y$-map cutouts. Additionally, we remove clusters that are closer than 1 degree to the edge of DES galaxy data (discussed in the Section \ref{sec:field_galaxy_data}). Thus, every cluster is surrounded by ample LSS information in all directions, necessary for the accurate determination of orientation.

We choose to limit the cluster sample to $\lambda>10$. This threshold is a trade-off between the disadvantage of small-number statistics when imposing a stricter cutoff and the advantages of a higher-richness sample. These advantages include (a) that the anisotropic Compton-$y$ signal from the LSS surrounding higher-$\lambda$ clusters is stronger, and (b) that higher-$\lambda$ cluster data are more pure. Point (a) is determined in Section \ref{sec:parameter_dependence} with simulations. As for point (b), past cosmology studies with \redmapper clusters have almost always excluded clusters with $\lambda<20$ due to their known impurities \cite[e.g.][]{Abbott2020, To2021, Costanzi2021}. These low-$\lambda$ clusters are more likely to be false detections from random fluctuations or line-of-sight projections in the galaxy field. They also suffer from more mis-centering \citep{Rykoff2016}. However, in our study, a $\lambda>20$ cutoff is not feasible. There are not enough $\lambda>20$ clusters available in the D56 sky region to achieve a detection of anisotropic thermal LSS, especially given that the Compton-$y$ signals beyond the cluster radius are much weaker than the internal signal, which is often the focus of cluster-stacking research. To choose the specific $\lambda$ threshold, we test $\lambda>10$, $\lambda>15$, and $\lambda>20$ on noiseless simulations (described below). The correlation described in (a) does not provide enough signal boost to offset the increase in random noise from limiting the cluster sample beyond $\lambda>10$. Since this cutoff results in the highest SNR, we apply it to the real data.

Effects from mis-centering are expected to be negligible in our study, as the center offsets are typically a fraction of the \redmapper cluster radius $R_\lambda$ \citep{Zhang2019} and we will examine signals beyond 1.5$R_\lambda$. However, the effects of false cluster detections may be non-negligible, and we would expect them to bias our results lower. We discuss this further in the conclusions.

\begin{figure}
    \includegraphics[width=0.45\textwidth
    ]{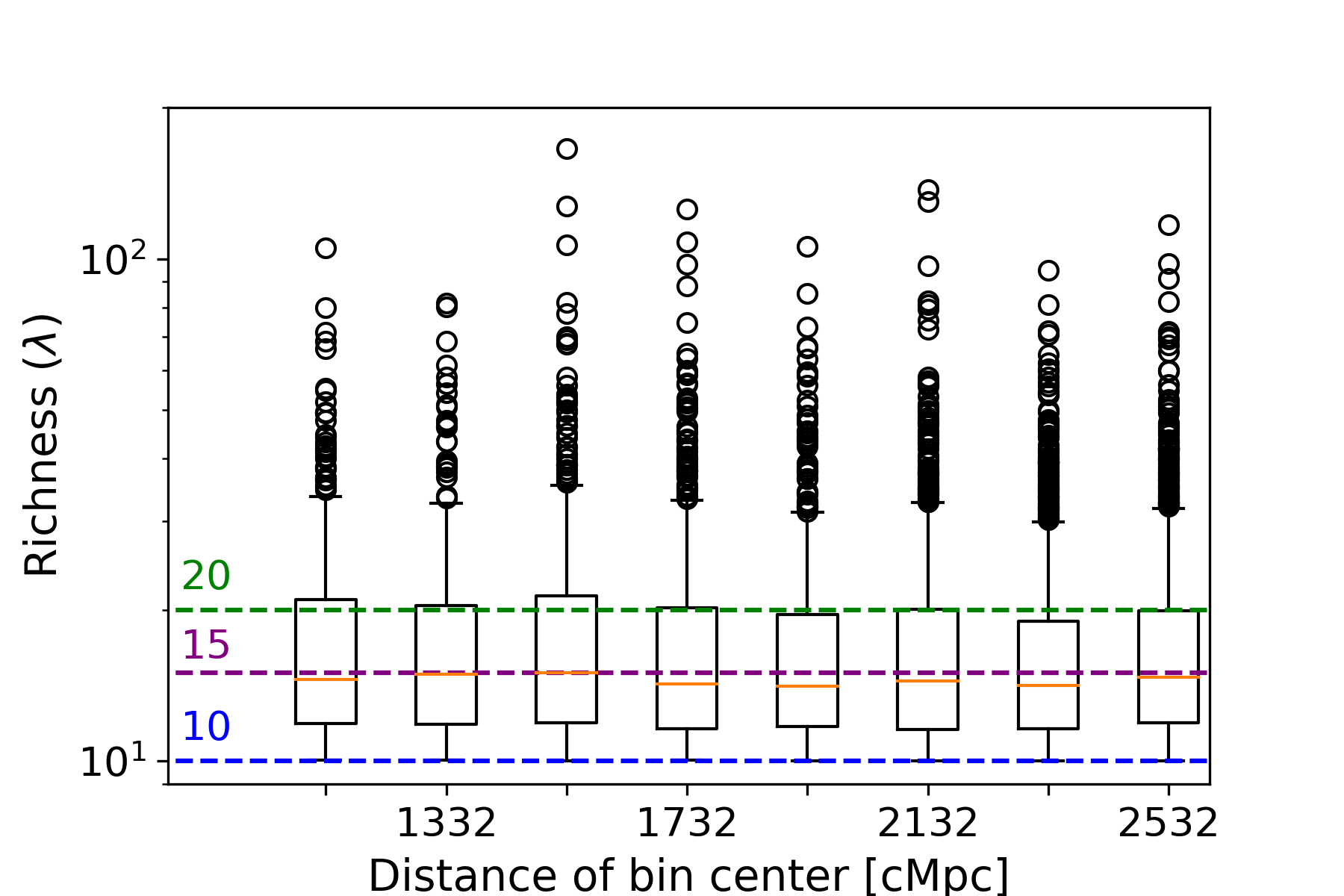}
    \caption{Box plots showing the richness distributions of selected $\lambda>10$ clusters after separating the sample into 200\,Mpc-thick slices in line-of-sight comoving distance. The central orange line in each box indicates the median richness in the bin, which is consistently $\sim15$. Each box is drawn from quartile 1 (Q1) to 3 (Q3) of the data, and the whiskers are drawn out to 1.5$\times$(Q1--Q3) beyond the box on either side. Black circles represent outliers. The richness is skewed towards low values but there are a fair number of high-richness outliers. The higher-richness clusters contribute a stronger tSZ signal from both the cluster itself and surrounding structure; however, lower-richness clusters must be included to achieve a sufficient SNR on extra-cluster gas. In section \ref{sec:parameter_dependence}, we use simulations to test 3 richness cutoff values (10,15,20), shown here in blue, purple, and green.}
    \label{fig:richness_boxplot}
\end{figure}

There are 5,494 clusters in the remaining sample. The median photometric redshift uncertainty is $\sigma_z/(1+z)=0.009$. As part of the stacking process described in Section \ref{Sec:Methods}, we divide the cluster sample into even slices in comoving distance along the line-of-sight, which are each 200 Mpc thick. Figure \ref{fig:richness_boxplot} shows the distribution of cluster richness for each distance slice of the $\lambda>10$ sample. The figure also shows three colored lines which represent the different richness thresholds tested in simulations in Section \ref{sec:parameter_dependence}.

\subsection{Galaxy data}\label{sec:field_galaxy_data}
 
We use galaxy data from DES to orient each Compton-$y$ map cutout with respect to the axis of surrounding elongated structure. The \redmagic algorithm selects Luminous Red Galaxies from photometric surveys using a matched-filter technique \citep{Rozo2016}. It has recently been applied to Y3 DES data over the full DES footprint, as detailed in \citet{Pandey2021}. The algorithm is designed to minimize errors in galaxy photo-$z$, resulting in an average scatter of $\sigma_z/(1+z)\sim0.013$. In this study we use the High Density catalog, which covers redshifts 0.1 to 0.7 with fairly consistent number density ($\sim[8.5, 10.5]\times10^4$ $h^3 \mathrm{Mpc}^{-3}$). The high density nature of this catalog is important for accurate orientation. The average halo mass of \redmagic galaxies is quite large, at $1.5\times10^{13}$ $h^{-1}$ \Msun \citep[$h=0.69$,][]{Pandey2021}. In the future, it would be interesting to explore stacking on every \redmagic galaxy rather than only \redmapper clusters, which would sample more points in the cosmic web and provide improved statistics.

% To supplement the RedMaGiC galaxies, we use the CMASS sample from the Baryon Oscillation Spectroscopic Survey \citep{Schlegel2009}. The CMASS sample consists mostly of galaxies between z$\sim$0.43 and z$\sim$0.7, each with spectroscopic redshifts, and are chosen to have fairly constant stellar mass centered at  $\log{M/\Msun}\sim11.25$ \citep{Maraston2013}.

\subsection{Buzzard Simulations} 
For comparison with the observational data, we make use of the Buzzard Version 1 simulations \cite[][hereafter D19]{DeRose2019}. The Buzzard galaxy and cluster catalogs were explicitly designed to provide a comparison with DES by replicating many of the survey's selection effects; therefore this simulation suite is the most straightforward choice for direct comparison with our ACT$\times$DES results. Buzzard assigns realistic galaxies to dark matter halos in $N$-body dark-matter only simulations using the ADDGALS method \citep{Weschler2021}. The implemented cosmology has $\Omega_m=0.286, h=0.7, \sigma_8=0.82$. This is slightly different from the \citet{Planck2016} cosmology which we use for radial and transverse distance calculations; however, the differences contribute at most a $\sim2\%$ error to the rescaling of cutout images which is later described in Section~\ref{Sec:Methods}. The galaxy catalogs are post-processed by the DES pipelines to create mocks of the \redmapper and \redmagic catalogs in the same footprint with the same selection effects. We apply the same redshift selections to Buzzard as we do to the observational data.

The \redmagic mock galaxies approximate the clustering in the real catalog well as shown in D19. However, the simulation struggles to match the \redmapper observables from DES. Key differences between the mass-richness relation and the cluster abundance in Buzzard and DES are shown in Figures 12 and 13 of D19. In Buzzard, the number of identified \redmapper clusters is a factor of 3--5 below the number of real \redmapper clusters for the redshifts and richnesses used in our work. This deficit is likely due to a reduced number of galaxies in the central regions of galaxy clusters and a reduced number of red galaxies in dense regions, both of which result in fewer richness selected clusters (D19, \citealt{Weschler2021}). Despite the lower cluster number density, the number of Buzzard clusters available to use for the noiseless theory cmoparison is $\sim5$ times larger than it is for DES. This is due to the fact that the Buzzard tSZ, cluster, and galaxy data overlap fully in the entire DES footprint, whereas which the ACT$\times$DES overlap is $\sim12$x smaller. Thus the random noise in stacks will be lower when using Buzzard. We further discuss the effects of the cluster abundance discrepancy in Section~\ref{subsec:systematics}.

To create a mock Compton-$y$ map, we paste pressure profiles from \citet{Battaglia2012b} on Buzzard halos from the same simulation run, then convert these to Compton-$y$. This follows the approach from \citet{Stein2020}, described in more detail in Sec. \ref{subsec:Websky}. We apply the model down to halo masses of $10^{12}$ \Msun, although halos at such low masses contribute very little to the overall tSZ signal. The result is a noiseless projected $y$ map. The map contains only signal from halos; because Buzzard is not a hydrodynamic simulation there is no prescription for ejecting baryonic material out of halos into the WHIM. Therefore, filaments in Buzzard only contain bound gas in halos.  We then convolve this map with a 1.6 arcminute beam and remove modes with $\ell<20$ to match the filtering of the ACT Compton-$y$ map.

Ideally, we would create many noisy versions of the Buzzard $y$ map by combining many realizations of the Buzzard simulation with many realizations of simulated ACT noise. Such maps would be useful to assess uncertainties and provide a direct comparison between simulations and data. This is unfeasible because there is only one readily available realization of the Buzzard simulation which has had all the post-processing steps applied to create galaxy and cluster catalogs as well as a $y$ map. Also, the relevant ACT noise has only been simulated in a small fraction of the Buzzard sky footprint. Instead, we estimate uncertainties by using spatial splits of the single Buzzard realization combined with many ACT noise simulations in D56; details are further described in Sec. \ref{subsec:uncertainties}. We also use Buzzard for a noiseless comparison to ACT results.

The Buzzard algorithm was recently improved in \texttt{v2.0} for validation of the DES Y3 results \citep{DeRose2021}; our follow-up paper will use the state-of-the-art for DES mock simulations.

\subsection{Websky Simulations}\label{subsec:Websky}
We wish to use simulations not only to make direct comparison with ACTxDES, but also to generate pure-theory expectations for the superclustering of thermal energy over time. The Buzzard \redmagic and \redmapper mock catalogs are not ideal for studying pure theory expectations because of the DES observational limits applied to them, i.e., the limited redshift range and sky coverage. The Buzzard suite also includes more extensive catalogs without DES limitations; however, we choose instead to make use of the full-sky Websky Extragalactic CMB Simulations \cite[][hereafter S20]{Stein2020}. The notable speed of the Websky algorithms will be useful in future work for generating alternative cosmologies to which we will apply oriented stacking.

% Although the post-processed Buzzard galaxy and cluster catalogs are useful simulations for direct comparison with ACT$\times$DES, their limited redshift range and sky coverage make them non-ideal for studying pure theory expectations.
The halo catalogs for Websky were produced via the Peak Patch algorithm, which rapidly generates halos from an initial density field using an ellipsoidal collapse model \citep{BondMyers1996, Stein2019}, excluding overlapping halos in the final list. Websky extends out to $z=4.6$ and was run with Planck 2018 cosmology: $\Omega_m=0.31, h=0.68$, and $\sigma_8=0.81$. These parameters are slightly different than those used for Buzzard, but this is unimportant as we will not directly compare the two simulations.
% Both the Websky dark matter halo catalog and Compton-$y$ maps have been extensively validated against multiple observational data sets, unlike Buzzard for which the majority of testing was done on the mock-\redmapper and \redmagic catalogs in comparison with DES.

The Peak Patch halo catalogs were transformed to sky maps of various probes of structure, including the tSZ effect. For the tSZ effect, the Websky simulation assigns each halo a thermal pressure as a function of redshift and mass. This pressure profile relationship was determined by the stacking of halos found in the large-scale structure simulations of \citet{Battaglia2012b}, which focused on clusters and galaxy groups and included AGN feedback. The exact prescription in Websky is given in equation 3.12 of S20. The Websky Compton-$y$ power spectra have been validated against multiple observational data sets.

% For this transformation, one must specify the relationship between the presence of a dark matter halo at a particular location and the behavior of the field (e.g., tSZ) at and around that point. We will refer to this relationship as a response function. The function depends on the halo mass and can depend on additional properties of the nearby halo field. 

In this paper, we will stack the Websky $y$ map on halo positions extending to $z\sim2$, which allows us to go well beyond the DES \redmagic limit of $z\sim0.7$. Because the full Websky halo catalog extends further (to $z=4.6$), for any stack, there are ample higher-redshift clusters contributing to a realistic uncorrelated tSZ background. As with the Buzzard map, we convolve the Websky $y$ map with a 1.6 arcminute beam and remove all power for modes $\ell<20$.

We select cluster-mass halos in a few mass ranges and use mass-weighted halos to provide information about large-scale structure, as further detailed in Section~\ref{sec:expectations_from_theory}.

% The CMASS galaxies are mocked by Peak Patch as follows.  By SED fitting, \citet{Maraston2013} found the mean log stellar mass of the CMASS galaxies to be $\sim11.3$  $[\log{M/\Msun}]$ with a dispersion of masses $\sim0.3$  $[\log{M/\Msun}]$. \citet{Sonnenfeld2019} use weak lensing data to find that for a subsample of CMASS galaxies with $\log{M_*/\Msun}>11$, the average log halo mass for a galaxy with $\log{M_*/\Msun}=11.4$ is $12.79\pm0.03$ $\log{M_*/\Msun}$. To mimic the CMASS galaxies in Peak Patch, we therefore select halos which fall within a Gaussian distribution in log mass, centered on 12.79 $[\log{M/\Msun}]$ to match the halo mass analysis with a dispersion of 0.3 $[\log{M/\Msun}]$ to match the stellar mass distribution (Figure \ref{fig:mass_hist}, lower plot). Then, for every redshift slice, halos are randomly drawn from this distribution until the sample consists of the same number as in the CMASS catalog in that slice.

\section{Galaxy field characteristics}\label{sec:galaxy_field_characteristics}

We begin by studying large-scale properties of the projected galaxy number density field. The bulk of this study is implemented with the Cosmology Object Oriented Package \citep[\texttt{COOP}\footnote{\url{https://www.cita.utoronto.ca/~zqhuang/work/coop.php}},][]{Huang2016}. Later, these properties will be used to constrain the \redmapper cluster sample and thus limit the stacks of $y$ map cutouts to special locations in the cosmic web. In particular, we are interested in the gas signal from superclusters, where the tSZ contribution from high and low mass clusters, lower-mass groups, and shock-heated  gas in filaments and other compressing configurations should be stronger than that of average aligned structure. Superclusters correspond with overdense and elongated regions of coarsely-smoothed galaxy maps \citep{Oort1983, Einasto1997, vdWetal:2008b}.

We search for regions which satisfy these criteria using projected galaxy overdensity maps in bins of redshift. The overdensity is defined as $\delta_g = n_g/\bar{n}_g - 1$, where $n_g$ is the two-dimensional number density of galaxies and $\bar{n}_g$ is its mean. In practice, each map is created in the Healpix\footnote{https://healpix.sourceforge.io/} pixelization scheme through the Python package \texttt{Healpy} \citep{Healpix2005, Zonca2019}. Each $n_g$ map is created with NSIDE=4096 by adding 1 to the appropriate pixel for every galaxy within the redshift bin. After transforming to $\delta_g$, we smooth the map by convolving with a 2D Gaussian function, creating a smoothed map $F_g$. Various choices of smoothing kernel would be valid, and the top-hat function is another that is frequently used in the LSS literature; we choose a Gaussian as it is most conveniently implemented in \texttt{COOP}. The Gaussian filter scale $R_G$ is related to the full-width at half-maximum by FWHM=$2\sqrt{2\ln{2}}R_G$. We vary the smoothing scale to observe the LSS properties at a range of scales. Specifically, we examine results for Gaussian smoothing with full-width at half-maximum (FWHM) ranging from 6 to 18 Mpc. The approximate equivalent range in top-hat radius $R_{\mathrm{TH}}$, if the maps were smoothed with a top-hat filter to produce similar field properties, is $R_{\mathrm{TH}}\sim 5-11$ Mpc. The conversion is done by enforcing an equal volume under the top-hat function with radius $R_{\mathrm{TH}}$ and height 1, and a Gaussian function with amplitude 1. The range is chosen to examine highly nonlinear structure beyond a typical cluster radius; smoothing scales are discussed further in Section \ref{Sec:Methods}. In the current section, we use a FWHM of 14 Mpc for demonstration purposes.

The primary property we use to determine superclustering is the field excursion $\nu$,
\begin{equation}
    \nu = F/\sigma(R),
\end{equation}
where $F$ is the field value at some position and $\sigma$ is its root mean square (RMS) after smoothing on some scale $R$ \citep{BBKS1986}. Given that the $\delta_g$ maps have a mean of zero, $\sigma$ is equivalent to the standard deviation, making $\nu$ a measure of signal-to-fluctuation-noise. Points with higher $\nu$ correspond to rarer overdensities. 

We also use the asymmetry of the field as a metric of superclustering. To measure the alignment and elongation at any point in the field, it is natural to consider using the tidal field, because the tidal shear in the early universe is key to producing filaments \citep{vdWetal:2008b}. In addition, it has the same power spectrum as the density. The derivative of a Gaussian-smoothed tidal field with respect to scale $R$ is the Hessian matrix of the Gaussian-smoothed density, which has frequently been used to characterize cosmic web phenomenology \citep[as reviewed in][]{Libeskind2018}. The Hessian is defined in 2D as

% We also use the asymmetry of the field as a metric of superclustering. The tide has the same power spectrum as the density, and it is natural to consider it for aligning. But any filtering could be used, not just top hat. The derivative of a Gaussian smoothed tide with respect to scale $R$ is the Gaussian-smoothed Hessian matrix of the density matrix. This identifies a characteristic shell region of the tide, which we are free to choose, but it is based on characteristic scales like that defining $\sigma_8$. For two dimensions:

\begin{equation} \label{eq:Hessian}
H = \begin{bmatrix}
\frac{\partial^2 F}{\partial x^2} & \frac{\partial^2 F}{\partial x\partial y} \\
\frac{\partial^2 F}{\partial y\partial x}  &\frac{\partial^2 F}{\partial y^2}
\end{bmatrix}
\end{equation}
for a field $F$, evaluated at some point. We use the Hessian to determine the asymmetry and alignment of the smoothed projected galaxy overdensity field at cluster positions. The choice of smoothing scale defines a characteristic radius from each selected field point at which the Hessian encompasses maximal information; this can also be thought of as a shell region of the tidal field.

\begin{figure}
    \centering
    \includegraphics[width=\columnwidth, trim={6cm 4cm 6cm 5cm}, clip]{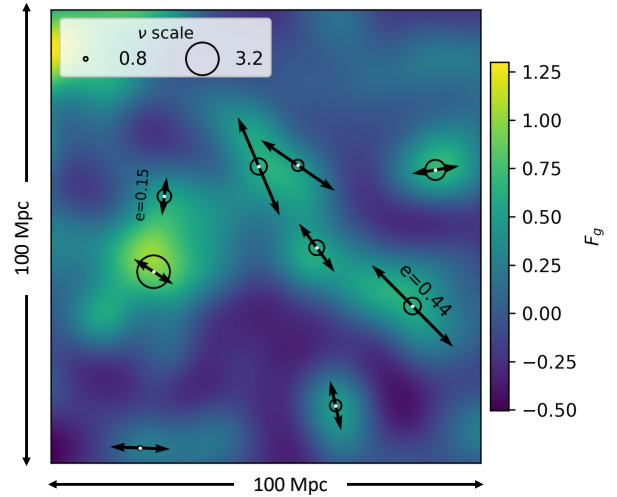}
    \caption{A visualization of $\nu$ and $e$ at cluster locations overlaid on a smoothed, projected \redmagic galaxy map $F_g$. A $3.5^{\circ}\times3.5^{\circ}$ patch of $F_g$ is shown, corresponding to $\sim 100$~Mpc at the center of the galaxy redshift bin ($0.35<z<0.5$). The smoothing was Gaussian with a FWHM=28' ($\sim14$ Mpc). \redmapper cluster locations are identified by white points. $\nu$ and $e$ are calculated at the cluster locations using $F_g$; because $F_g$ has been coarsely smoothed, they are properties of the large scale structure in which each cluster is embedded. Black circles have an area scaled by the $\nu$ value at the cluster point. Headless vectors have a length corresponding to $e$ and are oriented along the eigenvector $\vec{v_2}$ of the Hessian (Sec.~\ref{Sec:Methods}). We define supercluster regions as points where $\nu$ and $e$ are both large.}
    \label{fig:nu_e_vis}
\end{figure}

We adopt the notation and conventions of \citet{BE1987}, hereafter BE87, in defining dimensionless eigenvalues of the Hessian. BE87 first applied the study of these field properties to the CMB. At any field point, the Hessian has eigenvalues $-\lambda_i$ and corresponding eigenvectors. Note that with the negative sign, $\lambda_i$ are defined to be positive at peaks and negative at troughs. We order the eigenvalues as $\mid{\lambda_1}\mid>\mid{\lambda_2}\mid$, such that $\lambda_1$ corresponds to the eigenvector along which curvature is changing most rapidly: the `short axis' of curvature. Using the eigenvalues, we can define the ellipticity $e$:
 \begin{equation} \label{eq:e}
     e = \frac{\lambda_1 - \lambda_2}{2 (\lambda_1 + \lambda_2)}.
 \end{equation}
This follows the definition in BE87, which succeeded the 3D representations for galaxy fields in \citet{BBKS1986}, hereafter BBKS. The numerator describes how elongated the field is at a certain point by the difference in eigenvalues there. This is normalized by the trace of the Hessian to provide an equitable comparison between different-amplitude peaks/troughs. The $\nu$ and $e$ parameters are visualized in Fig.~\ref{fig:nu_e_vis}.

% Points for which $\lambda_i$ have opposite signs are saddle points.
In BBKS, the ellipticity was applied in the context of cluster-scale smoothing to the 3D galaxy density field. In this context, clusters lie at peaks in the field. At a location with negative curvature in both directions, such as a peak, the minimum value for $\lambda_2$ is 0, and consequently $e \leq 0.5$. However, for the larger scale (e.g. 14 Mpc) smoothing used in this work, most clusters do \textit{not} lie at peaks in the projected, smoothed galaxy overdensity maps. Many clusters exist where the gradient of $F_g$ is non-zero. These large-scale gradients point toward supercluster centers rather than toward the individual clusters that make up each supercluster. For some clusters, where the field has two equal-sign eigenvalues, $e\leq0.5$. Many clusters also lie at regions where the eigenvalues have opposite signs (which are saddle points if the background gradient is 0), where $e>0.5$.
% (As an aside, incorporating these large-scale gradients is of future interest as they contain information about the most massive regions toward which material in the cosmic web is flowing.)
Thus the distribution of $e$ values, while concentrated towards 0, extends well beyond 0.5 and can become very large when the denominator of equation \ref{eq:e} is small (in other words, when $\lambda_2\sim -\lambda_1$). We find that points with higher $\nu$, at rarer overdensities, are more likely to have $e<0.5$. Later, we choose to apply a minimum $e$ threshold to select for locations of the field that are highly elongated, which is a feature of superclustering. However, we do not limit our sample by applying a maximum $e$ threshold; $e$ is allowed to be arbitrarily large.

We also consider the field curvature excursion, $x$, related to the trace of the Hessian:
\begin{equation}
x = \nabla^2F/\sigma_2,
\end{equation}
where $\sigma_2$ is the root mean squared value of $\nabla^2F$. This property was defined in \citet{BE1987}; our definition differs by an absolute value sign such that our $x$ is allowed to be negative. Points with high $x$ have large curvature in one or both directions. $x\propto (\lambda_1+\lambda_2)$, which is the denominator of $e$, so points where $\lambda_1\sim-\lambda_2$ have small $x$ and large $e$. Due to the divergence of $e$ at $x\sim0$, we choose to show the non-normalized elongation $\vert x \vert e $ in upcoming figures for visual purposes.

To better understand these characteristics and how they may be used to find regions of strong superclustering in the universe, we compare the $\nu$, $\vert x \vert e$, and $x$ distributions of the galaxy overdensity field with that of a Gaussian random field we have constructed with the same power spectrum. We divide the Buzzard \redmagic galaxy catalog into 3 redshift bins: $0.15<z<0.35$, $0.35<z<0.5$, and $0.5<z<0.65$. Using measurements of the Buzzard galaxy autospectrum, $C_\ell^{gg}$ \citep{Pandey2021}, we generate a Gaussian Random Field (GRF) realization from the power spectrum by using the \texttt{Synfast} function from \texttt{Healpy}. We will refer to this map as the pseudo-galaxy map.

As we are interested in properties of the Buzzard field at the locations of mock-RedMaPPer clusters, it is necessary to identify RedMaPPer-like peaks in the GRFs. Accordingly, we identify peaks in the field of a similar angular size as RedMaPPer clusters for each $z$ bin. We determine the size by finding the approximate mass $M_{200\mathrm{m}}$ of a $\lambda=15$ cluster using the mass-richness relation from \citet{McClintock2019}. $M_{200\mathrm{m}}$ refers to the mass enclosed within a sphere of radius $R_{200\mathrm{m}}$ within which the density is, on average, 200$\times$ the mean matter density of the universe at that redshift. $\lambda=15$ was chosen because it is the median richness of our sample. Next, we convert $M_{200\mathrm{m}}$ to $R_{200\mathrm{m}}$ and find the Gaussian filter equivalent of a top-hat filter with that radius. (For discussion on top-hat to Gaussian conversion, see Section~\ref{subsec:smoothing}.) We smooth the pseudo-galaxy field with a Gaussian function with FWHM $\sim1.6$ Mpc. The equivalent angular size varies for each redshift bin. For the smoothed field, we find all peaks and sub-select them by applying a $\nu$ threshold on the small peak scale. We choose a $\nu$ threshold for each $z$ bin which results in the same number of GRF peaks as Buzzard mock RedMaPPer clusters. In summary, we perform approximate abundance-matching to find GRF peaks which are similar in size and amplitude to RedMaPPer clusters.

\begin{figure}
\includegraphics[width=0.45\textwidth]{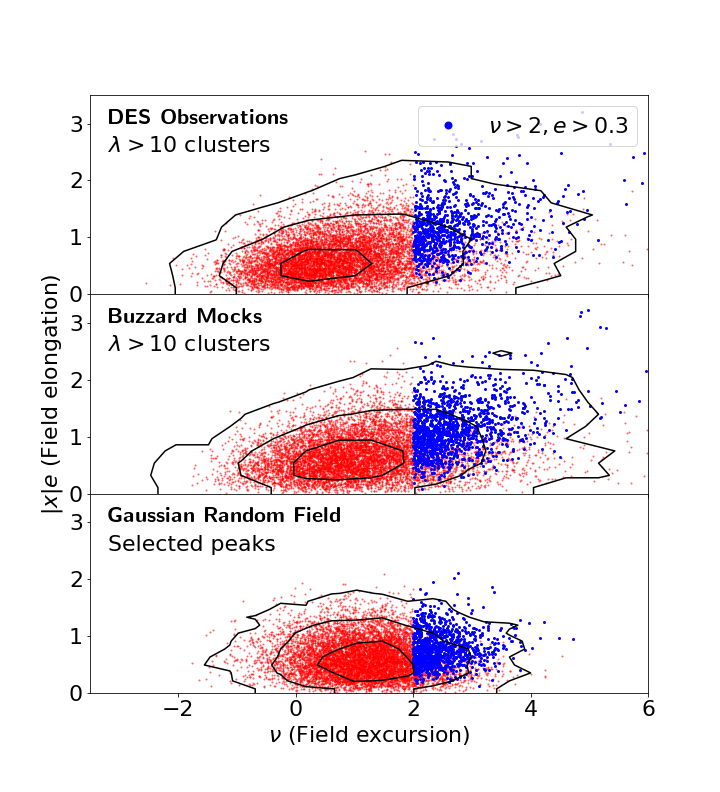}
\caption{The distribution of $\lvert x\rvert e$ (field elongation) versus $\nu$ (field excursion) for $\sim9,500$ points in the redshift bin $0.35<z<0.5$. From top to bottom, the distribution is shown for real RedMaPPer cluster positions in the RedMaGiC galaxy overdensity field, the same for Buzzard mocks of the DES data products, and the Gaussian Random Field with the same galaxy power spectrum ($C_\ell^{gg}$) as Buzzard. The maps have each been Gaussian smoothed with FWHM=14 Mpc. The entire cluster / peak sample consists of the red plus blue dots. The one, two, and three-sigma contours are drawn in black, and the points which are included after our chosen cuts of $\nu>2$ and $e>0.3$ are colored in blue. The cutoff is sharp at $\nu=2$ but less visually apparent for $\vert x \vert e$ because the threshold is in $e$ alone. These cuts select for high-superclustering regions in the real and mock galaxy fields.} \label{fig:nu_xe_comparison}
\end{figure}

\begin{figure}
\includegraphics[width=0.45\textwidth]{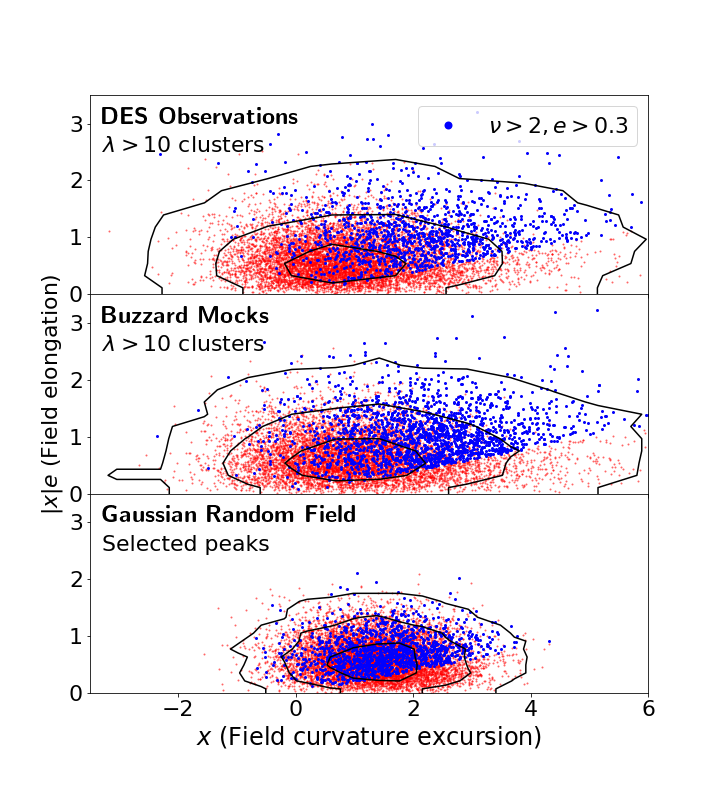}
\caption{The distribution of $xe$ (field elongation) versus $x$ (signal-to-noise of the curvature) for galaxy overdensity maps from DES observations (top), Buzzard mocks (middle), and a Gaussian Random Field (bottom). The chosen cuts which are used throughout much of this paper, $\nu>2$ and $e>0.3$, are shown in blue. The $\nu$ cut shifts the $x$ distribution rightwards because rarer overdensities also tend to have higher second derivatives. The $e$ threshold shows up here as a diagonal $\vert x \vert e$ threshold. \label{fig:x_xe_comparison}}
\end{figure}

Next, we examine the distributions of $(\nu, x, \vert x \vert e)$ for the galaxy overdensity fields smoothed on \textit{larger} scales. Figures \ref{fig:nu_xe_comparison} (\ref{fig:x_xe_comparison}) show the distribution for $\nu$ ($x$) versus $xe$ for the galaxy / GRF field smoothed at a 14 Mpc scale, at the chosen cluster / peak positions in the middle $z$ bin.

% Cluster locations with low values of $\nu$ are more likely to have high $e$ than points with high $\nu$. In other words, if a cluster lies at a large-scale dense region of the cosmic web, it is more likely to be at a peak (on that scale) and correspondingly have $e<0.5$. The less dense the region, the more likely a cluster will lie at a saddle point of the number density field. These characteristics may be able to be associated with large-scale matter flows (towards peaks and away from saddle points), the study of which will be the topic of future work.

In the real and mock galaxy fields, the clusters (red + blue points in the top two panels) display a $\nu$ and $x$ distribution which is more stretched compared to the GRF. The distributions are both skewed towards the high end, displaying more high-$\nu$ and high-$x$ points than the GRF which by definition has no skew. Additionally, the field elongation extends to larger values in the real and mock data than in the GRF.
We examine these distributions for a few galaxy field smoothing scales. For finer-grained smoothing, the distribution of GRF peaks is roughly the same but the skew of the real and simulated galaxy fields increases, especially in $\nu$. This is expected because with finer smoothing, the galaxy field is more non-Gaussian due to nonlinear structure formation. As the smoothing scale becomes coarser, the galaxy fields approach the GRF result.

We later demonstrate, in Section \ref{sec:parameter_dependence}, that enforcing a minimum $\nu$ and $e$ threshold for the RedMaPPer cluster sample enhances the supercluster gas signal in stacks. We choose $\nu>2, e>0.3$ as the optimal cuts (justified in Section \ref{sec:parameter_dependence}). Clusters satisfying this constraint are shown in blue in Figures \ref{fig:nu_xe_comparison} and \ref{fig:x_xe_comparison}. This selection furthers the distinction between the GRF and galaxy overdensity fields, as remaining points are more concentrated in the GRF in all properties. The field constraints select for clusters in highly overdense, elongated regions of coarse-grained galaxy maps: effectively, regions of strong superclustering. In Section~\ref{subsec:grf_comparison}, we examine the effect that these constraints have on the non-Gaussianity of oriented stacks.

\begin{figure}
    \centering
    \includegraphics[width=0.45\textwidth]{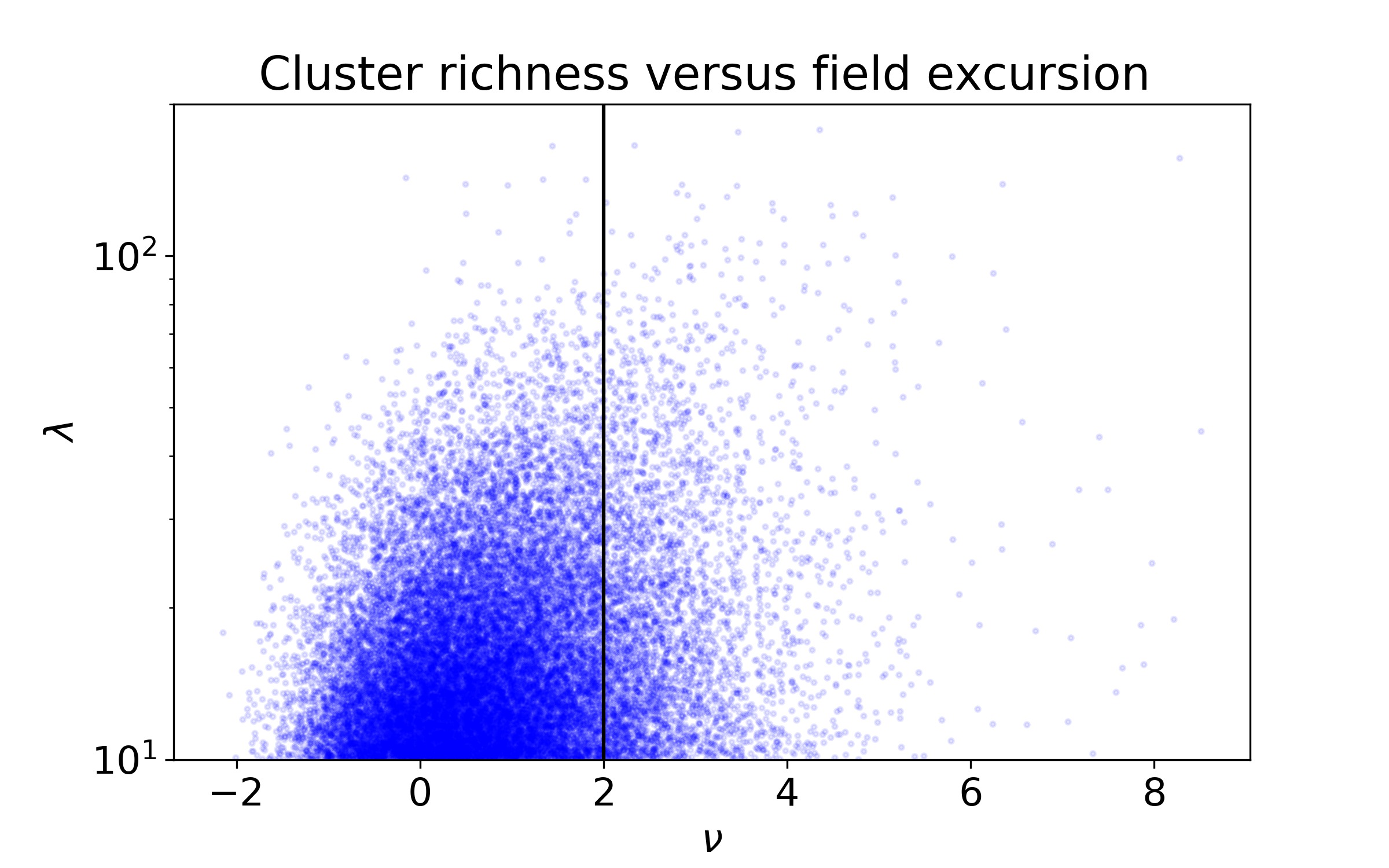}
    \caption{The relation between $\nu$, as measured on the galaxy overdensity field smoothed at 14 Mpc, and cluster richness $\lambda$. There is slight correlation, such that the $\nu>2$ clusters have a higher-skewed $\lambda$ distribution, but the constrained sample still features a range of $\lambda$.}
    \label{fig:nu_v_lambda}
\end{figure}

We also find that the $\nu$ and $e$ properties on the 14 Mpc scale are not highly correlated with cluster richness $\lambda$, a small-scale property. Figure \ref{fig:nu_v_lambda} demonstrates this: despite the weak correlation, the remaining clusters after the cut $\nu>2$ are still distributed across a wide range in $\lambda$. A higher-richness cluster is more likely to be in a higher large-scale overdensity, but low richness clusters may also belong to such regions. For example, a small cluster may lie on the edge of a supercluster and thus have a high $\nu$ value. For $e$ (not shown), there is a weak anti-correlation: clusters in very high-ellipticity regions tend to be lower-richness. This suggests that these small clusters are more likely to lie at saddle points in the field, such as between two massive overdensities, where $e$ is large due to its normalization. There is no correlation in the non-normalized $\vert x \vert e$ property.

\section{Stacking Methods}\label{Sec:Methods}
The stacking procedures were originally developed and implemented as part of the Cosmology Object Oriented Package \citep{Huang2016}, which we use in our stacking pipeline. This work is its first application to low-redshift objects (galaxies and clusters), whereas the program had been previously used for primary CMB analysis. 

 \subsection{Oriented Stacking with ACT and DES}
% Stacking is necessary to achieve a high signal-to-noise on superclustering gas.
In individual cutouts of clusters in the ACT tSZ data, only the most massive clusters are easily detectable by eye because the map is noise-dominated. The tSZ signal of lower-mass clusters, groups, and galaxies lies well below the noise and only emerges through stacking. In addition, stacking averages over the diversity of distributions and shapes of thermal energy along cosmic filaments. While the physics of individual superclusters is interesting in its own right, our goal is to measure the ensemble average of the anisotropic clustering of thermal energy around galaxy clusters.

Therefore, for each selected cluster sample, we stack cutout images from each Compton-$y$ map with orientation. Oriented stacking aligns and combines the gas signal from the most massive extended structures surrounding each cluster while driving down the noise (Figure \ref{fig:oriented_stacking_vis}).
% A visualization of the basic process is shown in Figure \ref{fig:oriented_stacking_vis}.

% \begin{figure}
% \gridline{\includegraphics[width=0.4\textwidth]{unoriented_stacking.png}{}}
% \gridline{\includegraphics[width=0.4\textwidth]{oriented_stacking.png}{}}
% \caption{A visualization of unoriented (top) versus oriented (bottom) stacking. When stacking cluster cutout images from the Compton-$y$ map with random orientation, the signal from external structure is averaged out such that the final image, as the number of cutouts increases, approaches perfect isotropy. In oriented stacking, we determine the direction of strongest alignment around each cluster using the curvature of galaxy overdensity maps. We rotate each cutout image before stacking such that the excess $y$ signal from superclustering is overlaid.}\label{fig:oriented_stacking_vis}
% \end{figure}

\begin{figure}
    \plotone{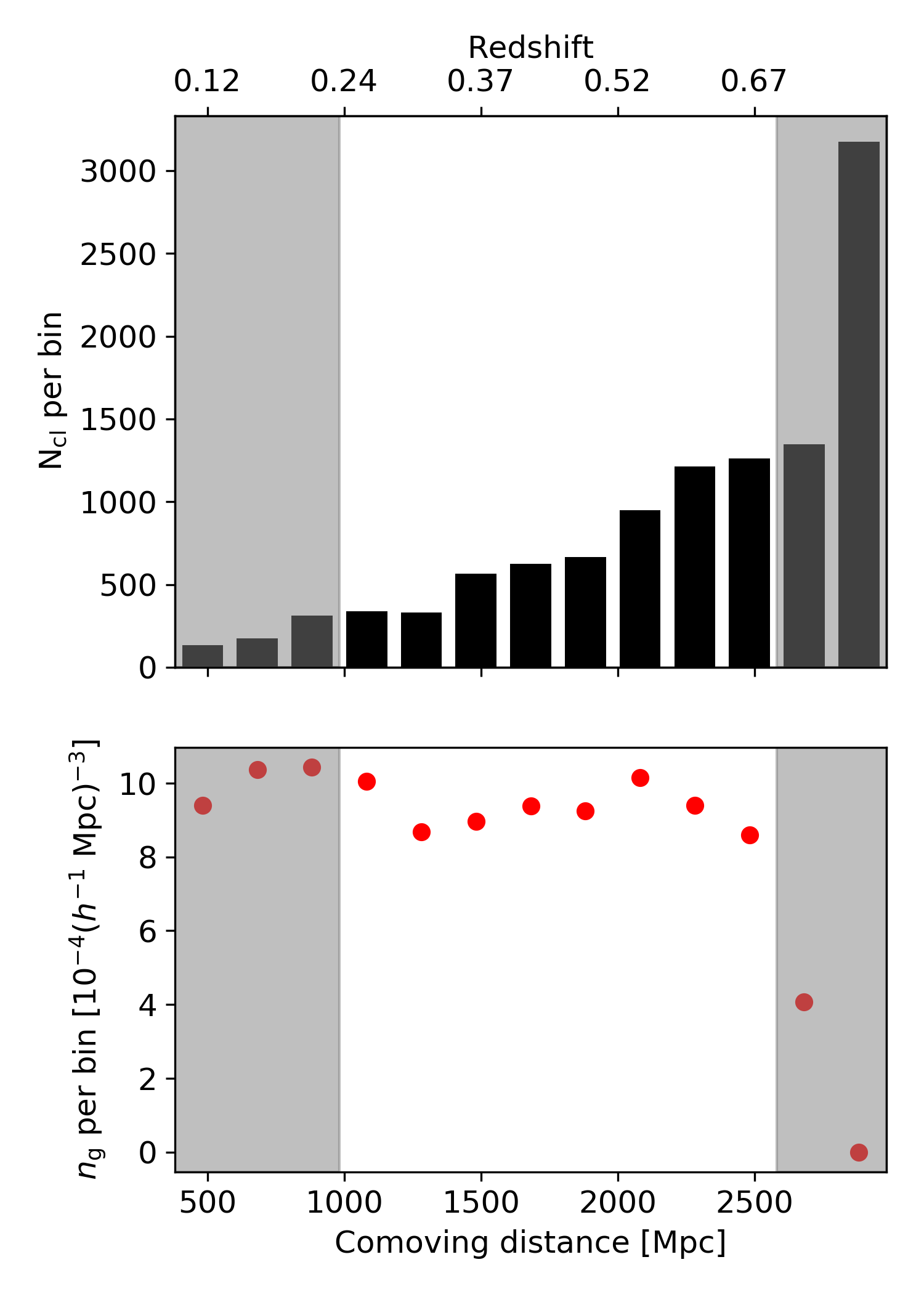}
    \caption{The number of RedMaPPer clusters (top) and projected comoving number density of redMaGiC galaxies (bottom) in each 200 Mpc-wide bin. Gray shaded regions cover data not used in this work. The total cluster number per bin is relevant because the signal-to-noise ratio scales as $\sqrt{N_{cl}}$ for a stack. $N_{cl}$ generally rises with redshift because the same sky area covers a progressively larger physical area. For galaxies, the comoving number density is the relevant quantity, as the available large-scale structure information around each cluster determines its orientation. The number density is relatively constant until $z\sim0.7$, the imposed edge of this volume-limited sample. We cut the data at the dropoff. We also cut the lowest-redshift data because 4x4 degree cutouts do not span a large enough physical extent to observe the full effects from superclustering at these redshifts.}
    \label{fig:redshift_dist}
\end{figure}

Applying the method to our observational data begins by dividing the selected \redmapper clusters and all \redmagic galaxies into equal-sized slices in comoving distance along the line-of-sight direction. Each slice is 200 Mpc thick. The thickness is chosen to minimize projection effects from uncorrelated structure, yet ensure that most clusters and galaxies are placed in the correct redshift slice given that the DES photo-$z$ uncertainties are $\sigma_z\sim0.01-0.02$ ($\sim30-60$ Mpc at the redshifts of interest). The distributions of the cluster and galaxy data within the slices are shown in Figure \ref{fig:redshift_dist}.

In each slice, we create a smoothed projected galaxy overdensity map as described in Sec.~\ref{sec:galaxy_field_characteristics}. The local Hessian matrix (Equation \ref{eq:Hessian}) of the galaxy overdensity at the position of each cluster provides information on the strongest axis of anisotropic clustering. The eigenvector $\boldsymbol{v_2}$ with eigenvalue $\lambda_2$ points along the axis of slowest change, which we define to be the superclustering axis. The choice of smoothing scale defines the characteristic radius at which this axis is identified; thus for the same galaxy map smoothed at different levels, the eigenvector for a given cluster can rotate. The vector will point along near-cluster structure, inter-cluster filaments, and superclusters as the map is increasingly smoothed.

For each cluster, we take a $4^\circ\times4^\circ$ cutout from the ACT Compton-$y$ map centered on the cluster (RA, dec). At the redshift range explored in this work, this corresponds to a coverage of $\sim80$ Mpc (closest slice) to $\sim120$ Mpc (furthest slice) on each side of the square cutout. Thus, if the central cluster is a member of a supercluster, the cutout should almost always contain the entire structure (see \citealt{Borgani1995} for a detailed discussion of supercluster scales). The square cutout is oriented along the superclustering axis, and each cutout is rotated so as to align the superclustering axis of all cutouts along the horizontal axis of the stacked image. After alignment, the final stack is the average of all the cutouts. The SNR of the final image is proportional to $\sqrt{N}$, where $N$ is the total number of stacked cutouts. The cutout, orientation, and stacking processes are implemented with the \texttt{GetPeaks} and \texttt{Stack} programs from the Cosmology Object Oriented Package\footnote{\url{https://www.cita.utoronto.ca/~zqhuang/work/coop.php}} \citep{Huang2016}.

% We weight all clusters equally in our stacking process, but future work could apply a richness weighting. 
% This process not only builds up signal from the gas in the central clusters, but also from the extended structure (neighboring clusters, galaxy groups, filament gas) along the horizontal axis. 

% Symmetry-breaking takes orientation one step further and not only aligns each patch along the long axis of curvature, but also identifies the asymmetry along this axis. The patch is then flipped so that the direction in which the majority of material lies is on the right side of every stack component; this produces a stronger signal on that side of the stacked image.

\subsection{Choice of smoothing scale} \label{subsec:smoothing}
Multiple smoothing scales are explored in Sections~\ref{sec:parameter_dependence}~and~\ref{Sec:Results}; they span a range of Gaussian FWHM from $6-18$ Mpc. The upper end of this range is motivated by recent cluster-pair stacking studies which revealed filament signal from pairs separated by a transverse distance of $6-10 h^{-1}$ Mpc \citep{Tanimura2019}. Considering that our stacking method centers on single clusters, the simplest way to enforce orientation at these typical scales would be to apply top-hat smoothing to the galaxy overdensity map with a radius of $R_{\mathrm{TH}}\sim6-10 h^{-1}$ Mpc. A simple conversion from top-hat to Gaussian filter involves equalizing the volume under each function; with a top-hat of height 1 and a Gaussian with amplitude 1, this leads to FWHM$\sim1.7 R_{\mathrm{TH}}$. Thus, for cluster pairs separated by, e.g., $\sim7 h^{-1}$ Mpc, Gaussian smoothing for which the Hessian would best encode alignment at that scale has FWHM$\sim12 h^{-1}$ Mpc or $\sim18$ Mpc.

We therefore choose 18 Mpc as our key scale to study with ACT$\times$DES data. We also select a range of smaller smoothing scales, down to FWHM=6 Mpc, to apply to both observational and simulated data in order to study aligned superclustering on highly nonlinear scales. Finer-grained smoothing causes the orientation to be determined primarily using inter-cluster filament galaxies and groups.
% The finest-grained smoothing, at FWHM=3 comoving Mpc (corresponding to $R_{\mathrm{TH}}\sim1.1$ physical Mpc in our redshift range), probes inside the interior of clusters. 

For each chosen comoving smoothing scale, the corresponding angular size is determined at the center of each slice. The angular size varies from map to map such that the FWHM in comoving Mpc is held constant across all redshifts.

\subsection{Redshift Slice Combination}
Ideally, it would be most interesting to observe the change in the average superclustering signal with cosmic time by comparing multiple redshift slices. We explore this in simulations in Section \ref{sec:expectations_from_theory}; however, we find that the signal-to-noise in the real data analyzed here is too low for each individual 200 Mpc slice. For the observed data we therefore combine stacks over nearly the full range of redshifts available with the DES data: $\sim0.25< z < 0.72$, or $\sim$1000--2600\,Mpc.

Each stack is the same angular size ($4^\circ\times 4^\circ$) by construction, and therefore stacks on structure in more distant slices span a larger transverse comoving distance. To properly combine multiple slices at different redshifts, we must first adjust all images to the same comoving size. We begin by calculating the transverse comoving size of the nearest image at the midpoint of the slice in the line-of-sight direction ($\sim1100$ Mpc). At this distance, $4^\circ$ spans $\sim80$ Mpc. For more distant slices, the stacks are cropped to the angular size which spans 80 Mpc at each slice midpoint and rescaled to the same pixelization via interpolation. Finally, all images are averaged together. By using the slice midpoints for the conversion from angular to transverse comoving size, we make the approximation that the slices are very thin, while in reality the conversion varies across the 200 Mpc slice thickness.

\begin{figure*}
\gridline{\includegraphics[width=0.33\textwidth, trim={2cm 0cm 1.2cm 0cm},clip]{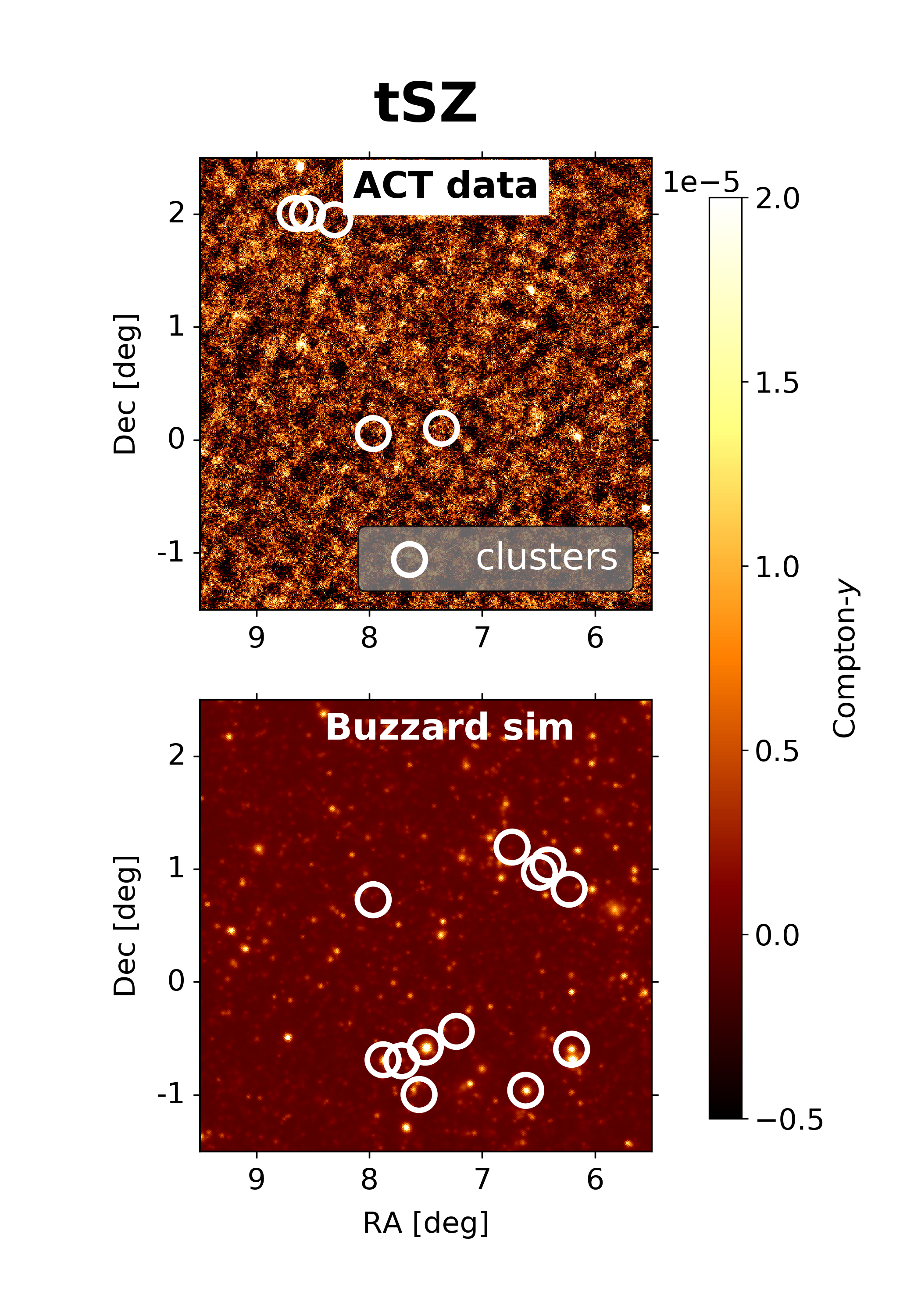} \includegraphics[width=0.33\textwidth, trim={4.3cm 0cm 1.35cm 0cm},clip]{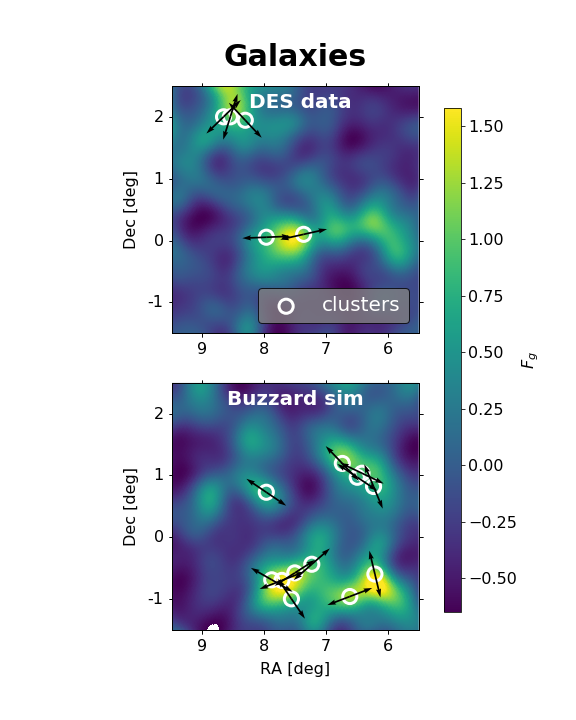}
\includegraphics[width=0.32\textwidth, trim={2.9cm 0cm .8cm 0cm},clip]{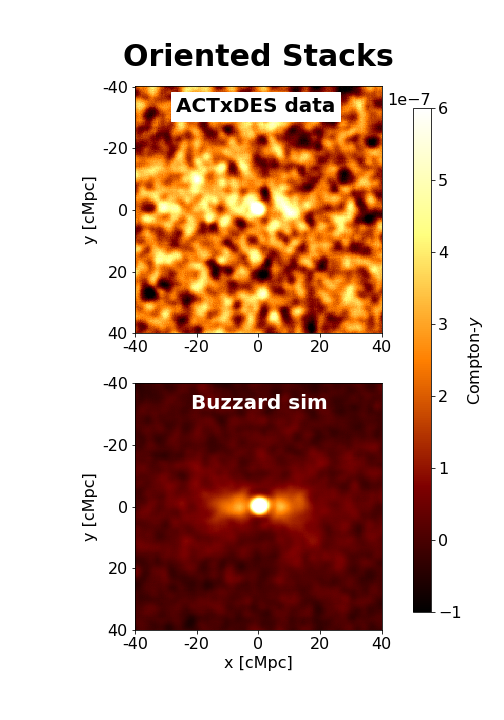}}
\caption{The ingredients for making an oriented stack in the slice [1632,1832] Mpc ($z$=[0.41, 0.47]). Left: a 4x4$^\circ$ sky patch from the observed (above) and simulated (below) Compton-$y$ maps. At this redshift, the corresponding physical size is $\sim180$Mpc. RedMaPPer cluster locations within the slice are circled in white. The ACT $y$ map is noise-dominated, whereas the Buzzard simulated map is noiseless. Both $y$ maps display the projected tSZ signal from all contributing sources along each line-of-sight. Because of this projection, as well as the dependence of tSZ strength on cluster mass, not all circled clusters are easily detectable by eye. Middle: Observed (above) and simulated (below) galaxy overdensity maps, Gaussian smoothed with FWHM=14 Mpc, in the same sky area. The same clusters are circled and black arrows indicate their orientation with respect to the surrounding large-scale structure, as determined by the Hessian matrix on $F_g$. Right: Oriented stacks made by a combination of multiple distance slices in the range 1032--2632 Mpc. Each stack is made by taking cutouts from the $y$ map centered on cluster locations and rotating it with the information from $\delta_g$ before stacking. This builds up signal-to-noise along the horizontal axis, showing the gas signal from superclustering. \label{fig:3-panel-figure}}
\end{figure*}

Figure \ref{fig:3-panel-figure} displays both the observational and simulated ingredients and outputs of the stacking process.
% The panels show a 16 square degree patch of the $y$ maps, the cluster positions within galaxy maps in a single distance slice, and a combined stack from the entire redshift range.

\subsection{Multipole Decomposition}
\begin{figure*}
    \centering
    \includegraphics[width=1\textwidth, trim={0cm 0cm 0cm 0cm}]{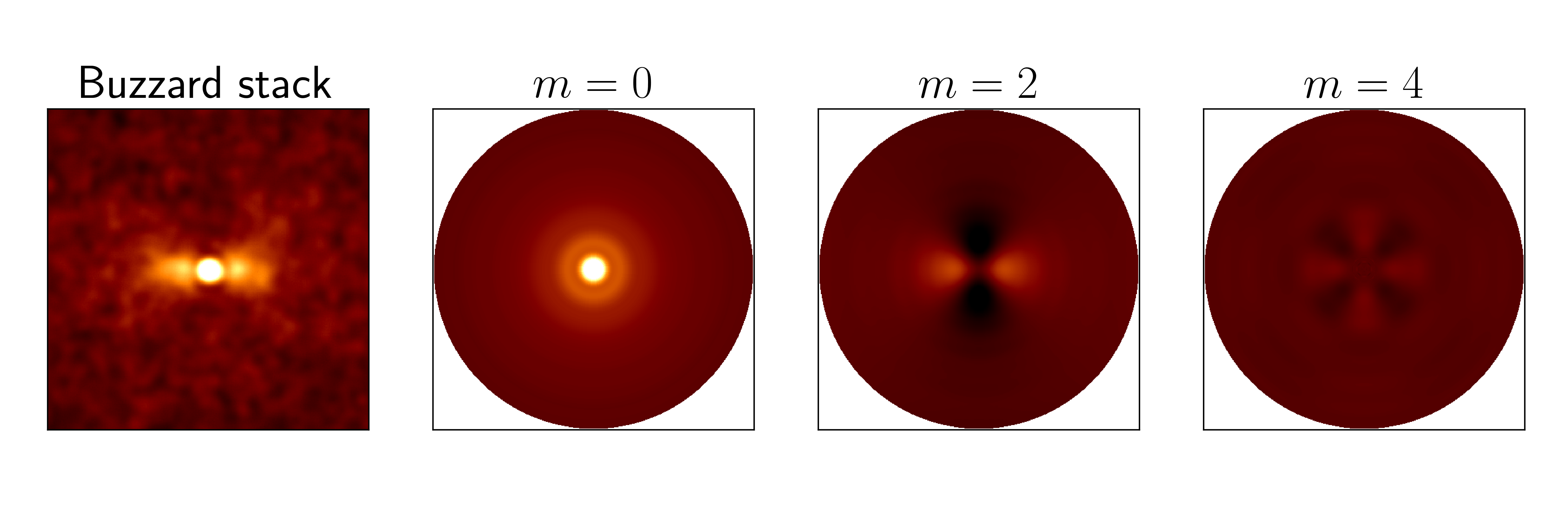}
    \caption{A representation of the three lowest even multipolar moments of an oriented stack from the Buzzard $y$ map (left). The stack image is passed as $I$ through Equation \ref{eq:multipole_moments} to get $C_m(r)$. The odd $m$ moments trend towards zero with increased numbers of stacked cutouts. For the even moments, we show the symmetrized representation $C_m(r) \cos{(m\theta)}$. Power in the $m=4$ moment comes from the horizontal wings of the stacked image, but gets evenly redistributed to four poles in this representation.}
    \label{fig:multipole}
\end{figure*}

To quantitatively compare the stacked images, we decompose each image into its multipole components. Each image $I$ can be deconstructed as
\begin{equation}
    I(\theta,r) = \sum_m \left (  C_m(r) \cos(m \theta)+S_m(r) \sin(m \theta) \right ).
\end{equation}
Since symmetry along the x-axis is enforced in the oriented stacking method, the odd components trend towards zero as the number of component images grows. The odd moments are consistent with zero in our results. For even $m$, due to the alignment along the x-axis, we are interested in only the cosine component. The sine term is like a noise term and fluctuates around zero when the number of stacked images is large.  The radial profile of the cosine component is taken by
\begin{equation}\label{eq:multipole_moments}
    C_m(r) = \frac{1}{X\pi}\int_0^{2\pi} d\theta F(\theta,r) \cos{(m \theta)},
\end{equation}
where X is 2 for $m=0$ and 1 for all other $m$.
The multipole decomposition is visualized in Figure \ref{fig:multipole}. Throughout the rest of the paper, we focus on a comparison of $C_2(r)$ and $C_4(r)$.

All $m=2$ and $m=4$ profiles shown in this paper feature a rise, peak, and fall. This feature is almost entirely dependent on the smoothing scale, as we later demonstrate in Section~\ref{sec:parameter_dependence}. In short, determining the orientation of each cluster with a galaxy overdensity map smoothed at a chosen scale enforces a radius at which structure is maximally aligned between all stacked images, translating to a peak at that radius.

Because the $m=4$ moment sums signal not only from the horizontal image axis, which contains signal by construction, but also from the vertical axis which contains only noise, $C_4(r)$ is expected to be noisier than $C_2(r)$.

\subsection{Uncertainties}\label{subsec:uncertainties}

Uncertainties in the stacked profiles are expected to stem from a variety of sources including random noise, contamination of the $y$ map from other components such as dust and the primary CMB, and photometric redshift uncertainties in the DES data.

We characterize the uncertainties by estimating the covariance matrix, $\Sigma$, of each $C_m(r)$ profile. If a profile is binned into radial bins $r_i$, each element $\Sigma_{ij}$ is the covariance between the signal in the $i$th and $j$th radial bins, and the elements along the diagonal of the matrix are the variances of the signal in each bin. We attempt two different methods to estimate $\Sigma$. Method 1 is sufficient for the simulated data but insufficient for the real data. In method 1, we split the clusters into $N_{reg}$ separate regions on the map where $N_{reg}$ ranges from 12 to 48 depending on the data set and cluster selection. Cluster samples in all the regions are approximately equal-sized and collected in (RA, Dec) space into patches $30-80$ deg$^2$ in area, depending on $N_{reg}$. The clusters in each region are stacked with orientation, where the orientation is given by the full $F_g$ map such that information beyond the region edges can be incorporated. Splitting the data into spatially separate regions is motivated by the fact that our measurement contains both spatially correlated noise from large-scale structure fluctuations (data and simulations) and long-wavelength noise from residual low-$\ell$ primary CMB contamination (data only). With large-area regions, each sample is reasonably independent from the rest, save for some inevitable LSS and long-wavelength noise overlap between neighboring regions.

The number of regions and the angular area per region depends on the data set. For Websky full-sky maps, we split the sample into 48 Healpix pixels. For the Buzzard and DES data, we use the \texttt{kmeans\_radec} algorithm\footnote{https://github.com/esheldon/kmeans\_radec} to split the clusters into approximately equal-sized regions on the sky. The full $\lambda>10$ cluster sample is split into 16 regions in DES and 48 regions in Buzzard. Adding in the field constraints $\nu>2$ and $e>0.3$ shrinks the number of clusters, so when stacking this constrained sample we reduce the number of splits to 12 for DES and 24 for Buzzard.

After stacking the clusters in each region, we have $N_{reg}$ stacks. We decompose each stack into multipole moments $m$ and measure the Compton-$y$ radial profiles $C_m(r)$. Each profile is binned in radius to make a data vector $\vec{C_m}=C_m(r_i)$, where $i\in(1,N_{bin})$. We combine the $N_{reg}$ profiles into a $N_{bin}\times N_{reg}$ matrix, called $X$. In calculating the covariance matrix, we weight each region by $w_p=N_{cl,p}/\overline{N_{cl}}$ where $N_{cl,p}$ is the number of clusters in the $p^{\mathrm{th}}$ region and $\overline{N_{cl}}$ is the average over all regions. If $X$ becomes the modified matrix $M$ by subtracting the weighted average of each bin across all regions (i.e., the row mean), so that element $M_{ip}$ is given by
% Sums and implicit summation combined version
% \begin{equation} \label{eq:mean_subtracted_matrix}
%     M_{ip} = X_{ip}-\frac{X_{ip'}
%     w_{p'p}}{\sum_{p'}^{N_{reg}}w_{p'}},
% \end{equation}
% using Einstein summation notation for the repeated $p'$.

% \begin{equation} \label{eq:covmat_regions}
%     \Sigma_{ij} = \frac{ M_{ip'} w'_{p'q}M^T_{qj}  \sum_{p'}^{N_{reg}}w_{p'}}{(\sum_{p'}^{N_{reg}}w_{p'})^2-\sum_{p'}^{N_{reg}}w_{p'}^2}.
% \end{equation}

% \begin{equation} \label{eq:mean_subtracted_matrix}
%     M_{ip} = X_{ip}-\frac{X_{ip'}
%     w_{p'p}}{w_{p'p'}},
% \end{equation}
% using Einstein summation notation for the repeated indices.

% \begin{equation} \label{eq:covmat_regions}
%     \Sigma_{ij} = ( M_{ip'} w'_{p'q}M^T_{qj}) \frac{ w_{\ell \ell}}{(w_{\ell \ell})^2-w_{\ell \ell}^2}.
% \end{equation}

\begin{equation} \label{eq:mean_subtracted_matrix}
    M_{ip} = X_{ip}-\frac{\sum_{p'=1}^{N_{reg}}X_{ip'}
    w_{p'}}{\sum_{p'=1}^{N_{reg}} w_{p'}},
\end{equation}
then the covariance matrix element between bins $r_i$ and $r_j$ is

\begin{equation} \label{eq:covmat_regions}
    \Sigma_{ij} = \left[ \frac{ \sum_{p'=1}^{N_{reg}}w_{p'}}{(\sum_{p'=1}^{N_{reg}}w_{p'})^2-\sum_{p'=1}^{N_{reg}}w_{p'}^2}\right] \sum_{p'=1}^{N_{reg}} (w_{p'} M_{ip'} M_{jp'}).
\end{equation}
The left bracketed term is a normalization by the degrees of freedom that would simply be $1/(N_{reg}-1)$ if there were no observation weights $w_p$. Expressing the rightmost unbracketed sum in words, the weight $w_p$ is applied to every radial bin in the zero-mean profile of the $p^{\mathrm{th}}$ region, and the result is matrix-multiplied with the transpose of $M$ (generally, $(M\times M^T)_{mn}=\sum_l M_{ml}M_{nl}$). In practice, we calculate $\Sigma$ with the \texttt{NumPy} \texttt{cov}\footnote{\url{www.numpy.org/doc/stable/reference/generated/numpy.cov.html}} function. $\Sigma$ as defined above represents an estimate of the covariance of a single region's $y$ profile which was `observed' $N_{reg}$ times, so we further divide $\Sigma$ by $N_{reg}$ to achieve our estimate of the covariance of the \textit{full} map data.

The map-split method works sufficiently well for the Buzzard and Websky simulations, which both can be split into a larger number of regions and have more clusters per region than in the real data. However, we find that there are not enough possible sub-regions of the smaller-footprint ACT $y$ map to achieve convergence of the covariance matrix. With only 12 regions and 3--5 radial bins, we find that the position and number of bins significantly affects the resulting $\chi^2$ and signal to noise estimates.

Due to the lack of convergence, we apply a different method to the observed data. In method 2, we assume that each covariance matrix for the final ACT $y$ profiles can be decomposed into the sum of components:

\begin{equation}
    \Sigma_{\mathrm{tot}} = \Sigma_{\mathrm{noise}} + \Sigma_{\mathrm{signal}},
\end{equation}
where $\Sigma_{\mathrm{noise}}$ refers to the covariance matrix from all non-signal components in the $y$ maps, and $\Sigma_{\mathrm{signal}}$ refers to the covariance matrix which would emerge from a noiseless $y$ map. 
%  For the latter, we assume that the Buzzard mocks reasonably approximate the real-data covariance matrix, modulo differences in the cosmology and gas prescription which are subdominant to the other contributions.

To estimate the former, and dominant, source of error ($\Sigma_{\mathrm{noise}}$), we use 120 simulated ACT Compton-$y$ maps described in Section VII of \citet{Madhavacheril2020}. Each map contains an independent realization of Gaussian $y$ signal generated with a tSZ power spectrum as well as independent realizations of the estimated noise contribution from all known contaminants. Each map covers the D56 footprint. The contaminants include the low-$\ell$ primary CMB, high-$\ell$ instrument noise, and high-$\ell$ residual foregrounds. We subtract the tSZ realization from each map so that only the noise contribution remains.
We estimate $\Sigma_{\mathrm{noise}}$ by generating a stack for each different map. The noise in the simulated maps is uncorrelated with true cluster and galaxy positions, so for each map we stack on a sample of random points which is the same size as the cluster sample and apply random orientation. We bin the data to sample the radial profile in positions of interest, and the resulting covariance matrix emerges from $N_{map}=120$ independent stacks. Because the number of images per stack is constant, Equations \ref{eq:mean_subtracted_matrix} and \ref{eq:covmat_regions} simplify to:

\begin{equation}
    M_{ip} = X_{ip} - \frac{\sum_{p'=1}^{N_{map}} X_{ip'}}{N_{map}};
\end{equation}

\begin{equation}
    \Sigma_{noise, ij} = \frac{\sum_{p'=1}^{N_{map}} M_{ip'} M_{jp'}}{N_{map}-1}.
\end{equation}

% We take cutouts from every simulated map, rotate them by a random angle, then stack them. We match the number of \redmapper clusters per slice and apply the rescaling. 

We estimate $\Sigma_{\mathrm{signal}}$, the covariance matrix of the signal component of the measurement, as being equal to the Buzzard map-splits covariance matrix (calculated by Eq.~\ref{eq:covmat_regions}) scaled by $N_{\mathrm{Buzz}}/N_{\mathrm{DES}}$ (the relative number of clusters in Buzzard versus DES). We compute $\Sigma_{\mathrm{signal}}$ for each separate smoothing scale that we apply to the galaxy maps; therefore it includes the correlation induced between radial bins by smoothing. This is an imperfect estimate of the true signal variances and covariances due to the known inaccuracies in Buzzard's mass-richness relation and our approximate, prescriptive approach to adding gas to the simulation. We compute $\Sigma_{\mathrm{tot}}=\Sigma_{\mathrm{signal}}+\Sigma_{\mathrm{noise}}$ and examine which component contributes more to the total. $\Sigma_{\mathrm{noise}}$ dominates the variances of the signal in each radial bin, as its diagonal is $\sim3-6\times$ larger than the diagonals of $\Sigma_{\mathrm{signal}}$. However, the off-diagonal covariances are similar in magnitude. Therefore, although the noise contribution is more important overall, any inaccuracies in the Buzzard simulation may have a non-negligible impact on $\Sigma_{\mathrm{tot}}$. Future work using expanded ACT sky coverage will avoid this concern by estimating uncertainties through data only.

When working with simulations to determine general theory expectations in Section~\ref{sec:expectations_from_theory}, we use arbitrarily fine binning for the radial profiles $C_m(r)$. However, when making robust comparisons between ACT$\times$DES and Buzzard in Section~\ref{Sec:Results}, we choose the binning more carefully. We determine the convergence of $\Sigma_{\mathrm{noise}}$ for $n$ bins by examining the matrix and its inverse as the number of contributing maps increases from 60 to 100 to 120. We also examine the condition number (the ratio of the largest to smallest eigenvalue) of $\Sigma_{\mathrm{tot}}$. We find instability and a high condition number for 5 or more bins when using the constrained cluster sample, because the Buzzard covariance matrix was calculated using only 24 separate map regions. Therefore we select 3 bins for stable results and useful placement along the radial profile. It may be possible to achieve a more stable finely-binned covariance matrix with alternative uncertainty methods, but we leave that for future work with larger ACT$\times$DES sky coverage.

% Since this covariance matrix is not only necessary for comparison with the Buzzard results, but is also a component of the estimated ACT$\times$DES covariance where it contributes most of the off-diagonal correlations for $m=2$ and $m=4$, binning more finely causes fluctuations in the SNR and $\chi^2$ estimates for both simulations and data. 

We note that the DES photo-$z$ uncertainties are expected not only to contribute to the overall uncertainty of the anisotropic stacked $y$ signal, but also to bias the signal in a redshift-dependent manner. The contribution to uncertainty is accounted for within method 2: $\Sigma_{\mathrm{signal}}$ includes variance across small samples of different photo-$z$ realizations for galaxies and clusters within Buzzard. Buzzard reproduces the photo-$z$ uncertainties of the DES catalogs well, shown in D19.

The redshift-dependent bias, however, is unaccounted for in the uncertainties. Due to photo-$z$ error, for any redshift slice, some objects near the edge are not included and some interlopers just outside the slice are included. This results in a distribution of the true redshifts of objects which is concentrated toward the middle of each slice with tails extending beyond the slice edges. On average, $\sim25\%$ of galaxies and $\sim20\%$ of clusters are interlopers in a given slice. The combination of the peaked distribution and the inclusion of interlopers has a non-trivial effect on the determination of orientation, as it both increases the correlation of objects within the slice but also adds in distant, uncorrelated structure. We test the effect this has on the anisotropic stacked $y$ signal using the Websky simulations. For each redshift slice, we transform the true Websky halo redshifts to photometric redshifts by drawing from a Gaussian $z$ distribution with the average $\sigma_z$ of galaxies / clusters in DES. We then compare the $m=2$ radial $y$ profiles from a stack using photo-$z$s versus a stack using true $z$s. We find that the profile can be boosted, decreased, and/or shifted in radial units due to the use of photo-$z$s, depending on the redshift slice. The most significant effect is in the most distant slice, where there is a decrement of $\sim25\%$ in the maximum height of the $y$ profile. To make physical inferences from the values of our observational results, these effects must be fully characterized and corrected for, which will be one of the goals of our succeeding paper. However, they do not impact the conclusions of this paper as we limit our comparison to only Buzzard vs. ACT$\times$DES. Due to Buzzard's accurate reproduction of the photo-$z$ uncertainties from DES, oriented stacks using Buzzard will be biased in the same manner. We do not attempt to compare Websky directly with Buzzard or ACT$\times$DES.

\subsection{Comparison with a Gaussian Random Field}\label{subsec:grf_comparison}

\begin{figure*}[htbp!]
    \centering
    \gridline{\includegraphics[width=0.33\textwidth]{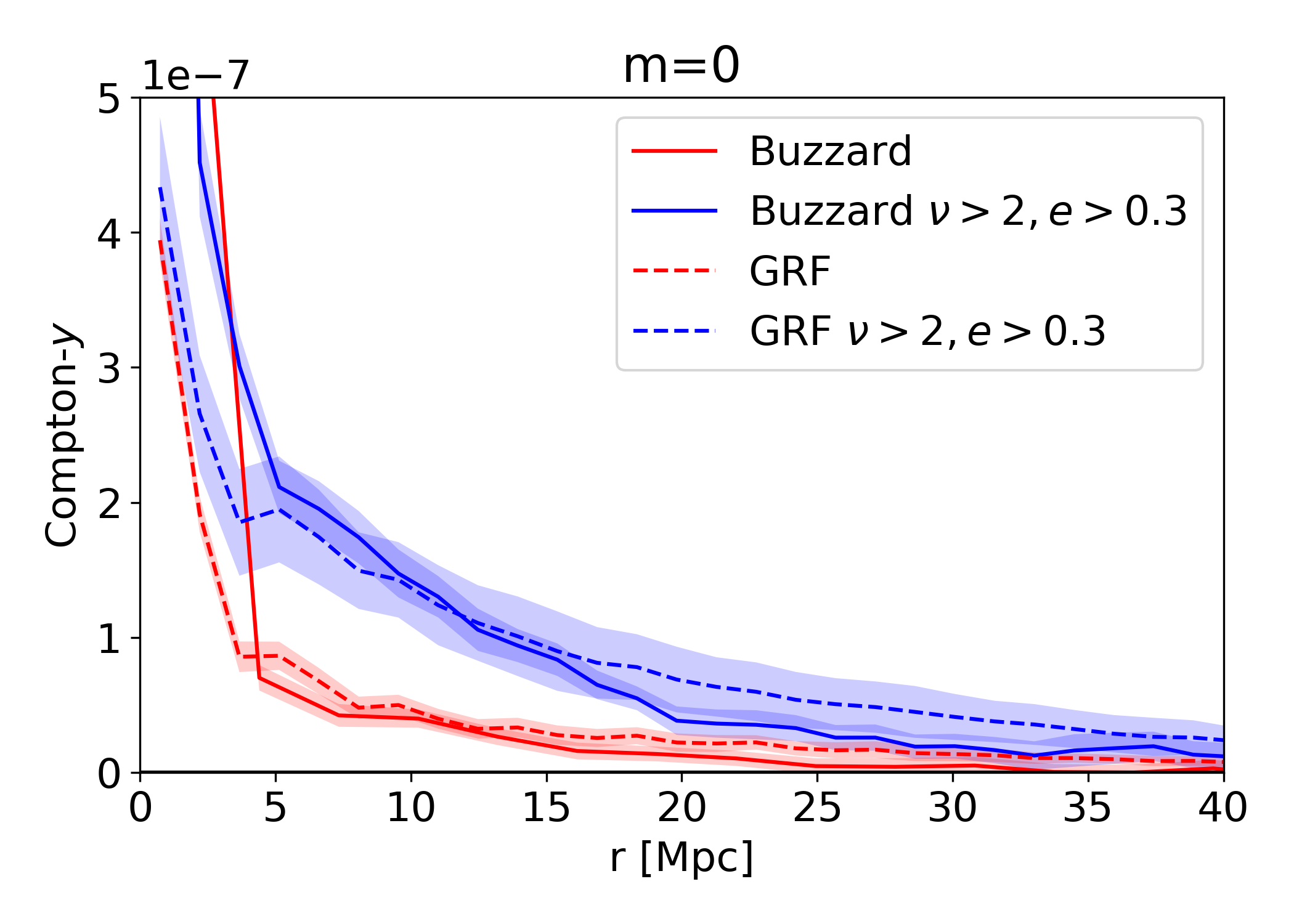}{}
    \includegraphics[width=0.33\textwidth]{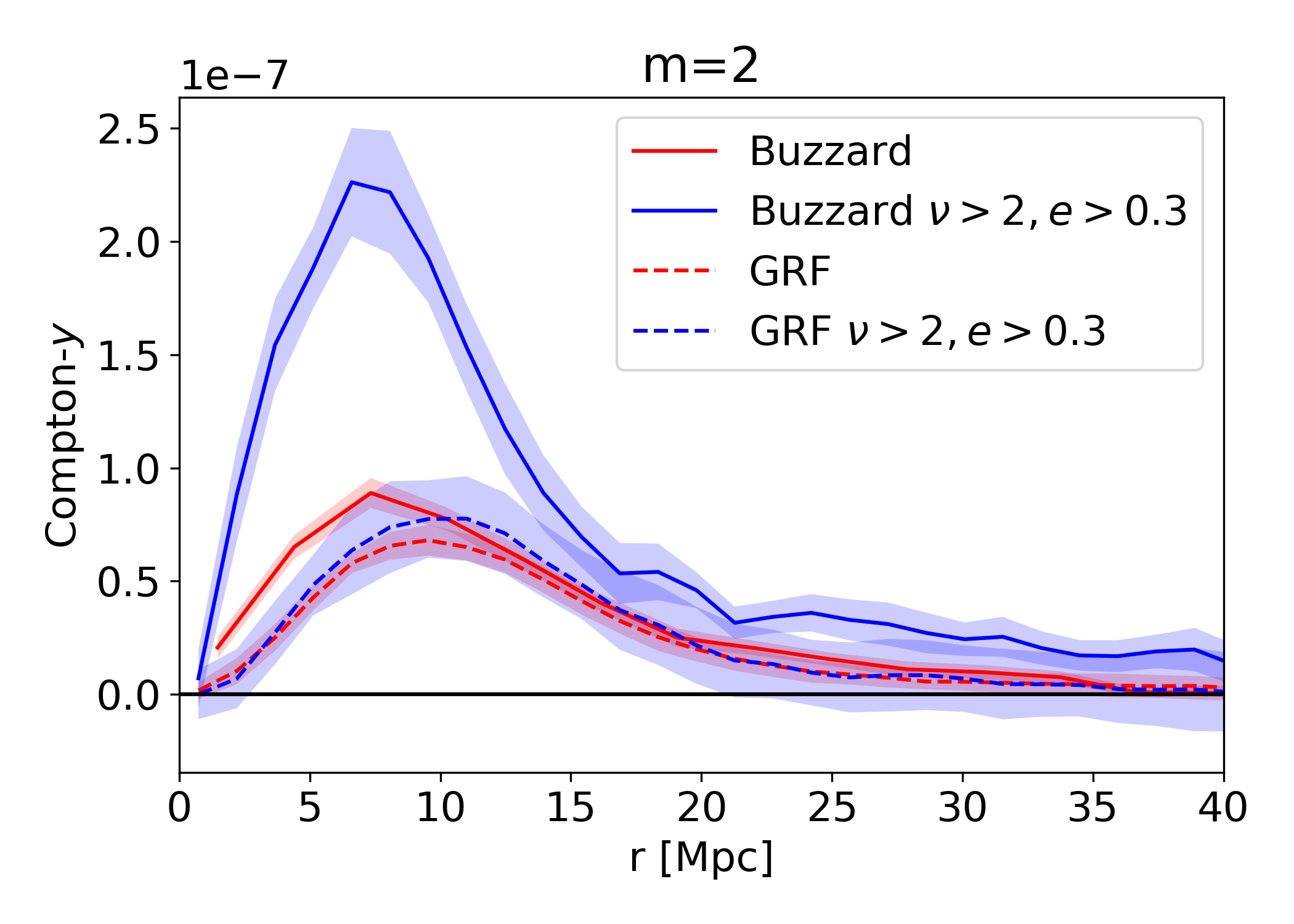}
    \includegraphics[width=0.33\textwidth]{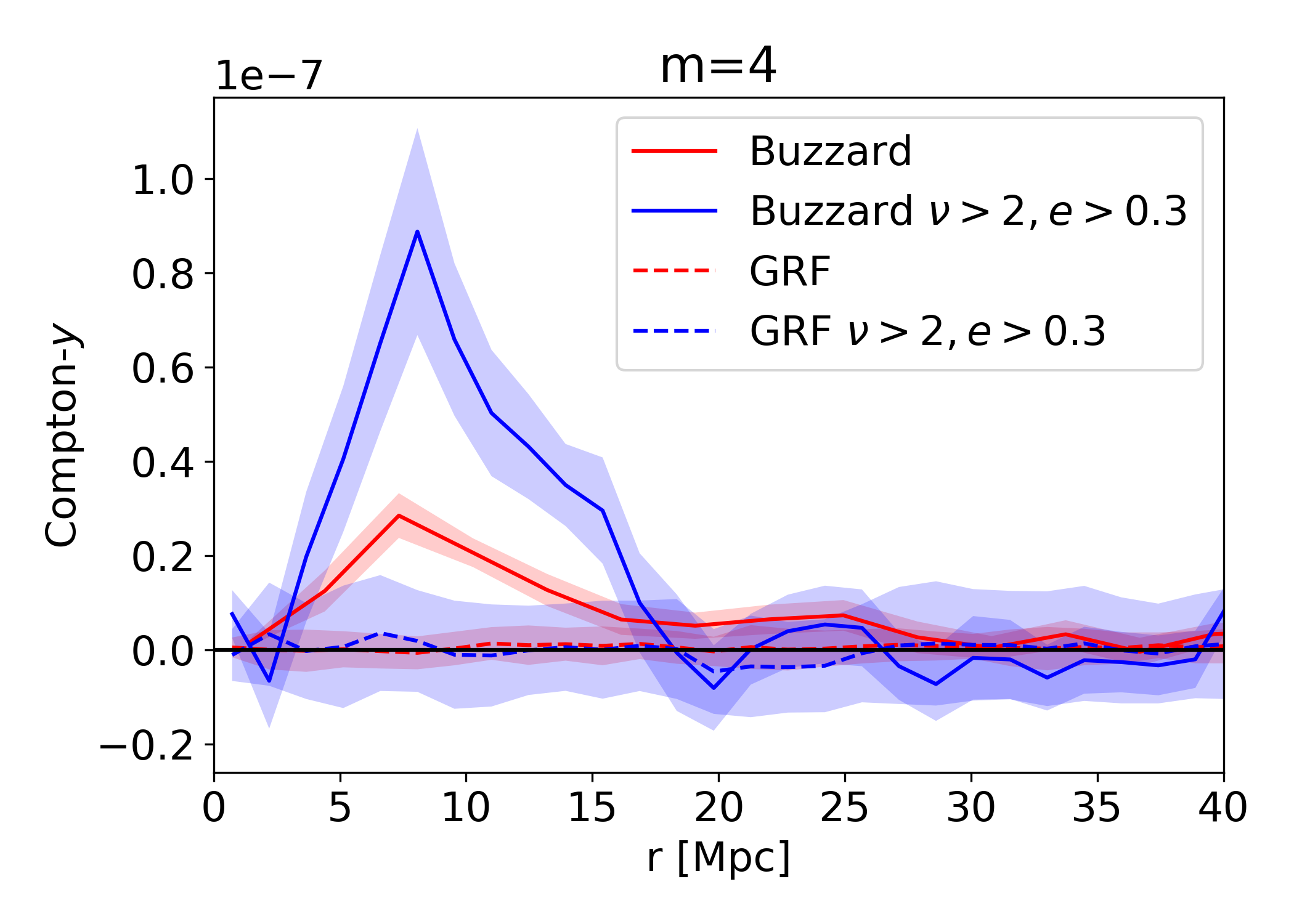}}
    \gridline{\includegraphics[width=0.5\textwidth,trim=1.5cm 0cm 3cm 1.3cm, clip]{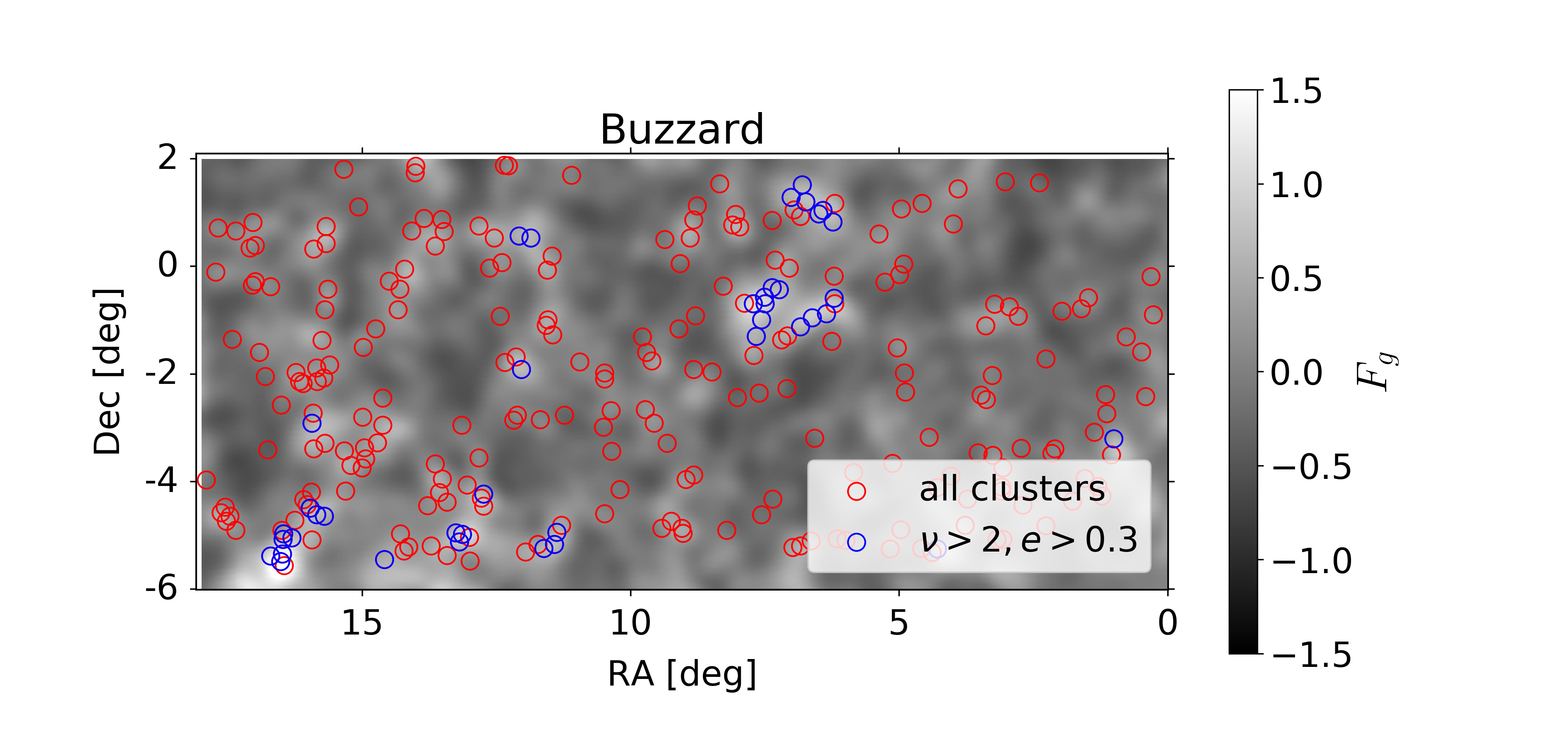}{}
    \includegraphics[width=0.5\textwidth,trim=1.5cm 0cm 3cm 1.3cm, clip]{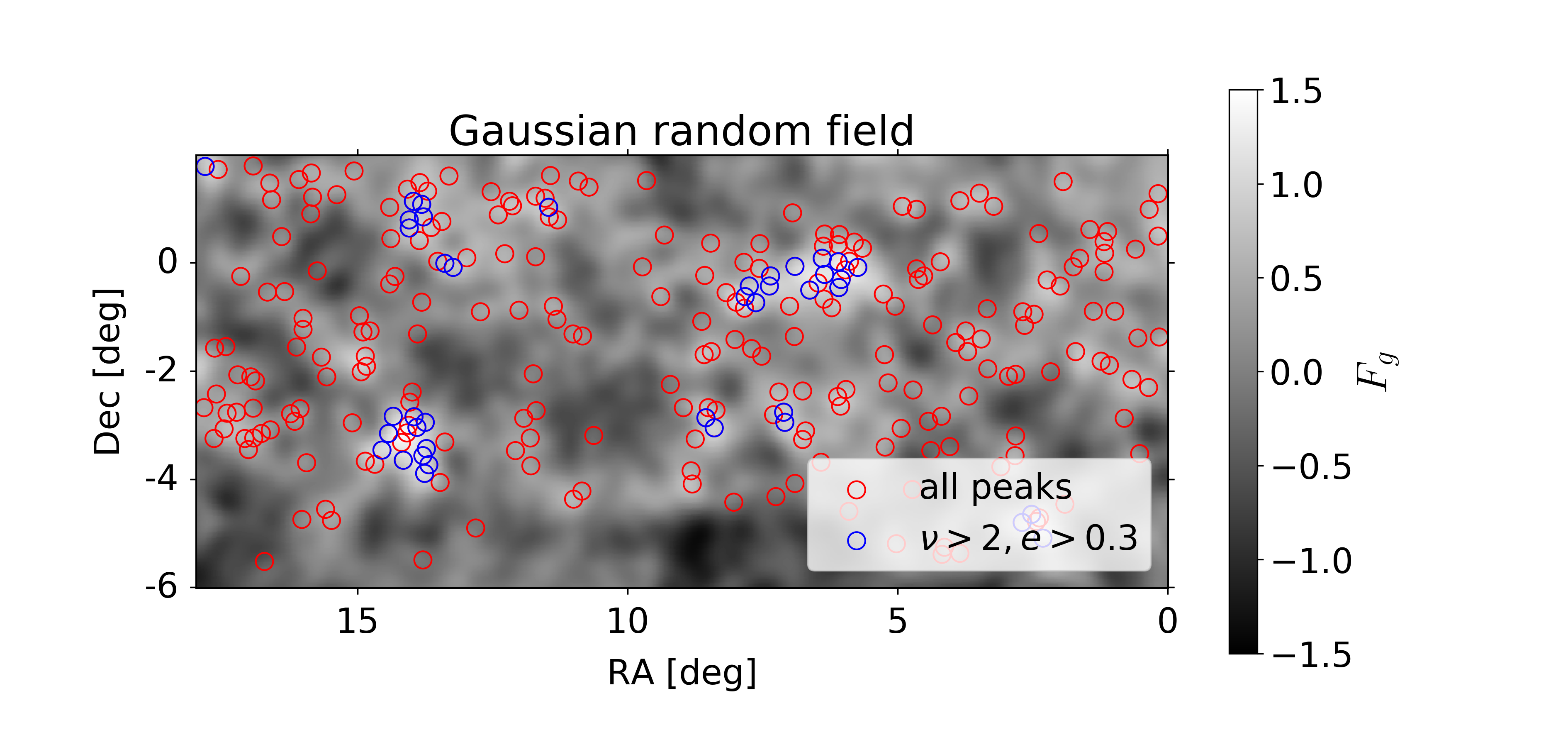}{}}
    
    \caption{Top: a comparison of the $m=0$, $m=2$, and $m=4$ moments of oriented stacks on Buzzard clusters (solid lines) and Gaussian random field peaks (dashed lines); note the three different $y$ axes. Red indicates that all clusters were used in the stack, whereas blue is the sample that remains after a constraint on $\nu$ and $e$ at large scales. The $m=0$ moment is not well matched near the cluster interior, but at larger radii the Buzzard and GRF profiles are consistent both before and after large-scale field constraints are imposed. For $m=2$, the red profiles are similar to each other, showing that the quadrupole does not strikingly display signs of non-Gaussianity when the stack includes all clusters. However, the field constraints induce a much larger boost in the Buzzard signal than the GRF (blue dashed versus continuous line). The $m=4$ moment is notably consistent with zero for the GRF, whereas in Buzzard the profile peaks near 10 Mpc, which also is boosted with the field constraints. We interpret this sign of non-Gaussianity to be related to the flatness of filaments, arising from nonlinear growth of structure at late times. Bottom: A cutout of the galaxy (Buzzard, left) and pseudo-galaxy (GRF, right) fields that were used for the orientation and field constraints. Both fields are Gaussian smoothed with FHWM=14 Mpc. Red circles are plotted around all cluster locations. Many clusters are cut by imposing $\nu>2, e>0.3$ and the remaining ones have blue circles over-plotted (in other words, blue circles also belong to the red sample). The remaining clusters are themselves clustered. Clumps of blue clusters in the left field are more aligned than in the right. This demonstrates how the $\nu$ and $e$ thresholds select clusters in high-superclustering, highly non-Gaussian regions.}
    \label{fig:GRF_v_Buzzard_stack_comparison}
\end{figure*}

Can oriented stacking be used to examine non-Gaussian structures in the late-time universe which result from non-linear evolution? In the cold dark matter model, gravitational instabilities cause the primordial matter field, which is Gaussian or very nearly Gaussian, to cluster. At early times, the amplitudes of the matter fluctuations grow linearly at almost all scales. As the universe evolves, small-scale overdensities begin to exceed a critical threshold for collapse. The increase in clustering on small scales causes non-Gaussianity in the matter field and non-linear deviations in the matter power spectrum. Over time, as the overall power spectrum grows, increasingly larger scales pass the threshold, leading to the collapse of the first stars, then galaxies, then clusters. These non-linear effects are important at the scales and redshifts studied in our paper, e.g., non-linearity increases the amplitude of the matter power spectrum $P(k)$ by $\sim1.5\times$ at $z=1$ and $k=1\,h\,\mathrm{Mpc^{-1}}$, and by $\sim6\times$ at the same scale at $z=0$ \citep{Illustris2018}. Accordingly, we search for a characteristic signal of late-time non-Gaussianity in the stacked maps by comparing the Buzzard simulations to a purely Gaussian random field.
% Specifically, non-linearity is defined when the dimensionless power spectrum $\Delta^2(k)$ reaches or exceeds 1. 

We generate GRFs using the measurements of the galaxy autospectrum, galaxy-$y$ cross-spectrum, and $y$ autospectrum in the Buzzard simulations \citep{Pandey2021}. These measurements were made for three wide redshift bins in the galaxy simulation, rather than thin 200 Mpc slices which are used elsewhere. For each bin, the three power spectra contain all necessary information to generate two correlated GRFs which represent the galaxy field and $y$ field without non-Gaussianities.

Next, we stack on `peaks' in the GRF that roughly match the size of $\lambda>10$ galaxy clusters in Redmapper, as described in Section \ref{sec:galaxy_field_characteristics}. We pass the maps and peak positions through the same oriented stacking pipeline such that cutouts from the pseudo-$y$ map are rotated based on information from the pseudo-galaxy map, then stacked.

The comparisons between the resulting GRF stacks versus the Buzzard stacks are shown in Figure \ref{fig:GRF_v_Buzzard_stack_comparison}. The lower plots show the smoothed projected galaxy overdensity field for both Buzzard (left) and the GRF (right). These figures provide a visualization of the spatial distribution of the full cluster sample (red circles) and the clusters constrained by $\nu$ and $e$ limits (over-plotted blue circles; all blue circles belong to the red sample as well). The radial profiles for stacks on these points are shown above, for three multipoles, with the same color scheme for the unconstrained and constrained cluster samples. First we examine the isotropic signal, $m=0$, which is equivalent to what it would be in an unoriented stack. It is consistent beyond $\sim$5 Mpc for both the full (red) and constrained (blue) cluster sample, demonstrating that large-scale differences between the GRF psuedo-$y$ and Buzzard $y$ maps are indistinguishable from unoriented stacking alone. For $r<5$ Mpc, the Buzzard profiles are significantly higher than the GRF profiles. The discrepancy at this small scale is expected because the realistic Buzzard $\delta_g$ and Compton-$y$ fields have high skew at small scales from the collapse of massive, rare peaks in the density field and the strong $Y-M$ relation for massive clusters. A GRF, by definition, has no skew at any scale. Thus by generating the GRF, we redistribute Buzzard power evenly to both low and high values in the psuedo-galaxy and pseudo-$y$ maps. This redistribution of power causes there to be fewer high-$y$ peaks in the GRF. Some additional discrepancy may come from the imperfect peak-finding in the GRF, namely, the selection of peaks at only one scale.

However, in the anisotropic components $m=2$ and $m=4$, significant distinctions appear. Generally, oriented stacking with the GRF results in some anisotropic signal because, as in the real universe, the field around any peak has a preferred alignment. The imposed correlation between the pseudo-galaxy and pseudo-$y$ fields results in an aligned pseudo-$y$ signal. We find that the $m=2$ moment is similar in shape and peak height for the full cluster sample between the GRF and Buzzard, although not perfectly in agreement. Notably, though, imposing the field constraints makes a significant difference. The Buzzard profile rises by $\sim2.5\times$ while the constrained GRF stack remains nearly the same as before the constraints were applied. This demonstrates that imposing $\nu>2$ and $e>0.3$ on the Buzzard galaxy field selects for cluster locations in the $y$ map where the local anisotropy is highly non-Gaussian. Finally, the $m=4$ moment demonstrates the most striking signal of non-Gaussianity. The profile of the GRF stack for both peak samples is consistent with zero, while the Buzzard profile shows a significant peak which increases with the imposed field constraints. We interpret the $m=4$ moment as a sign of filament structure, flattened from small-scale gravitational effects in the late-time universe.

To support this claim, in the following section we examine the $m=4$ signal in oriented stacks from the Websky simulation.

\section{Expectations from Theory}\label{sec:expectations_from_theory}

\subsection{Websky}
We use the Websky simulations to give pure theoretical results which are not subject to the selection effects of DES and the Buzzard mocks. Two key questions are:
\begin{itemize}
    \item How does superclustering depend on redshift?
    \item How does the average superclustering signal from hot gas depend on the mass of the clusters being stacked?
\end{itemize}

Websky is useful for addressing the former question because the simulations extend to redshifts of $z=4.6$. In addition, answering either question requires splitting data into smaller sub-samples, which increases noise in the stacks. Due to the full-sky coverage of this simulation, there are enough clusters that the SNR of sub-sample stacks is sufficient to distinguish between the different results.

We create the analog to a galaxy overdensity map by using all Websky dark matter halos in the mass range [$1.5\times10^{12}, 1\times10^{15}$] \Msun. This range incorporates most Redmagic-galaxy-hosting halos \citep{Clampitt2017, Pandey2021}, and also includes most halos that would host galaxies from the Sloan Digital Sky Survey BOSS-CMASS sample \citep{BOSS2013, Sonnenfeld2019, Maraston2013}. The lower limit is slightly higher than the Websky minimum halo mass ($\sim1.2\times10^{12}$ \Msun) because cutting out the lowest-mass halos saves computational time. Nevertheless, the wide halo range paints a near-complete picture of the large-scale structure, while the \redmagic galaxy data only form a subset of the galaxies which trace the underlying matter. We create mass-weighted halo number-density and overdensity maps using a weight of $M_h/10^{12}$ \Msun. These maps are roughly proportional to galaxy overdensity maps because the number of galaxies in a halo is approximately linearly proportional to the mass \citep{Kravtsov2004}. With smoothing scales far larger than a cluster, the missing details of the subhalo distribution are unimportant.

\begin{figure}[htbp!]
    \centering
    \includegraphics[width=0.45\textwidth, trim=0.3cm 0cm 0.5cm 0cm, clip]{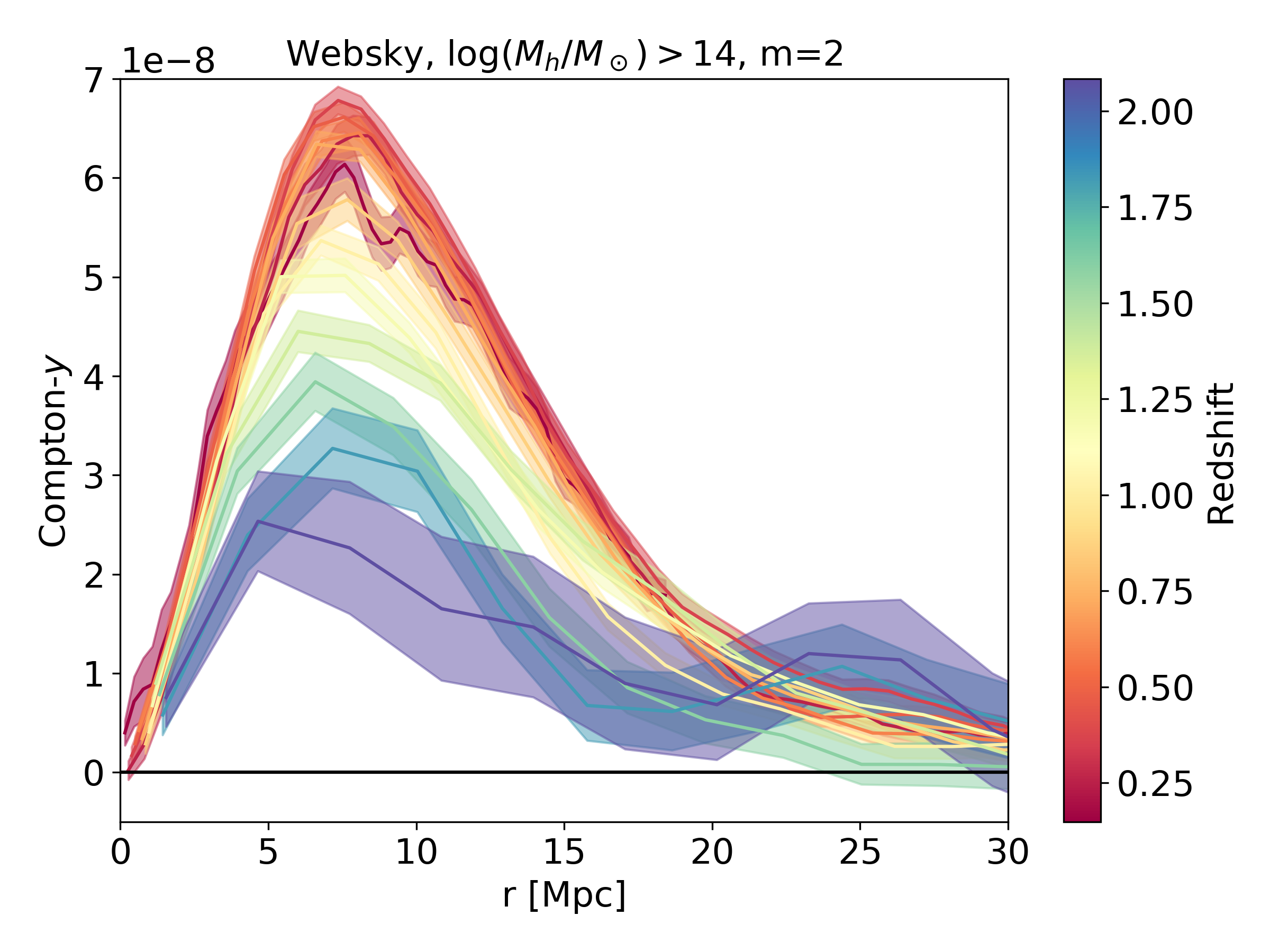}\\
    \includegraphics[width=0.45\textwidth, trim=0.0cm 0cm 0.5cm 0cm, clip]{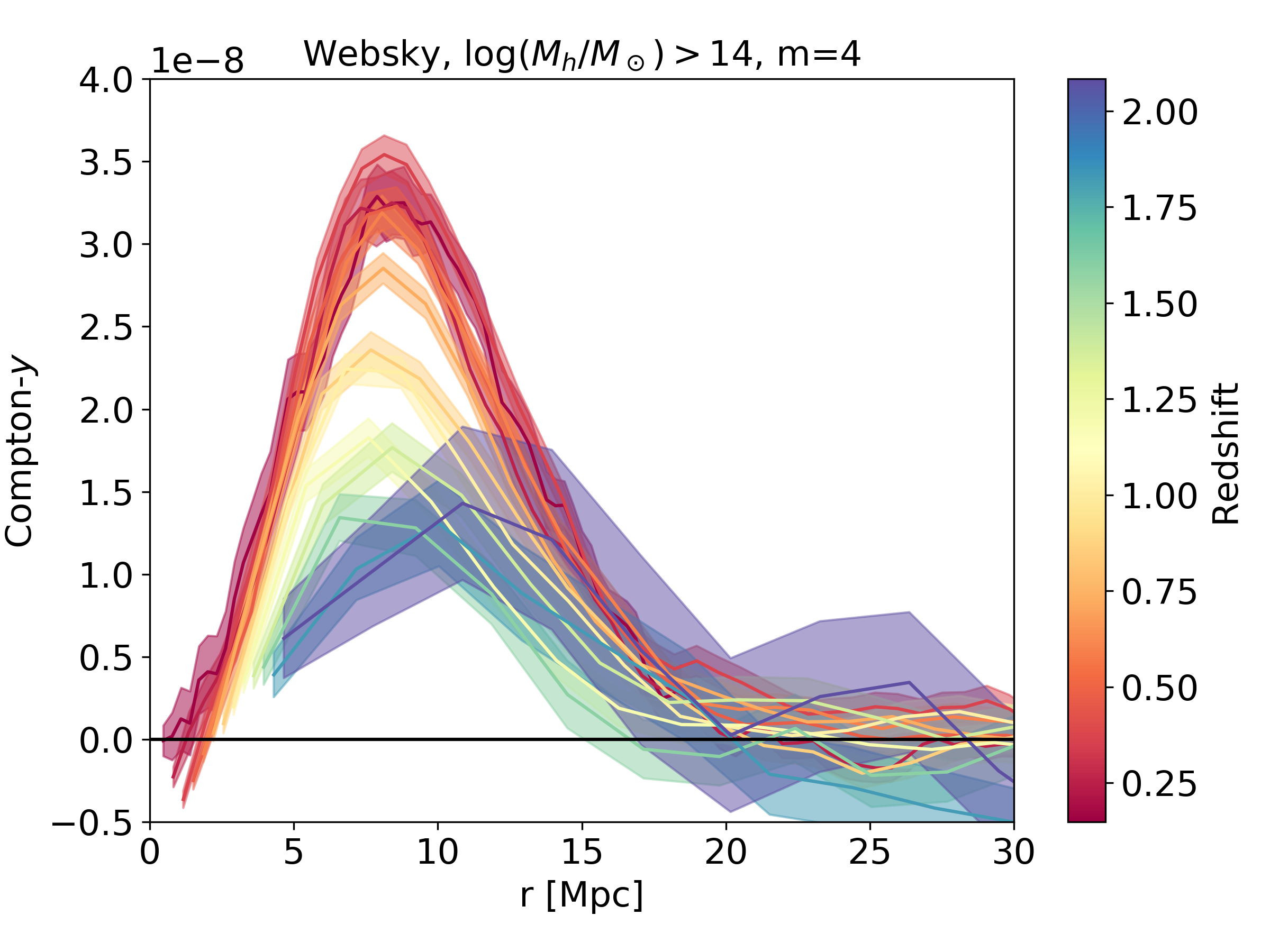}
    \caption{The $m=2$ (top) and $m=4$ (bottom) radial profiles for Websky oriented stacks in progressive 400 Mpc slices in line-of-sight distance. These stacks of massive clusters ($M>10^{14}$M$\odot$) were oriented using a mass-weighted halo density field smoothed at the scale of FWHM=14 Mpc. The profiles are binned equally in arcminutes, thus the bin sizes and centers in Mpc vary across the different redshifts. We remove the first $m=4$ bin to account for uncertainty in the angular integration near the center of the stacked image. In both moments, the peak location is fairly consistent with redshift and there is a trend towards lower overall signal at higher redshifts. The profiles at lower redshifts have smaller uncertainties because of the progressively larger number of massive halos at later times. In the range available in DES ($0.25<z<0.7$), the profiles are roughly consistent within the shaded 1$\sigma$ error regions. This suggests that the evolution of superclustering on these scales slows down during these redshifts, due in part to the onset of $\Lambda$ domination at $z\sim1$.
    \label{fig:redshift_bin_profiles}}
\end{figure}

We test how the extended gas around massive clusters varies with redshift by stacking on halos with $M>10^{14}$ \Msun~from $z\sim0.25$ to $z\sim2$. We rescale the stacked images for pairs of consecutive 200 Mpc slices to the same physical size and combine them. The results are shown in Figure \ref{fig:redshift_bin_profiles}, demonstrating that the anisotropic tSZ signal increases from a low level at high redshifts (early times) until $z\sim0.75$, after which it becomes fairly stable. As time evolves past $z=0.75$, the peak exhibits a slight increase and subsequent decrease. However, as most of these low-$z$ profiles are consistent within the 1$\sigma$ error regions, we leave a detailed study of the low redshifts to future work.

These results can be interpreted as a combination of the halo mass function evolution as well as the evolution of superclustering in the late-time universe. As the simulation evolves, the cosmic web structure in which clusters are embedded becomes more pronounced. Halos merge to form larger and more massive halos, which are prescribed larger tSZ profiles in post-processing. This prescription mimics the theoretical and observational understanding that, over time, galaxies merge along filaments and flow towards clusters. These mergers are expected to heat up the gas in galaxy groups. In addition, mergers and AGN feedback can also heat up intergalactic gas.  The effect of feedback and mergers are de facto included in the Websky simulations, encoded in the gas response to the presence of the evolving halo population, but there are many possibilities that are not accounted for. When galaxies fall into clusters, they undergo ram-pressure stripping and the clusters gain mass and heat up through shocks. Overall, the tSZ signal from the contributions of clusters, groups, and intergalactic gas should grow over time and thus the anisotropic superclustering signal should be stronger at lower redshifts. Our results match this expectation.

\begin{deluxetable}{cccc}[htbp!]
\tablewidth{0pt}
\tablecaption{Websky halo mass bins.\label{table:pp_halo_massbins}}
\tablehead{
\colhead{Bin Title} & \colhead{Mass range ($M_{200\mathrm{m}}$ [\Msun])} & \colhead{Equiv. $\lambda$} & \colhead{$N$}}
\startdata
Low-Mass & $[1\times10^{13},  5\times10^{13}$] & [3, 11] & 3,017,917\\
Mid-Mass & $[5\times10^{13}, 10^{14}]$ & [11,18] & 314,389\\
High-Mass & $[10^{14}, 2.6\times10^{15}]$ & [18,200] & 169,316\\
\enddata
\tablecomments{The equivalent cluster/group richness in the third column is calculated from Eq. 52 in \citet{McClintock2019} for a redshift of 0.5, with the caveat that the mass-richness relation shouldn't be extrapolated to masses as small as the low-mass bin. The final column gives the number of halos in a 200 Mpc slice at $z\sim0.4$.}
\end{deluxetable}

\begin{figure*}[htb!]
\includegraphics[width=0.9\textwidth, trim={0.5cm 2cm .5cm 2cm},clip]{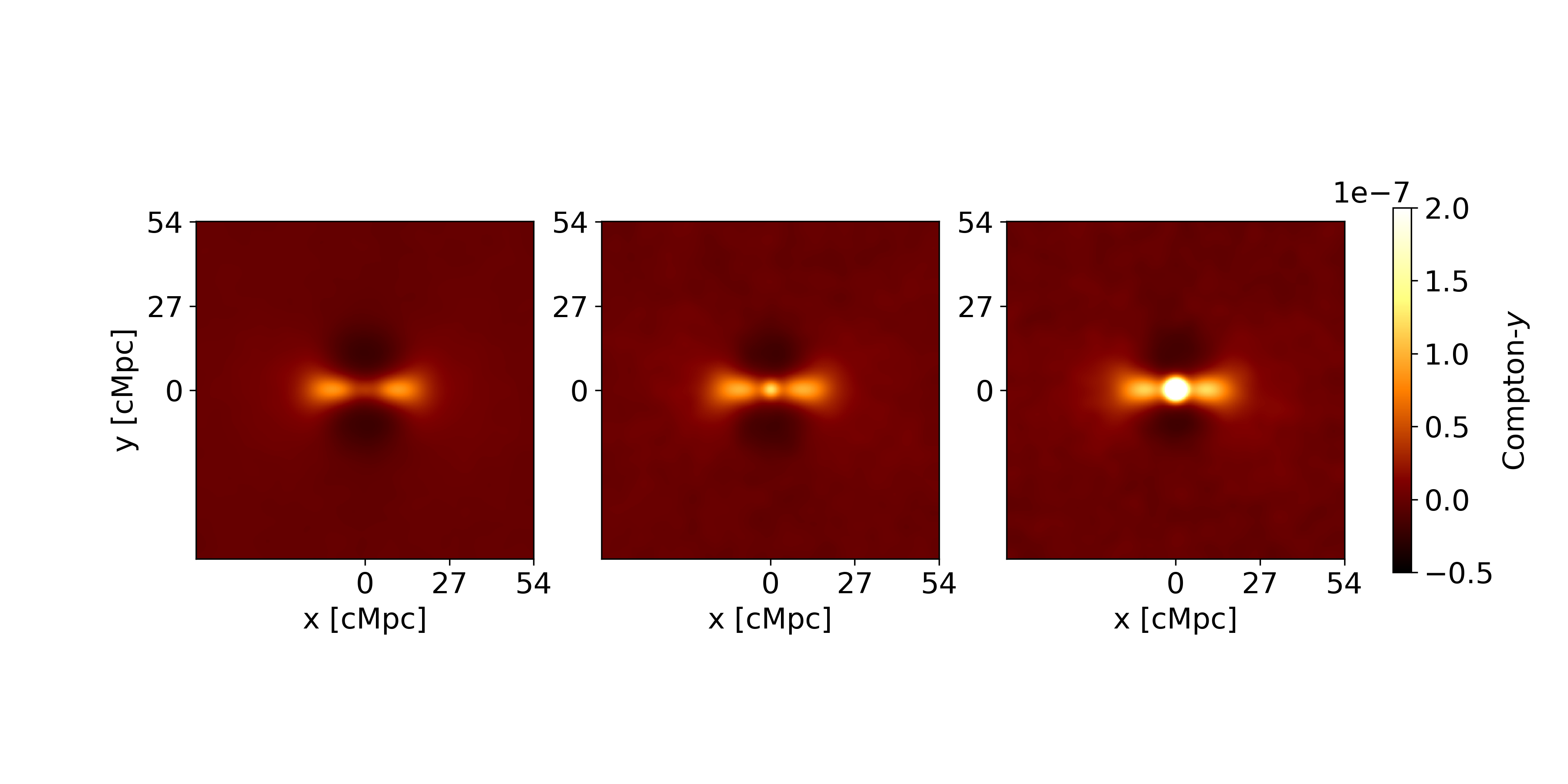}
\caption{Stacks of the Websky Compton-$y$ map on dark matter halo locations. From left to right, stacks are on the low-, mid-, and high-mass bins of Websky halos described in Table \ref{table:pp_halo_massbins}. Each stack combines halos in the distance range 1232--1832 Mpc, or $z\sim0.4$. \label{fig:mass_bin_figure}}
\end{figure*}
\begin{figure}[htbp!]
    \centering
    \includegraphics[width=0.4\textwidth, trim={.5cm 0cm .5cm 0.5cm},clip]{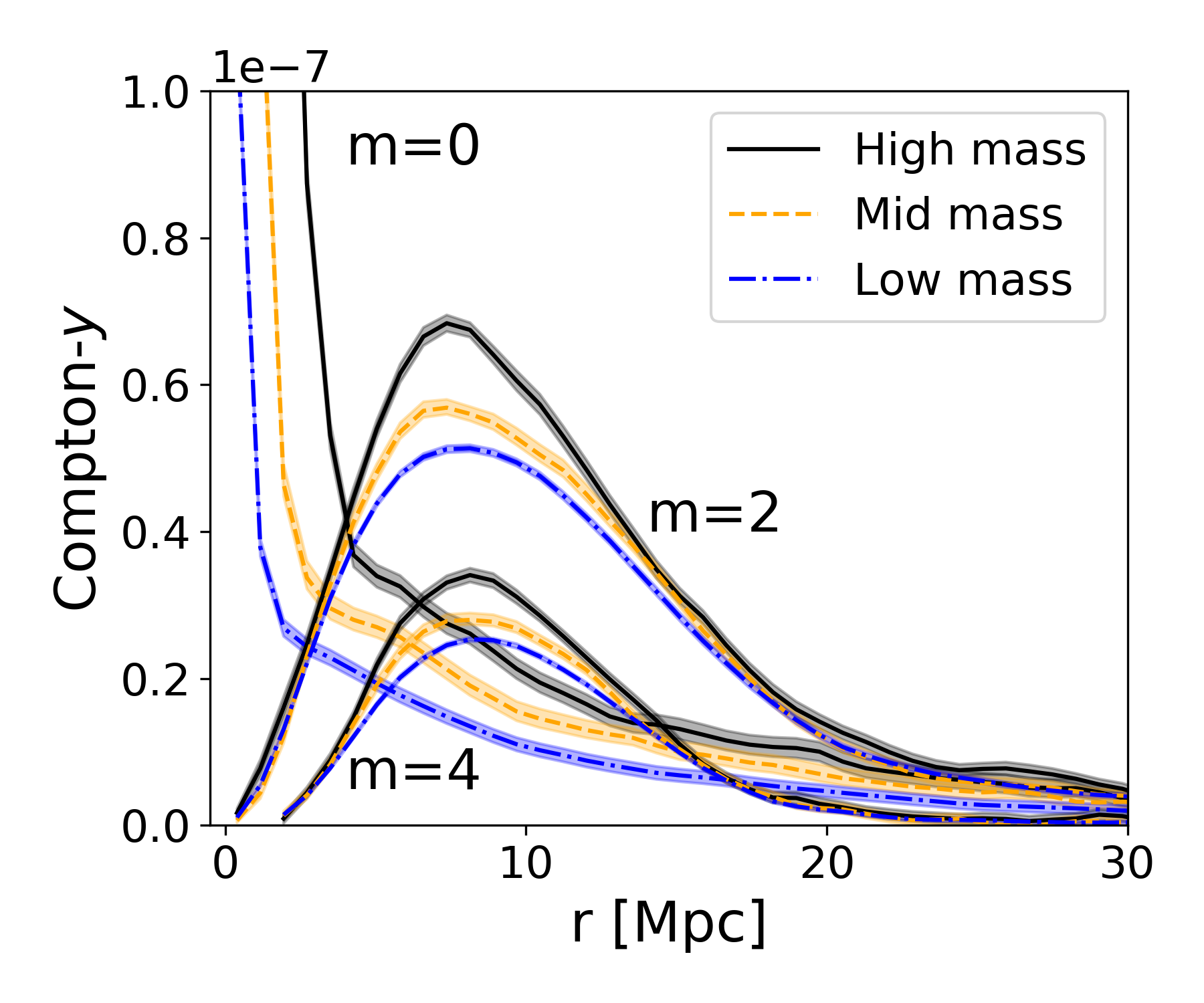}
    \caption{$m=0,2,4$ radial profiles for the oriented stacks of three mass bins of Peak Patch halos shown in Fig \ref{fig:mass_bin_figure}. The highest-mass bin has the largest signal in all components, demonstrating that not only is the isotropic gas signal higher for larger clusters (as prescribed), but the surrounding supercluster structure also has a stronger signal. Notably, the $m=2$ and $m=4$ components rise above the $m=0$ component at their peaks, demonstrating that outside of the central stacked cluster, the anisotropic structure contributes a larger tSZ signal than the isotropic component.}
    \label{fig:mass_bin_profiles}
\end{figure}
We next explore how the strength of the gas signal from superclustering varies with the mass of the stacked clusters. We divide the data into three bins, the ranges for which are shown in Table \ref{table:pp_halo_massbins}. The table also shows the equivalent DES cluster richness range ($\lambda$) at $z=0.5$ assuming the mass-to-richness relation from \citet{McClintock2019}. After stacking the cluster cutouts, we combine stacks from 5 slices from 1200--1800 Mpc to get the images shown in Figure \ref{fig:mass_bin_figure}. Figure \ref{fig:mass_bin_profiles} shows the $m=0, 2, 4$ moments of the stacks in each mass bin. Larger halos have higher isotropic tSZ profiles, as the $y$ signal is prescribed in the simulations to be proportional to $M^{5/3}$. Our results demonstrate that additionally, the anisotropic $y$ signal is stronger for structures surrounding higher-mass halos.

This stronger signal could result from a combination of effects. Because the Websky simulations are hydrodynamical only in the limited-spatial-range response to the presence of halos, albeit of all masses, we emphasize that this dependence is \textit{not} from temperature differences in intergalactic filament gas, since these are not included in these  simulations. The contributing tSZ sources are clusters and groups along the alignment axis. The increase in the $m=2, 4$ and signal around more massive clusters is due to their being embedded in overall denser filaments on average. These regions would have more halos along the alignment axis contributing tSZ signal. In addition, massive clusters are known to have higher connectivity to the cosmic web \citep{AragonCalvo2010, Codis2018}. In other words, more massive clusters are connected to a higher number of filaments. This could contribute to some of the correlation between mass and tSZ signal in $m=2$ and 4; massive clusters may have multiple filaments partially-overlapping along the line-of-sight of the alignment axis, boosting the signal. Both effects could be contributing simultaneously.
% Future work could incorporate observations to further explore the correlation between halo mass and anisotropy of the surrounding large-scale structure, as well as halo mass and filament thickness.

We attempt to address the same questions with the observational data. However, we find that splitting the small amount of available cluster data into richness bins increases the noise in the stacks enough that the results for different bins are all consistent within the errors. In addition, the \redmagic high-density galaxy catalog extends only to $z=0.7$. Within this range, the Websky results indicate that the superclustering signal is expected to remain fairly constant. We determine that the error bars in ACT$\times$DES and even the Buzzard mocks are too large to observe a redshift evolution in this range, and any apparent evolution may be only a function of the data selection functions.

\begin{figure*}[ht]
\setlength{\tabcolsep}{-7pt}% Default value: 6pt  %horizontal (column) spacing
\renewcommand{\arraystretch}{-20} % Default value: 1 vertical (row) spacing %\renewcommand{\arraystretch}{0.01}
\begin{tabular}{lll}
\begin{turn}{90}
\begin{minipage}{0.2\linewidth}
\vspace{-30pt}\hspace{40pt}\Large{$\mathbf{m=0}$}
\end{minipage}
\end{turn}
\hspace{-15pt}\includegraphics[width=0.36\textwidth,trim={0.1cm 0.80cm 0.1cm 0.55cm},clip]{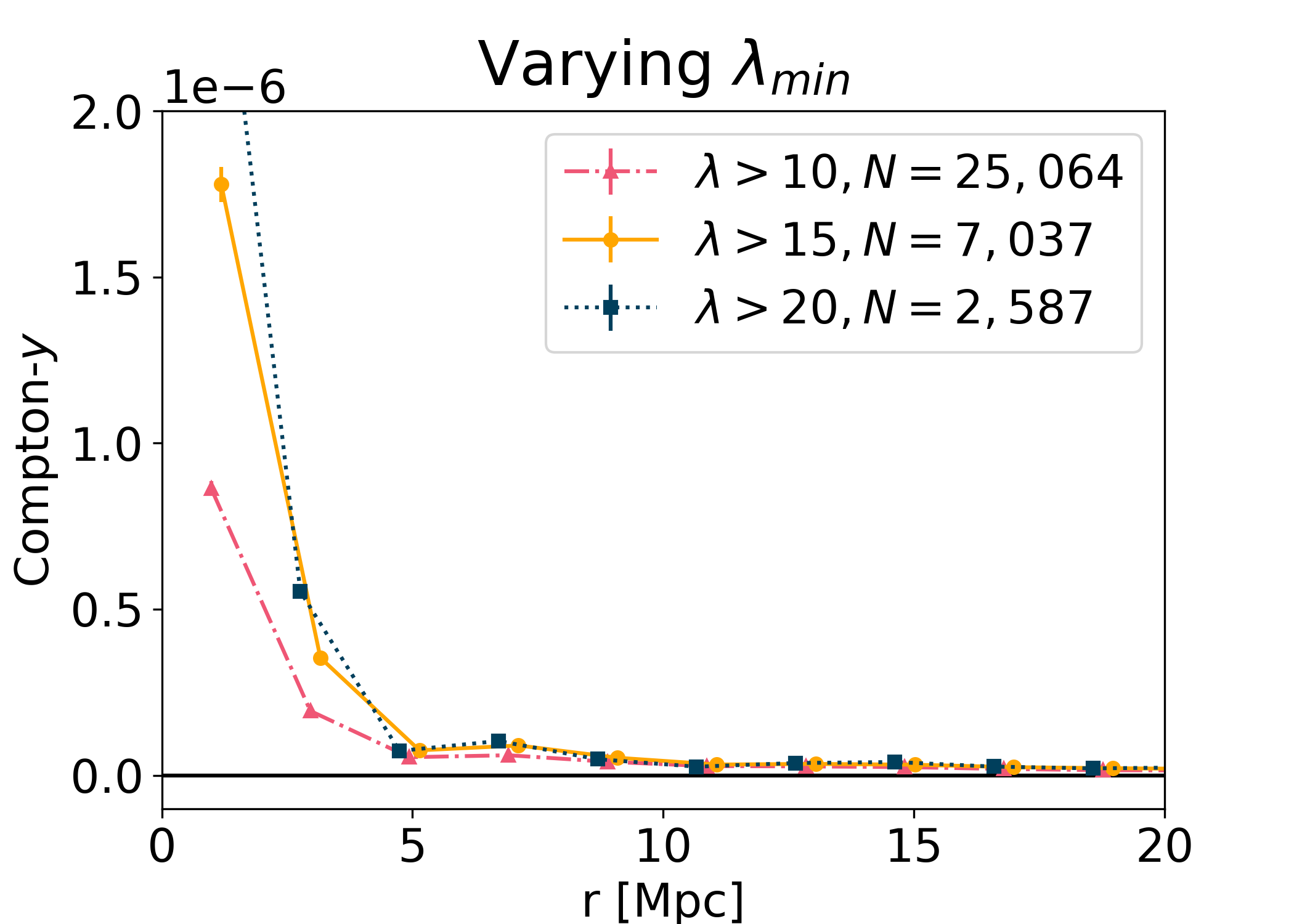} & \includegraphics[width=0.36\textwidth,trim={0.1cm 0.65cm 0.1cm 0.25cm},clip]{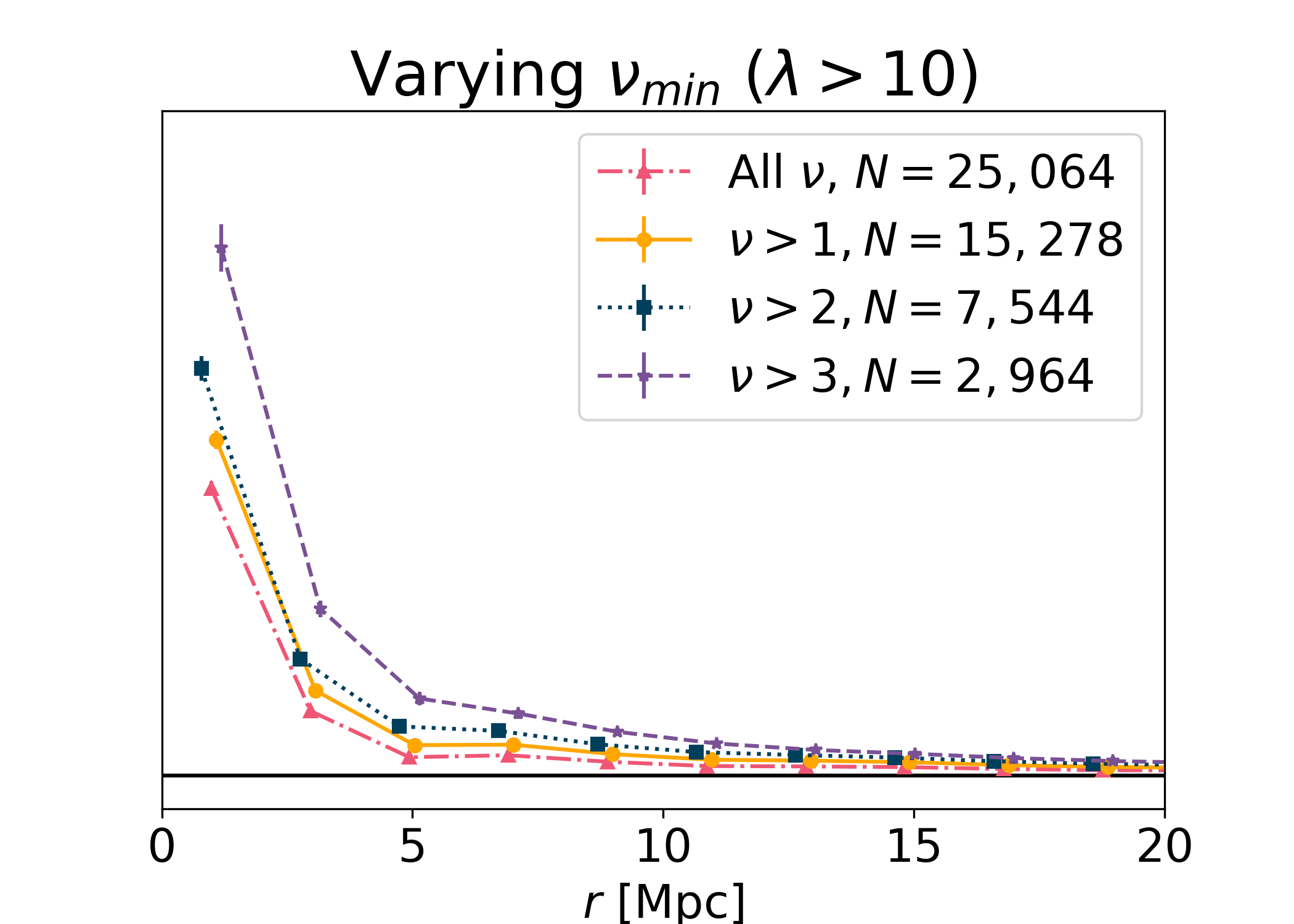} & \includegraphics[width=0.36\textwidth,trim={0.1cm 0.65cm 0.1cm 0.25cm},clip]{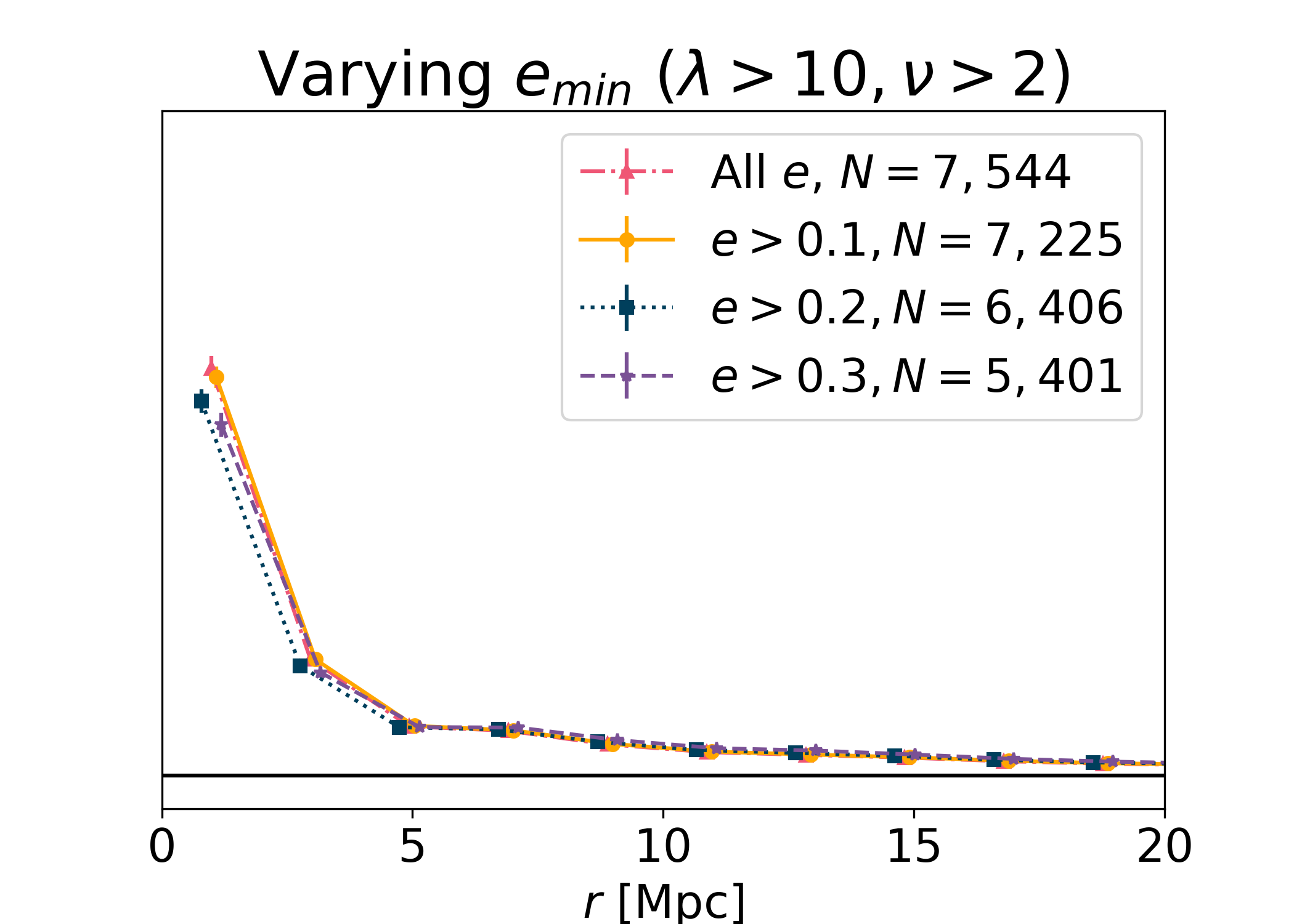}\vspace{0pt} \\
\begin{turn}{90}
\begin{minipage}{0.2\linewidth}
\vspace{-30pt}\hspace{40pt}\Large{$\mathbf{m=2}$}
\end{minipage}
\end{turn}
\hspace{-15pt}\includegraphics[width=0.36\textwidth,trim={0.0cm 0.65cm 0.1cm 0.8cm},clip]{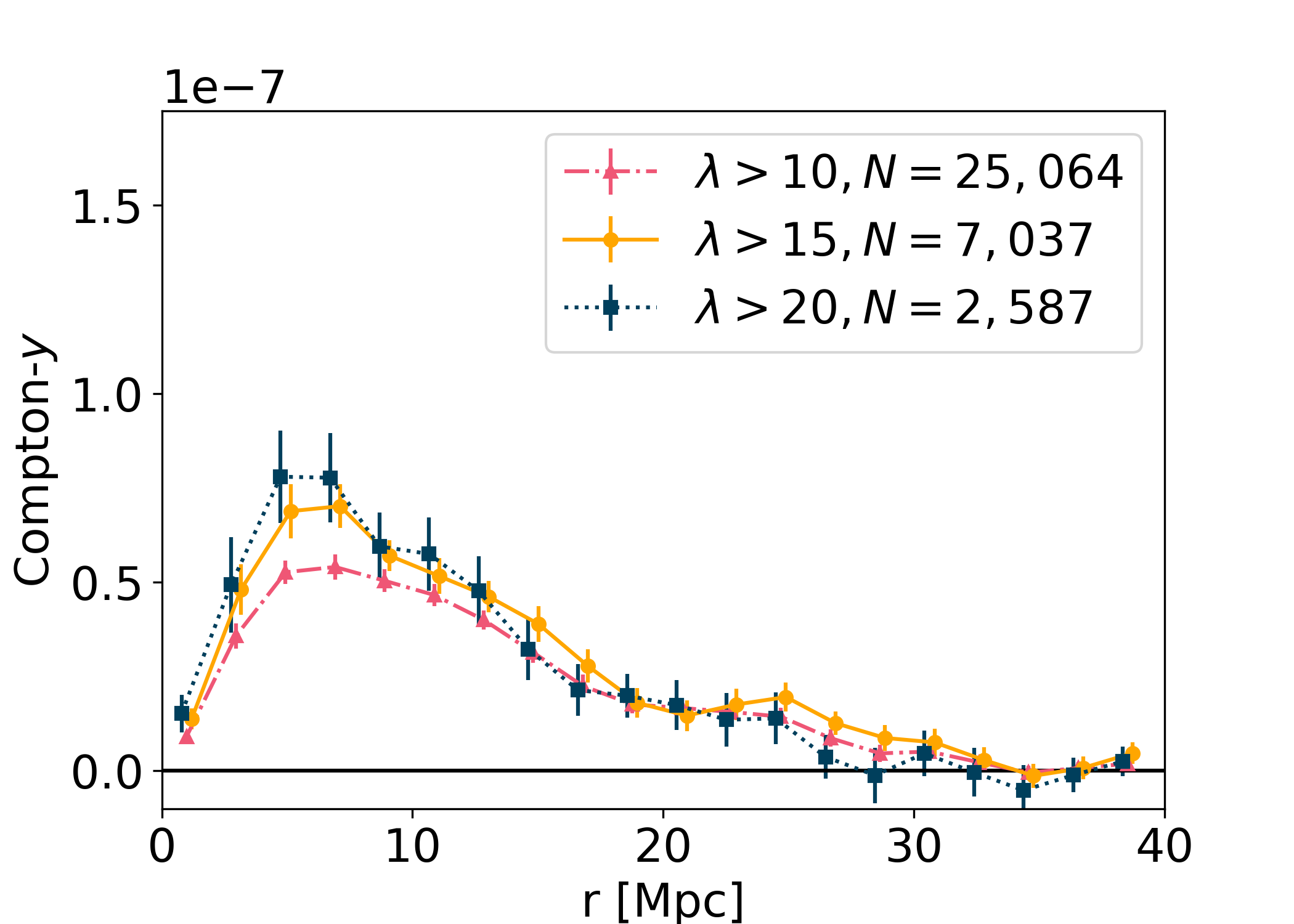}
& \includegraphics[width=0.36\textwidth,trim={0.1cm 0.65cm 0.1cm 0.5cm},clip]{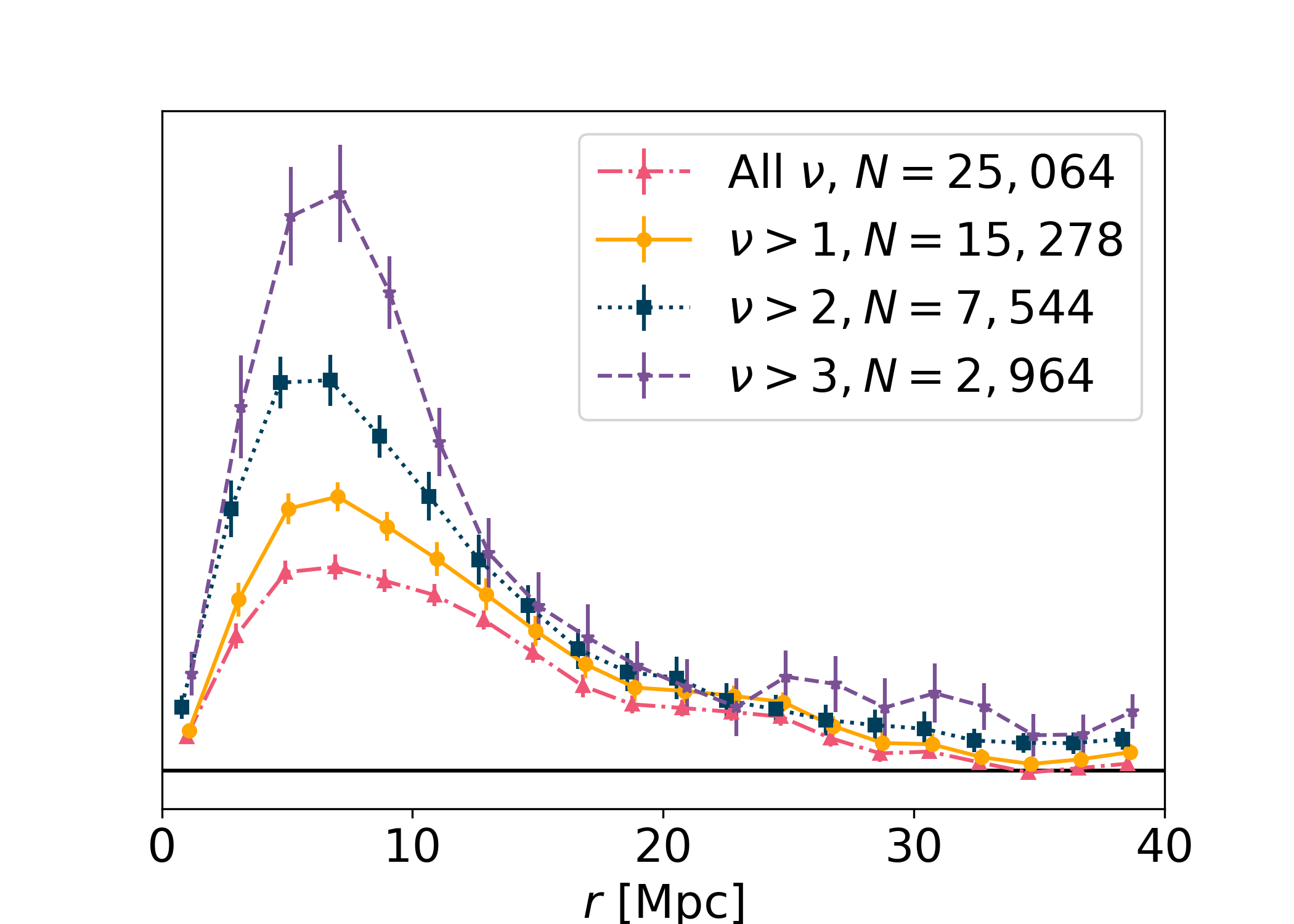}  & \includegraphics[width=0.36\textwidth,trim={0.1cm 0.65cm 0.1cm 0.6cm},clip]{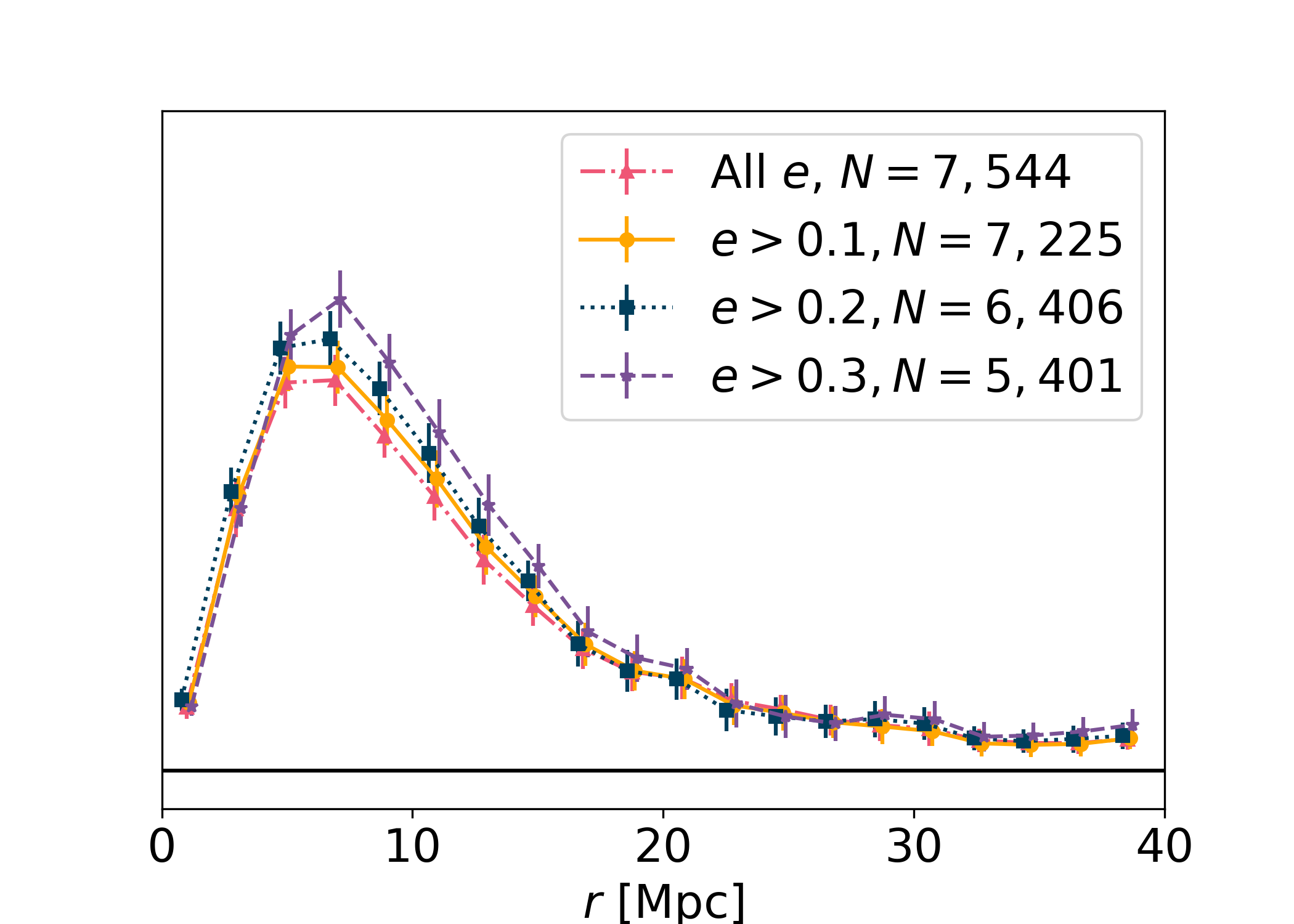} \\
\begin{turn}{90}
\begin{minipage}{0.2\linewidth}
\vspace{-30pt}\hspace{47pt}\Large{$\mathbf{m=4}$}
\end{minipage}
\end{turn}
\hspace{-15pt}\includegraphics[width=0.36\textwidth,trim={0.cm 0cm 0.1cm 0.5cm},clip]{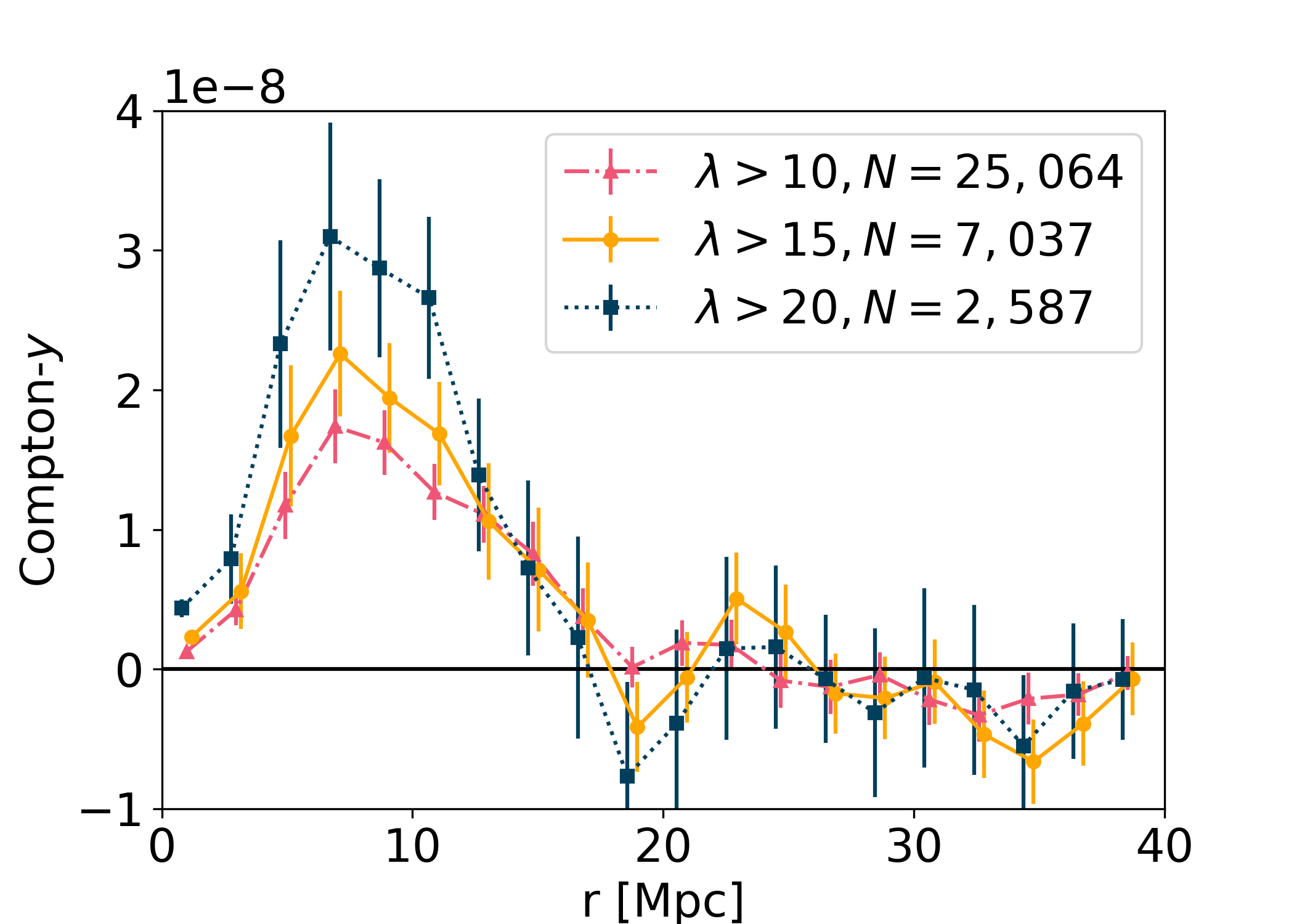} & \includegraphics[width=0.36\textwidth,trim={0.1cm 0cm 0.1cm 0.5cm},clip]{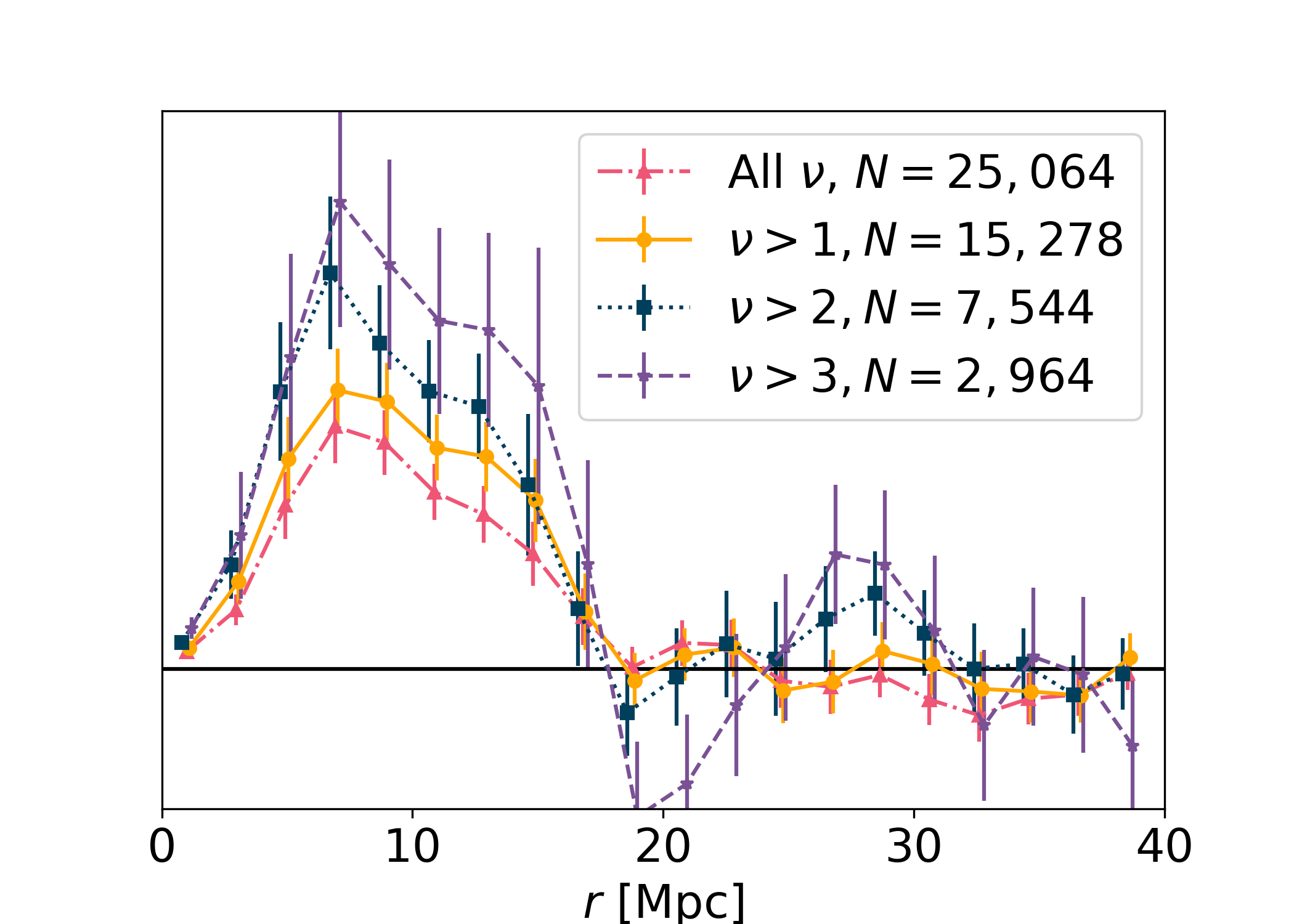} & \includegraphics[width=0.36\textwidth,trim={0.1cm 0cm 0.1cm 0.5cm},clip]{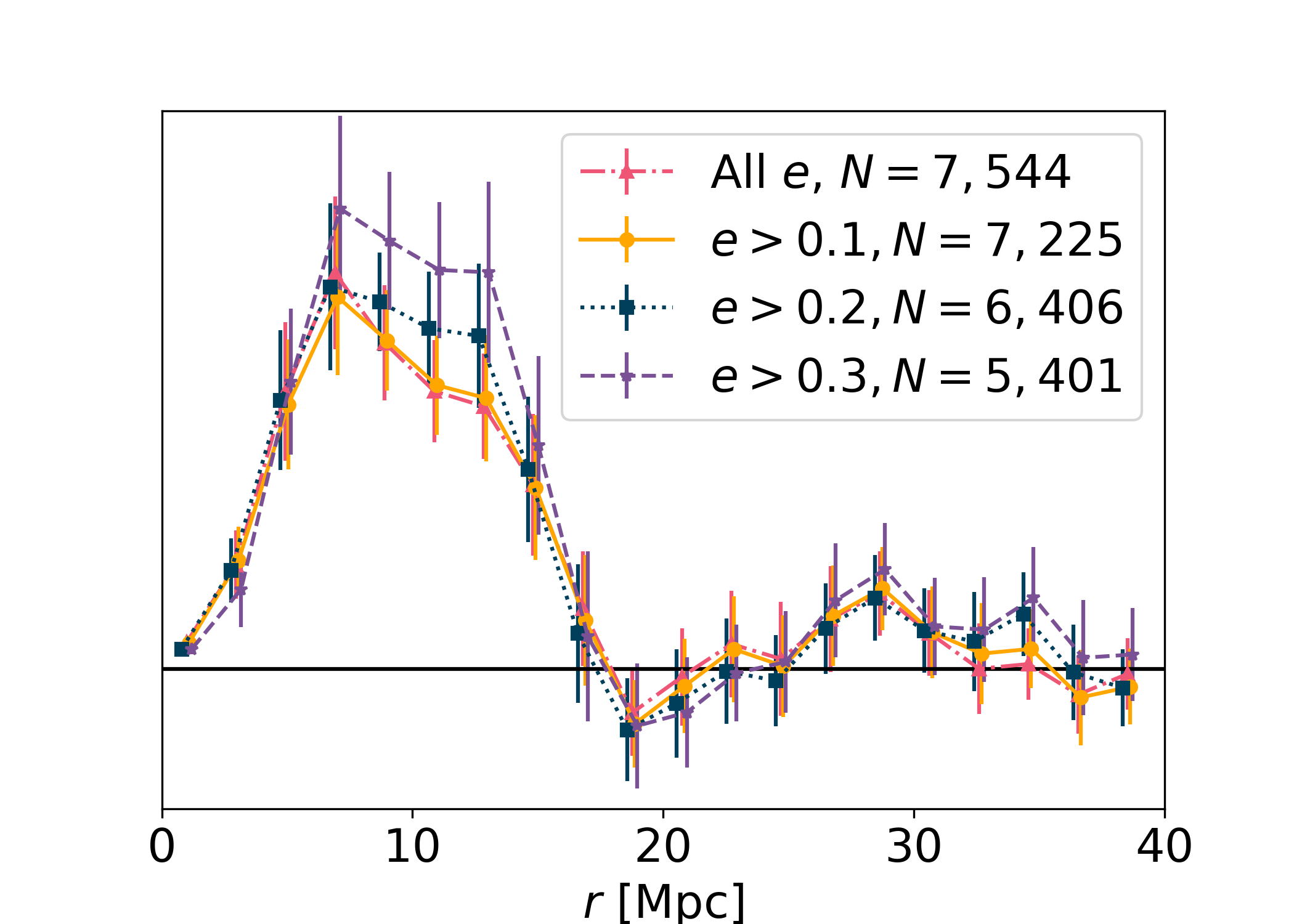}
\end{tabular}
\caption{Results of changing the lower threshold for cluster richness ($\lambda$, left column), field excursion on a 14 Mpc scale ($\nu$, middle column), and field elongation on a 14 Mpc scale ($e$, right column). For $\nu$ changes, the $\lambda$ threshold is fixed at 10; for $e$ changes, the $\nu$ threshold is fixed at 2. From top to bottom, the plots are of $m=0,2,4$. $1\sigma$ error bars are taken from the diagonal of the covariance matrices calculated through cluster sample splits. Increases in $\lambda_{min}$ and $\nu_{min}$ augment the signal in all three moments, while $e_{min}$ only has an impact on the anisotropic moments. The  strongest effect to $m=2$ comes from changes to $\nu_{min}$. We adopt ($\lambda>10, \nu>2, e>0.3$) as the combination of constraints to apply to observed ACT$\times$DES data, as they increase the anisotropic signal without overly depleting the cluster sample.} \label{fig:parameter_changes}
\end{figure*}
\renewcommand{\arraystretch}{1}

\subsection{Dependence on parameters} \label{sec:parameter_dependence}
We examine the dependence of superclustering $y$ signal on cluster richness $\lambda$, field excursion $\nu$, field ellipticity $e$, and smoothing scale using the Buzzard mocks. Figure \ref{fig:parameter_changes} shows the effects of imposing minimum limits on $\lambda$, $\nu$, and $e$ on the stacked Compton-$y$ signal. The $\nu$ property has the strongest impact, demonstrating that clusters embedded in large-scale highly overdense regions tend to be members of more massive filaments and superclusters.

\begin{figure}[htbp!]
    \centering
    
    \includegraphics[width=0.4\textwidth, trim={9cm 3cm 10cm 4cm}, clip]{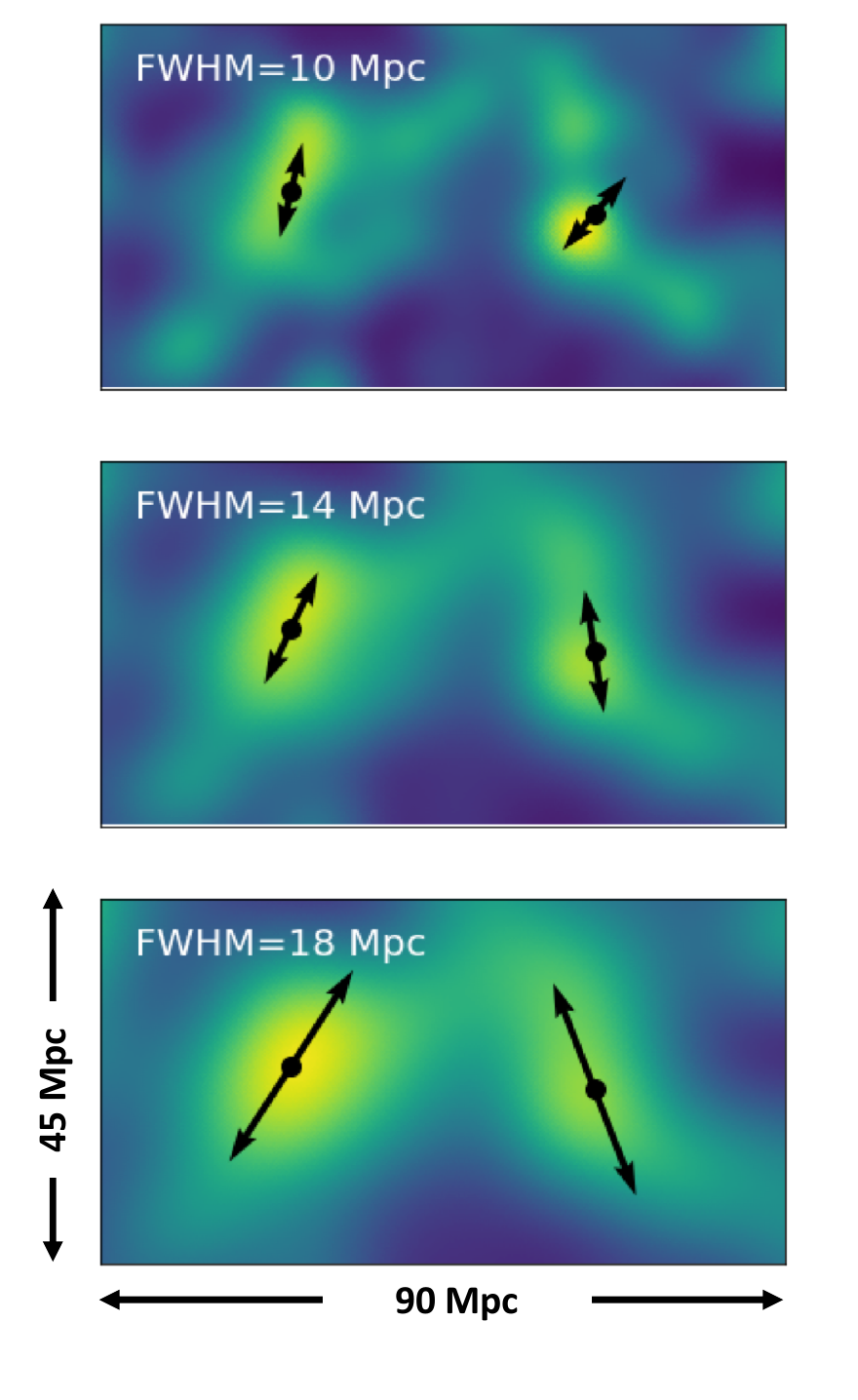}
    \caption{Overdensity maps smoothed at three scales and the respective orientations determined for two clusters. For the left cluster, smaller-scale surrounding structure is fairly aligned with larger-scale structure so the orientation is only slightly rotated between the different maps. The right cluster lies at a kink in the surrounding large-scale structure, so the orientation angle depends more strongly on the smoothing scale. All three maps are smoothed at scales larger than a cluster, so the small-scale peaks corresponding to the two clusters are not visible in the maps.}
    \label{fig:three_orientations}
\end{figure}

Certain cuts to the cluster sample significantly boost oriented $y$ signal, yet a reduction of factor $N$ to the number of stacked images augments the random noise by $\sqrt{N}$. For the different parameter values tested, we assess this trade-off by examining the signal-to-noise of the maximum bin of $m=2$. We find that the full $\lambda>10$ sample has the highest signal-to-noise. If including the field constraints, a combination of ($\lambda>10, \nu>2, e>0.3$) is optimal.

\begin{figure}[htbp!]
\centering
    \includegraphics[width=0.45\textwidth,trim={0.3cm 0cm 0.3cm 0.2cm},clip]{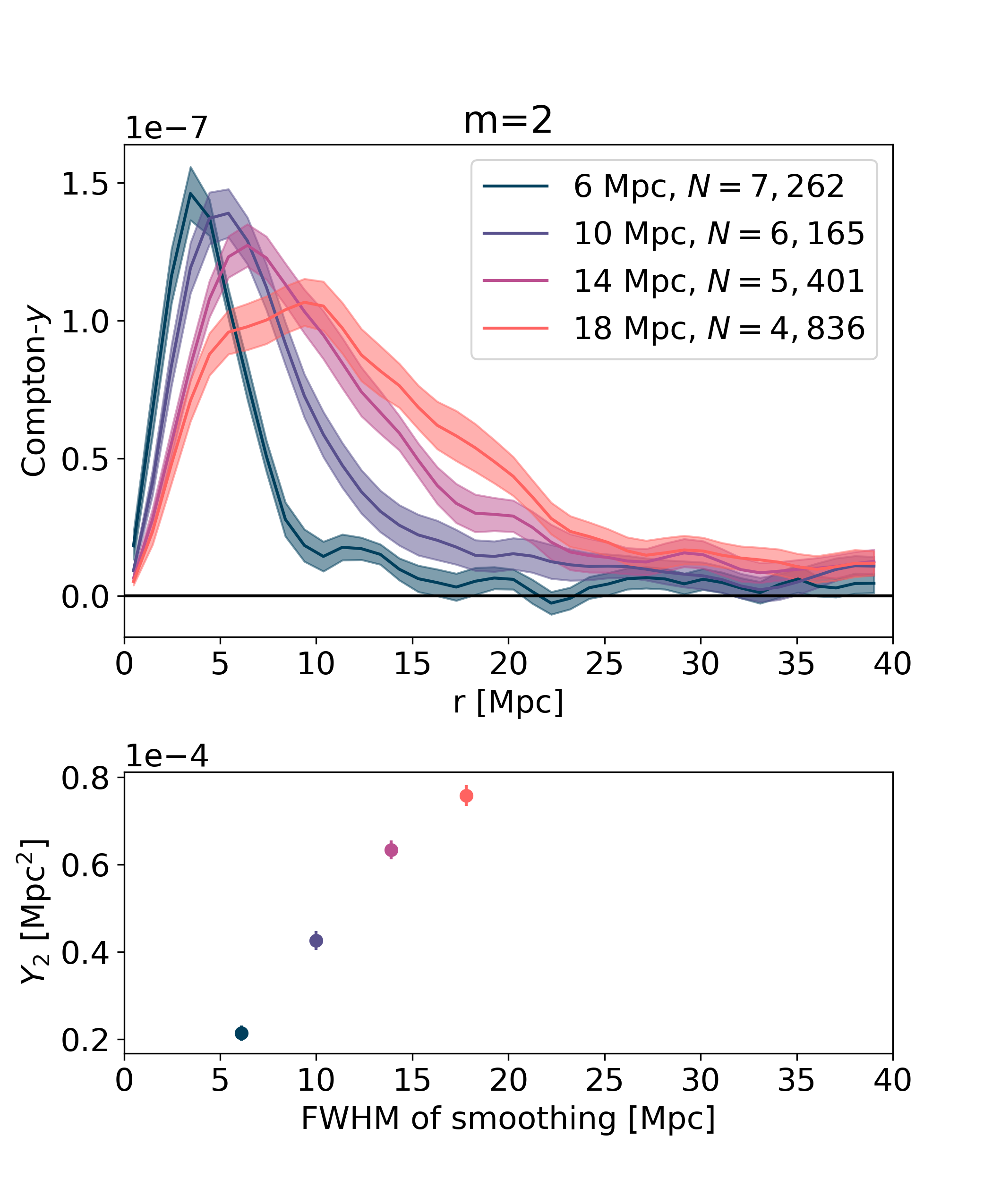}
    \caption{Effect of changing the smoothing scale applied to $\delta_g$. Top: changes to the $m=2$ radial profile; bottom: changes to the integrated $m=2$ Compton-$y$ signal out to 40 Mpc. All profiles are constrained by $(\lambda>10, \nu>2, e>0.3)$, where the latter two parameters are calculated post-smoothing. The smoothing scale thus impacts both the selection and orientation of clusters. The location $r$ of the maximum highly depends on the FWHM and thus contains little physical information.}
    \label{fig:sigma_changes}

    \includegraphics[width=0.45\textwidth,trim={0.0cm 0cm 0.8cm 0.2cm},clip]{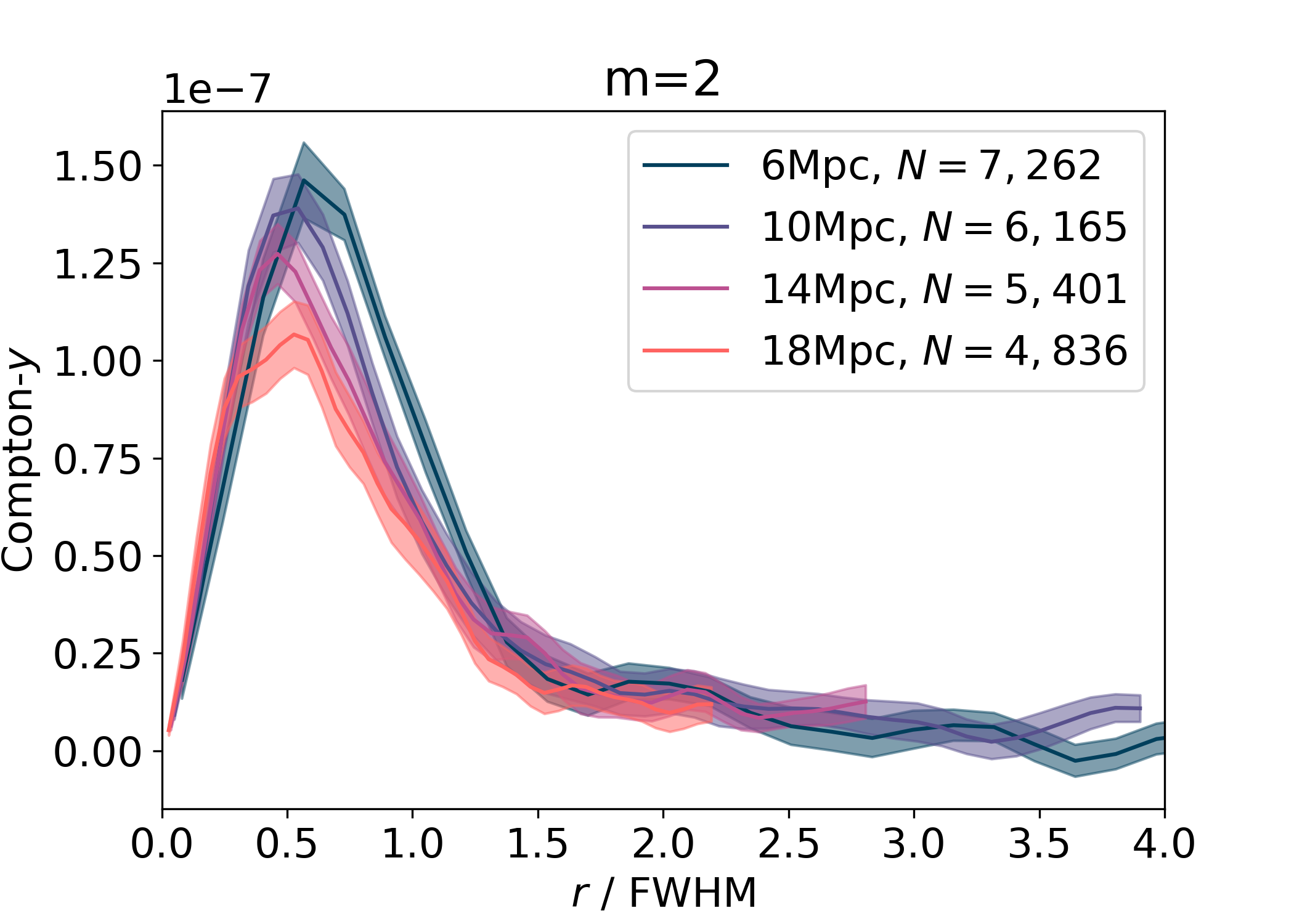}
    \caption{The same figure as Fig \ref{fig:sigma_changes} (upper), but with $r$ scaled by the FWHM of the Gaussian to create a unitless radial quantity. This scaling brings nearly all curves into alignment, demonstrating that the location of the peak of $m=2$ should not be interpreted as the true position of maximum anisotropic gas signal from the cluster, but rather a property which depends almost exclusively on the smoothing scale.}
    \label{fig:sigma_changes_rescaled}
\end{figure}

Next we examine changes to the smoothing scale that is applied to the $\delta_g$ maps. As Figure \ref{fig:three_orientations} visually demonstrates, finer-grained smoothing causes the orientation for each cluster to be more dependent on local features such as nearby filament galaxies, whereas coarser smoothing makes larger-scale features (like the nearest-neighbor clusters) more important to the determination of curvature. With the Buzzard simulations, we perform oriented stacking with a Gaussian-smoothed map of FWHM=[6, 10, 14, 18] Mpc. The results are shown in Figure \ref{fig:sigma_changes}. We only show $m=2$ for brevity, but $m=4$ displays similar effects.

We find that the location of the peak of the $m=2$ profile scales nearly linearly with the smoothing scale. This is because the smoothing scale sets the radius from the cluster at which orientation is determined, and thus by construction, anisotropic structure is maximally aligned along the horizontal image axis of the stack at that radius. We will call this the `radius of maximal alignment.' Figure \ref{fig:sigma_changes_rescaled} demonstrates this concept by rescaling the $x$-axis by the Gaussian FWHM, bringing nearly all peaks into alignment. $m=4$, not shown, behaves similarly. We emphasize that the peak location therefore contains little to no physical information. The \textit{height} of the profile at the radius of maximal alignment, however, is physical, as it is determined by the average temperature and density of aligned structure at that scale. We will further analyse the meaning of peak height and its relationship to physical properties in a subsequent paper.

Another useful quantity is the total integrated Compton-$y$ signal for each moment,
\begin{equation}
    Y_m = \int_0^R C_m(r) r dr,
\end{equation}
where $C_m(r)$ is defined in Equation \ref{eq:multipole_moments}. For $m=0$, this is similar to the unitless angularly-integrated Compton-$y$ parameter, with the difference that our $Y$ has units of Mpc$^2$. $Y_2$ is shown in the lower panel of Figure \ref{fig:sigma_changes}. $Y_2$ depends not only on the gas properties at the radius of maximal alignment, but also on how coherent the structure is within and beyond that radius. It also depends on the galaxy field constraints $\nu$ and $e$, which limit the cluster sample in a smoothing-dependent manner. As the smoothing becomes coarser, we observe that the $m=2$ $y$ profiles broaden and $Y_2$ increases. This may suggest that LSS is more coherently aligned along the direction of orientation determined for larger scales than for smaller scales. However, because of the nontrivial impacts of smoothing on the galaxy field constraints, we leave a more robust physical interpretation for the succeeding paper.

The flexibility of our method to smoothing demonstrates that it can be applied to scales as small as individual clusters, to study the alignment of galaxies and cluster gas, and as large as the longest superclusters.

\section{Comparison with Observational Data}\label{Sec:Results}
We apply the method to ACT$\times$DES data and the Buzzard mocks using a few combinations of smoothing scales and field constraints. We will highlight results for 18 Mpc smoothing, a scale which roughly corresponds to inter-cluster filaments (as motivated in Sec.~\ref{subsec:smoothing}). We later show all three smoothing scales. Each result figure shows the observational $y$ measurements in 3 radial bins, as motivated in Sec~\ref{subsec:uncertainties}. Each figure also shows the Buzzard results, which demonstrate what a ‘pure’ measurement would look like across the full DES footprint, with errorbars due to variance in the large scale structure but not instrumental noise. The Buzzard results are binned into the same 3 radial bins to assess the consistency with observations, while the figures also display the continuous Buzzard profiles for visual purposes.

\begin{figure}[b]
    \includegraphics[width=0.45\textwidth, trim={0cm 0cm 1cm 0cm},clip]{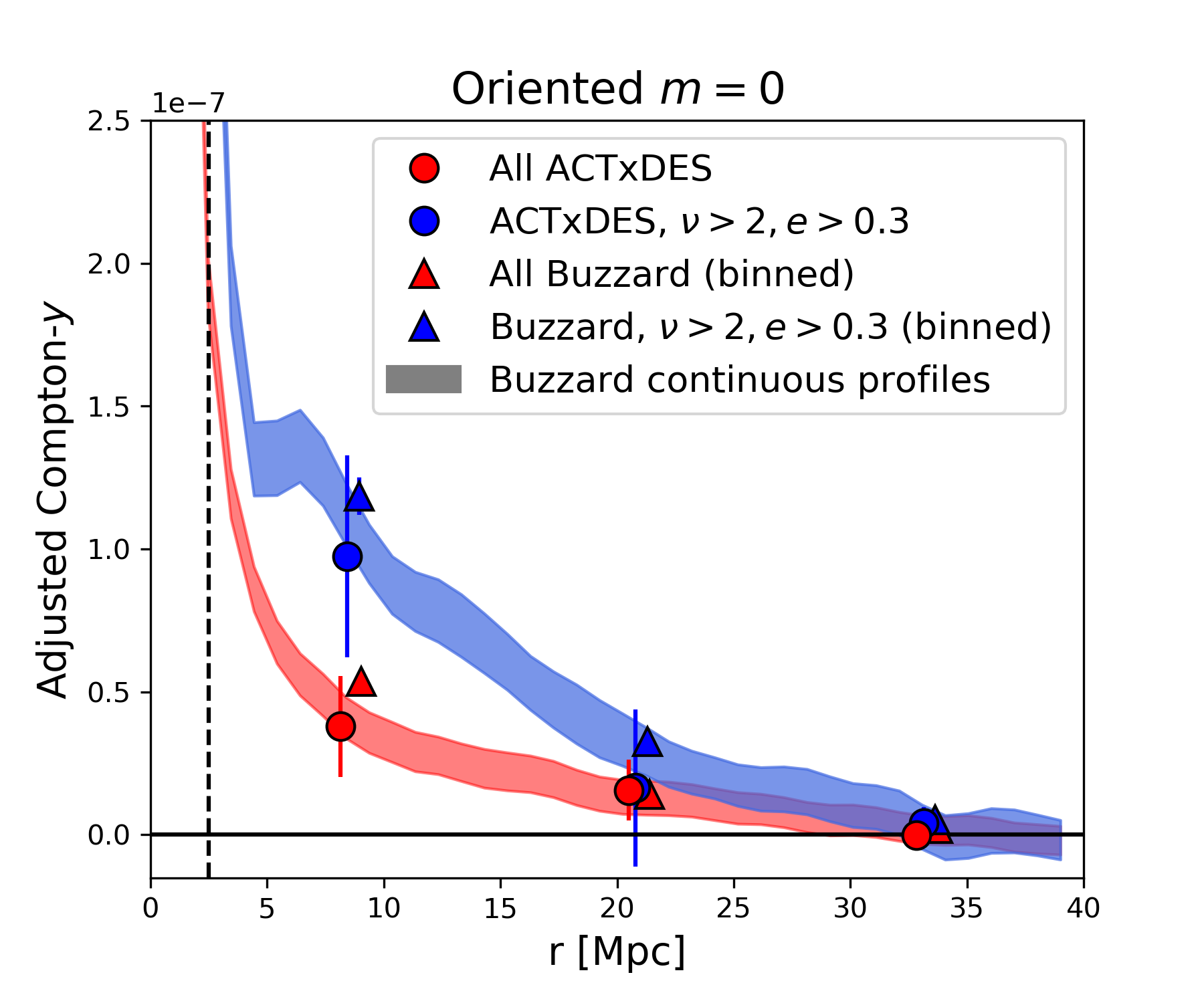}
    \caption{Comparison of the binned $m=0$ component of a Buzzard oriented stack (triangles) versus ACT$\times$DES (circles), with continuous Buzzard profiles also shown in the background for reference. Red indicates the stack using the full $\lambda>10$ cluster sample, and blue indicates the sample constrained by $\nu$ and $e$ thresholds for 18 Mpc smoothing. The data are binned evenly, including only data beyond the cluster radius (black vertical dashed line). Each profile has been adjusted by subtracting its tail, the average value from 33 to 40 Mpc. The binned profiles represent the average thermal SZ signal from large-scale structure beyond the central stacked cluster. $\nu$ thresholding boosts this signal within the first $\sim20$ Mpc.}
    \label{fig:buzzard_act_comparison_18_m0}
\end{figure}

To begin, we briefly examine the isotropic ($m=0$) component of the stacked images in Figure~\ref{fig:buzzard_act_comparison_18_m0}. This is identical for an oriented and unoriented stack and will not provide specific information about filaments. Nevertheless, comparing the $m=0$ Compton-$y$ signal beyond the stacked cluster radius between simulations and observations is informative of how well the simulations reproduce the large-scale clustering and gas content of halos. 

Because our focus lies beyond the cluster interior, we exclude the majority of central-cluster tSZ contributions to the radial $y$ profile by choosing an inner cutoff radius. Our choice is $R_c = 1.5 R_\lambda$, where $R_\lambda$ is the \redmapper cluster radius \citep{Rykoff2014}. $R_\lambda$ is given by $(\lambda/100)^{0.2} h^{-1}$ physical Mpc, which we calculate at the median richness in our sample, $\lambda=15$. After conversion, this results in a comoving radius of $\sim2.5$ Mpc for the average redshift of the cluster sample. Therefore, in both the simulations and real data, we begin binning the signal beyond 2.5 Mpc.

The raw ACT$\times$DES profiles each have a constant positive offset with respect to the Buzzard profiles (not shown in any figure). This is primarily due to long-wavelength noise in the $y$ map from residual low-$\ell$ primary CMB contamination which does not average down with more stacked clusters. Stacks on random points in simulations of the ACT $y$ map described in Section \ref{subsec:uncertainties} each result in a different constant offset depending on different realizations of the low-$\ell$ primary CMB. To account for the offset in the real ACT $y$ map and each simulation, we subtract the average value of $C_0(r)$ from 33 to 40 Mpc (the tail of each profile) from the full profile. This is similar to performing aperture photometry, a technique typically applied to unoriented stacks to subtract the noise calculated in an annulus from the signal within some inner radius.

The adjusted $m=0$ profiles for 18 Mpc smoothing, after subtraction of the tail, are shown in Figure \ref{fig:buzzard_act_comparison_18_m0}. There is a nonzero signal in the first two bins which is boosted when the $\nu>2$ constraint is applied. (The $e$ constraint has little to no effect on the isotropic profile, as previously demonstrated in Figure \ref{fig:parameter_changes}). The $\nu$ and $e$ constraints reduce the cluster sample by a factor of $\sim5$, causing an increase in noise of $\sim \sqrt{5}$. We can determine the signal-to-noise of the results by comparing them to a null profile. This null result corresponds to the unphysical hypothesis that there is no tSZ signal outside of the average cluster radius. This would occur in a universe where every cluster were separated by $>40$ Mpc from the nearest object containing hot gas. We compute the reduced-$\chi^2$ of the observational data vector $\boldsymbol{y}_\mathrm{obs}$ (the coarsely binned $C_0(r)$ profile) with respect to null:
\begin{equation}
    \chi^2_{\mathrm{red}} = \boldsymbol{y}^\mathrm{T}_\mathrm{{obs}} \boldsymbol{\Sigma}^{-1} \boldsymbol{y}_\mathrm{{obs}}\big/ N_{\mathrm{bins}},
\end{equation}
where $\Sigma^{-1}$ is the inverse covariance matrix of the data. Next, we find the probability that a truly null vector could exceed that $\chi^2$. This Probability to Exceed (PTE) is measured by drawing 1 million random vectors from a normal distribution with mean zero and the covariance matrix from the data. We compute the $\chi^2$ of each vector with respect to null, then find the fraction of the sample for which the $\chi^2$ exceeds that of the real data vector. This fraction is the PTE; lower values indicate that the random vector is unlikely to exceed the true data vector, providing stronger evidence for a detection of non-zero signal in the data. We then relate this to a signal-to-noise (SNR) estimate which is the number of Gaussian sigmas away from null,
\begin{equation}
     (1-\mathrm{PTE}) = \mathrm{erf}(\mathrm{SNR}/\sqrt{2}),
\end{equation}
where erf is the error function. We also assess the goodness-of-fit to the Buzzard results by finding the reduced $\chi^2$:
\begin{equation}
    \chi^2_{\mathrm{red}} = (\boldsymbol{y}_\mathrm{{obs}} - \boldsymbol{y}_{\mathrm{sim}})^\mathrm{T} \boldsymbol{\Sigma}^{-1}_{o+s} (\boldsymbol{y}_\mathrm{{obs}}-\boldsymbol{y}_{\mathrm{sim}})\big/ N_{\mathrm{bins}},
\end{equation} 
where $\Sigma_{o+s}$ is the summed covariance matrix for observational data and simulations, and $\mathbf{y}_\mathrm{sim}$ refers to the coarsely binned Buzzard profile.

The bins are highly correlated for $m=0$, so the SNR is only 1.7 and 2.3 for the profiles without and with constraints, respectively. Both profiles agree well with Buzzard, with $\chi^2_{\mathrm{red}}<1$. The low SNR of this extended isotropic signal, and the requirement to subtract a constant offset, are both arguments that extended structure is not probed well by unoriented stacking. Notably, the same constant offset does not appear in $m=2$ or $m=4$. In addition, we achieve a higher-SNR detection with $m=2$.

\begin{figure}
\centering
\includegraphics[width=0.45\textwidth,trim={0.1cm 0cm 0cm 0cm},clip]{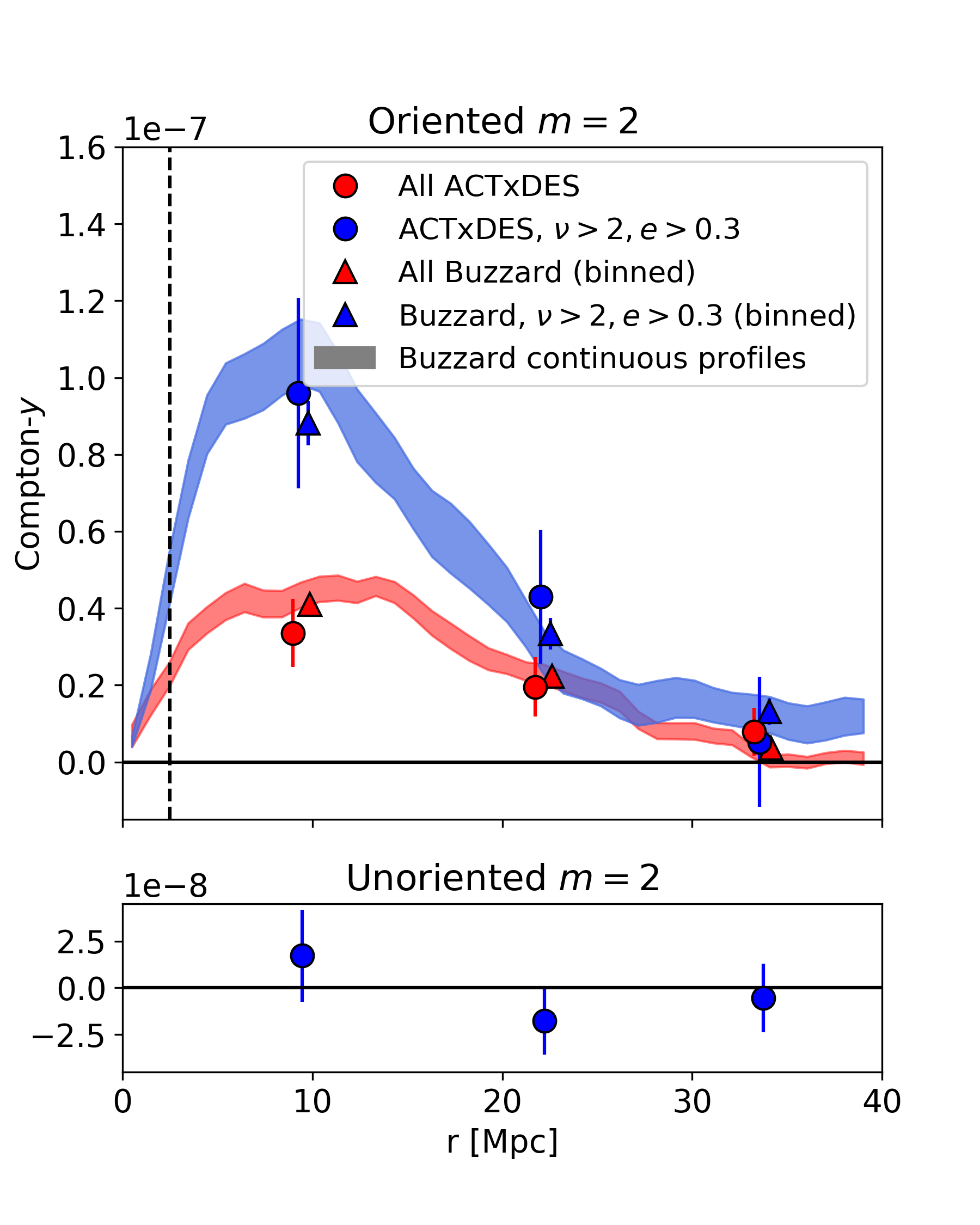}\\
    \includegraphics[width=0.47\textwidth,trim={0.5cm 1.5cm 0cm 2.5cm},clip]{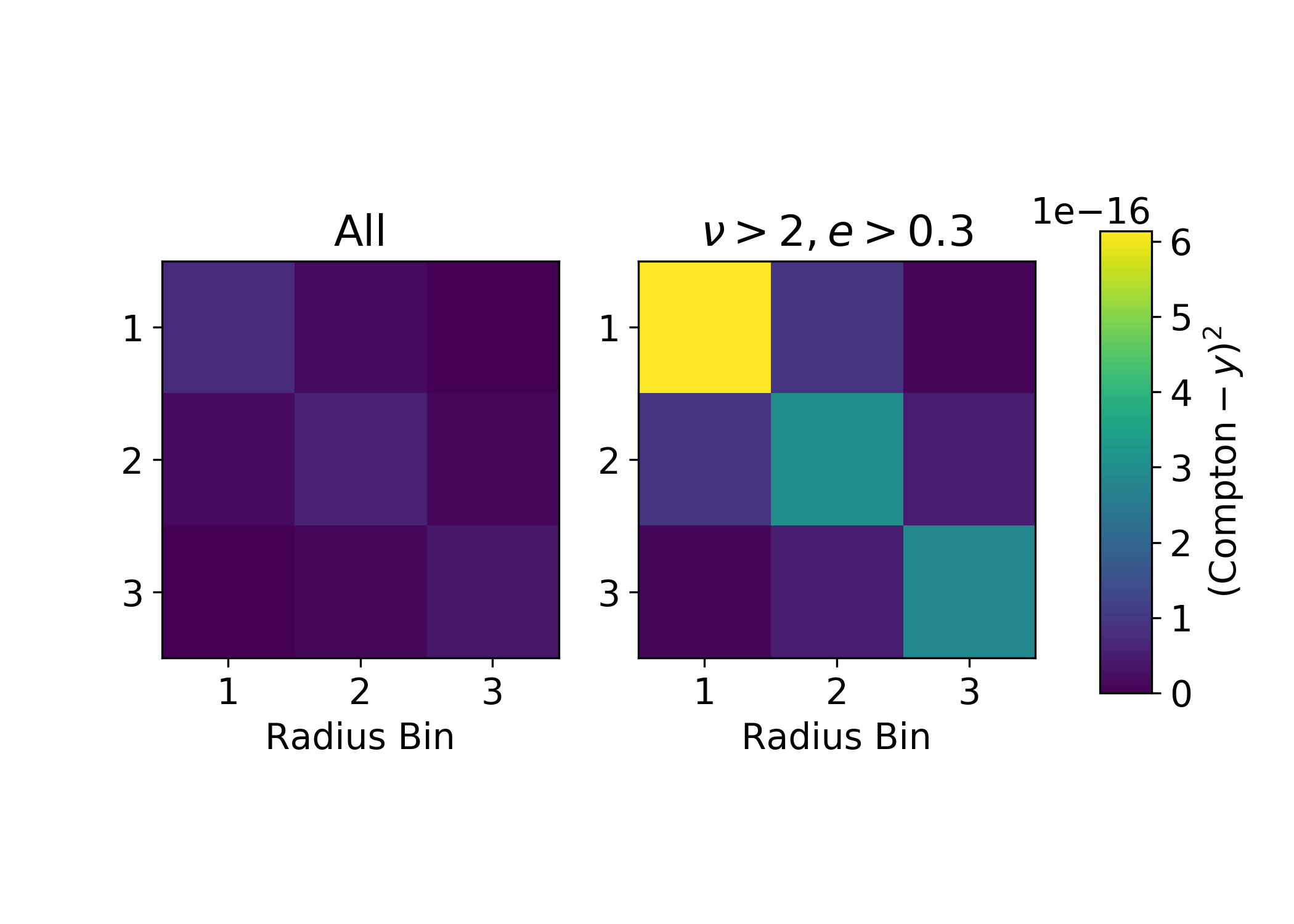}
    \caption{Top: Comparison of the $m=2$ component of the Buzzard (triangles) versus ACT$\times$DES (circles) oriented Compton-$y$ map stacks. The orientation and $\nu$ and $e$ selection were performed using a Gaussian smoothed galaxy field with FWHM=18 Mpc. The background shaded curves surround 1$\sigma$ regions above and below the continuous mean profile from Buzzard; the triangles correspond to coarse binning of this profile for direct comparison with real data. The first triangle falls below the curve because of the wide bin size which includes lower values on either side of the peak. Points are artificially spaced in $r$ for visual distinction. The attached lower panel shows a null test for comparison: the quadrupole profile for an ACT$\times$DES \textit{unoriented} stack using the same selected superclustering points. Bottom: The covariance matrices for each ACT data vector. The discrepancy between the visual appearance of the significance of the real data and the values reported in Table~\ref{tab:summary_stats} is due to the correlations between bins.}
    \label{fig:buzzard_act_comparison_13.9_m2}
\end{figure}

The $m=2$ moment of each stacked image, for the same 18 Mpc smoothing scale, is presented in Figure \ref{fig:buzzard_act_comparison_13.9_m2}. We again remove the inner cluster region when coarsely binning the data to avoid cluster mis-centering effects. The binning is slightly uneven, designed to place the first bin at the location of the Buzzard continuous-profile peak. The signal for all $\lambda>10$ clusters in Buzzard peaks at $y=4.5\times10^{-8}$. Enforcing ($\nu>2, e>0.3$) raises the signal at the peak by a factor of $\sim2.3$. The Buzzard results are in strong agreement with ACT$\times$DES. We also show a null test, the $m=2$ of an `unoriented' stack in which each cluster cutout was randomly rotated before stacking. This profile is consistent with zero as expected. Additionally, the figure shows the covariance matrices for both ACT$\times$DES data vectors.

We compare the $m=2$ signal to a completely null profile -- corresponding to a perfectly isotropic stack -- to determine its significance. Having no signal in $m=2$ would indicate that either (a) the average thermal energy distribution surrounding clusters is highly isotropic, (b) the galaxy distribution is uncorrelated with the gas distribution, or (c) our oriented stacking method does not effectively align large-scale structure. Table \ref{tab:summary_stats} presents the reduced-$\chi^2$, PTE, and SNR values. Without constraints on the galaxy field, the ACT$\times$DES oriented stacks have 3.5$\sigma$ level evidence for signal in the $m=2$ moment. Introducing the field constraints does not change the significance, as the larger errorbars compensate for the boost in signal. Table \ref{tab:buzzard_summary_stats} presents similar summary statistics for the Buzzard results, which have much higher significance (12$\sigma$ and 10$\sigma$) because of the lack of instrumental noise and larger cluster sample.

\begin{figure}
\centering
\includegraphics[width=0.45\textwidth,trim={0.1cm 0cm 0cm 0cm},clip]{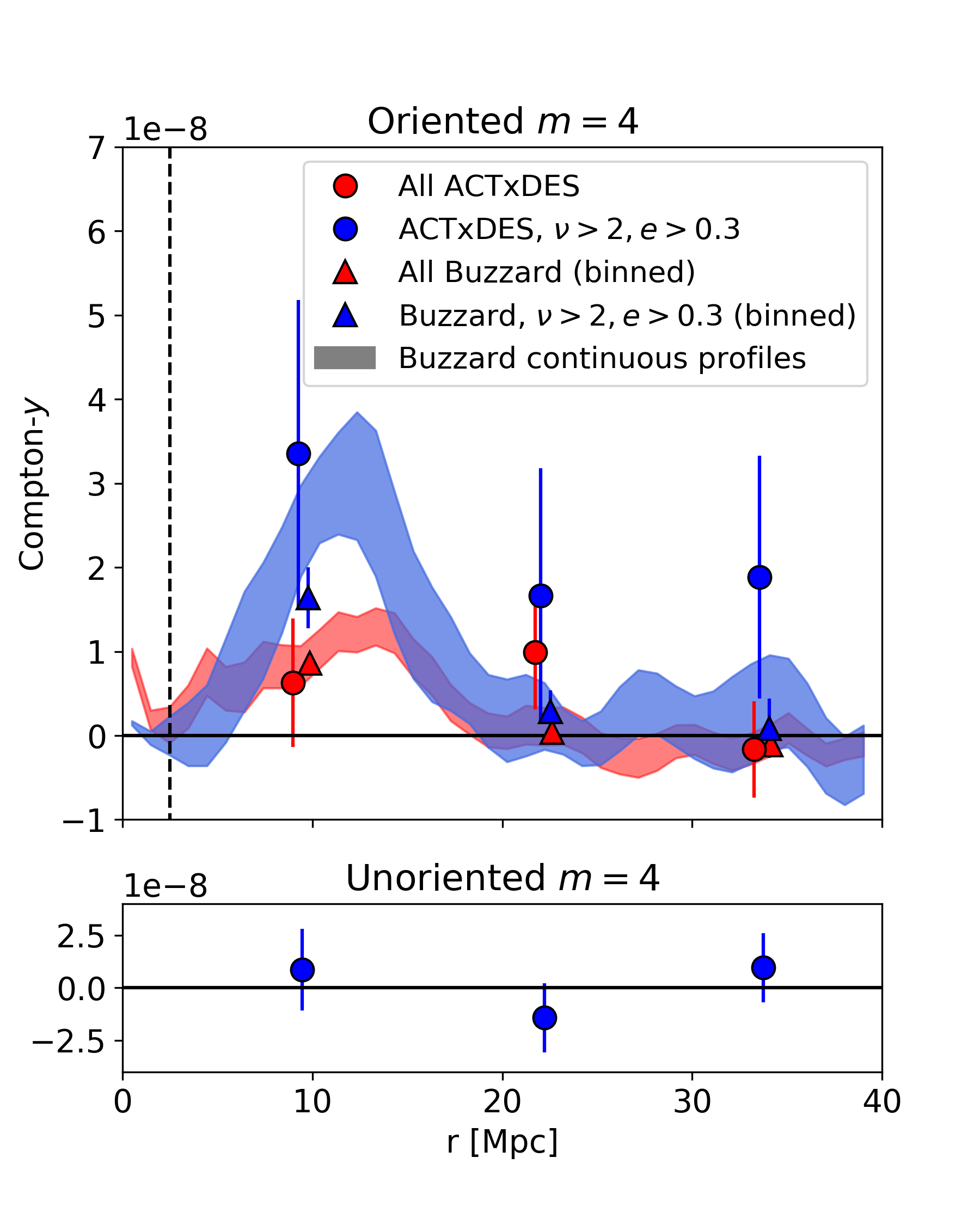}\\
\includegraphics[width=0.47\textwidth,trim={0.5cm 1.5cm 0cm 2.5cm},clip]{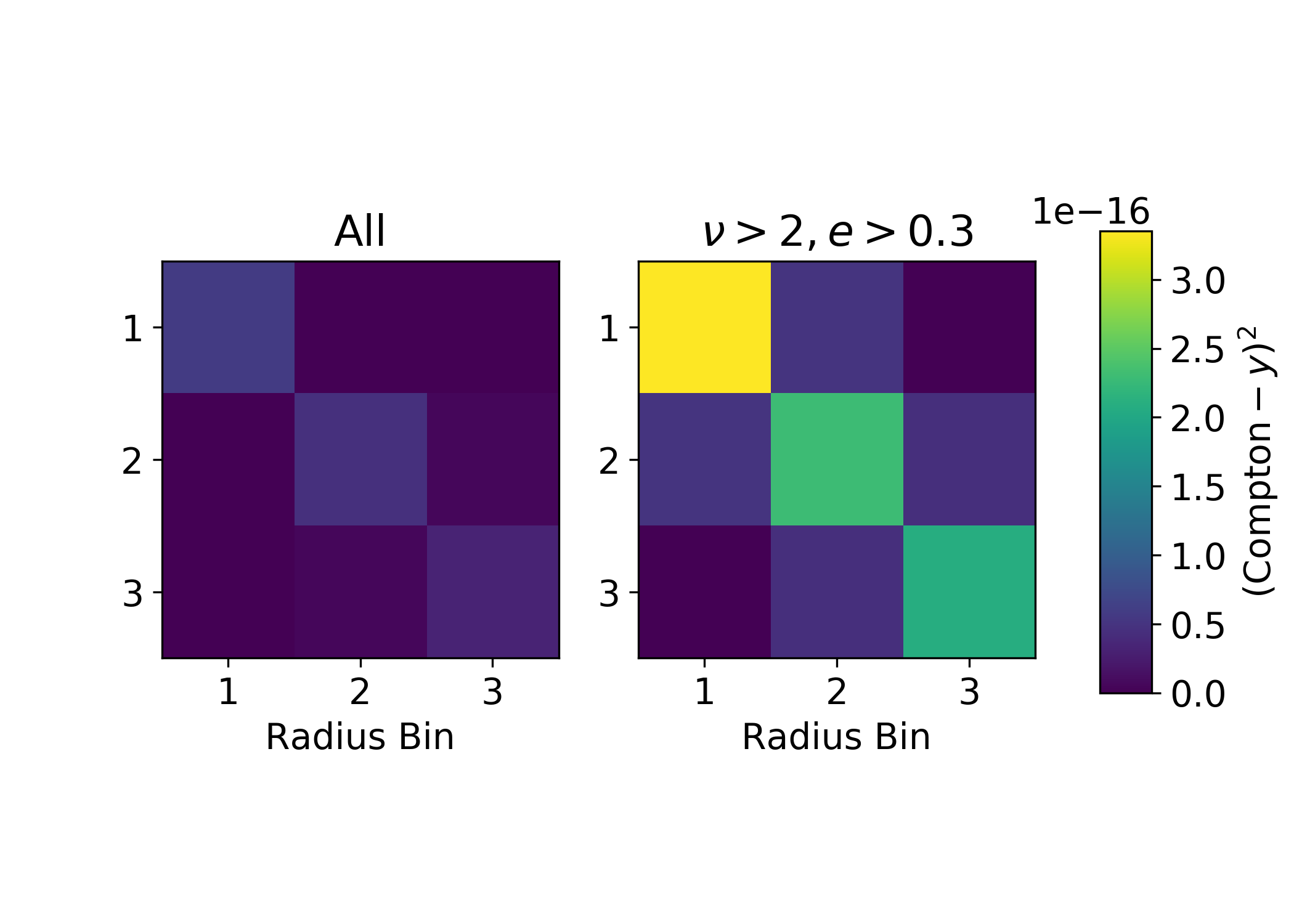}
    \caption{Top: Comparison of the $m=4$ component of Buzzard versus ACT$\times$DES for smoothing with FWHM=18Mpc. The attached lower panel shows the $m=4$ profile for an unoriented stack as a null test reference. The second red point and third blue point are significantly (1.3$\sigma$ and 1.1$\sigma$, respectively) higher in the real data than in the simulation. Because the tension is not extreme, this may be due to random fluctuations in the ACT $y$ map at these cluster locations; larger data sets will address whether there is a true physical difference in the average profile. Overall, however, the Buzzard and ACT$\times$DES profiles are in statistical agreement. Bottom: The covariance matrices for each ACT data vector. }
    \label{fig:buzzard_act_comparison_13.9_m4}
\end{figure}

Finally, the $m=4$ results are shown in Figure \ref{fig:buzzard_act_comparison_13.9_m4}. The same bins are applied to $m=4$ as $m=2$. The Buzzard profiles peak at $\sim4$ times smaller $y$ values than $m=2$. For this component, a null profile could correspond to any of the (a), (b), and (c) possibilities listed for $m=2$ or the signal from a perfectly Gaussian random field. As shown in Tables \ref{tab:buzzard_summary_stats} and \ref{tab:summary_stats}, the Buzzard $m=4$ component from the constrained cluster sample is at $2.8\sigma$, but the corresponding ACT$\times$DES result is only 1.5$\sigma$. Therefore, we can not claim evidence for the $m=4$ component in observations. The $m=4$ profiles are in agreement with Buzzard.

% The smaller, noisier signal is unsurprising, since the poles along the horizontal image axis contain signal while the poles along the vertical axis contribute only noise to $m=4$. ML note as to why I deleted this sentence: I'm not sure this should make it smaller. The division in eq 7 is the same for both 2 and 4

\begin{deluxetable}{ccccccc}
\tablewidth{0pt}
\tablecaption{Summary of the ACT$\times$DES results.\label{tab:summary_stats}}
\tablehead{
\colhead{m} & \colhead{Cuts (Y/N)} & \colhead{$N$} & \colhead{PTE} & \colhead{SNR} & \multicolumn{2}{c}{$\chi^2_{\mathrm{red}}$  (d.o.f. = 3)} \\ \cline{6-7} \colhead{} & \colhead{} & \colhead{} & \colhead{} & \colhead{} & \colhead{absolute} & \colhead{data-sim}}
\tablecolumns{7}
\startdata
\cutinhead{FWHM of smoothing = $10\, \mathrm{Mpc}$}
2 & Y & 1190 & 0.06 & 2.6 & 1.8 & 0.9\\
4 & Y & 1190 & 0.3 & 2.4 & 3.9 & 1.2\\
\cutinhead{FWHM of smoothing = $14\, \mathrm{Mpc}$}
2 &  Y & 1103 & 0.004 & 3.2 & 5.2 & 0.5\\
4 &  Y & 1103 & 0.2 & 1.5 & 1.8 & 1.1\\
\cutinhead{FWHM of smoothing = $18\,\mathrm{Mpc}$}
2 & N & 5494 & 0.0004 & 3.5 & 6.0 & 0.1\\
4 & N & 5494 & 0.4 & 0.9 & 1.0 & 0.2\\
2 & Y & 975 & 0.0005 & 3.5 & 5.9 & 0.07\\
4 & Y & 975 & 0.1 & 1.5 & 1.8 & 0.3\\
\enddata
\tablecomments{From left to right, $m$ is the multipole moment, `Cuts' refers to whether or not the cluster sample has been constrained by $\nu>2, e>0.3$; $N$ is the number of stacked clusters, PTE is the probability to exceed, SNR gives the number of Gaussian $\sigma$ from null, and $\chi^2_{\mathrm{red}}$ is the reduced $\chi^2$ value using 3 degrees of freedom (d.o.f.). $\chi^2_{\mathrm{red}}$ is shown for the absolute value with respect to 0 as well as the value with respect to Buzzard. The highest signal-to-noise detections of extended signal come from the $m=2$ components of the $14\,\mathrm{Mpc}$ and $18\,\mathrm{Mpc}$ smoothed stacks. Data and simulation are generally in agreement, with $\chi^2_{\mathrm{red}}$(data-sim) consistently near or below 1.}
\end{deluxetable}

\begin{deluxetable}{ccccc}
\tablewidth{0pt}
\tablecaption{Summary of the Buzzard results.\label{tab:buzzard_summary_stats}}
\tablehead{
\colhead{m} & \colhead{Cuts (Y/N)} & \colhead{$N$} & \colhead{SNR} & \colhead{$\chi^2_{\mathrm{red}}$  (d.o.f. = 3)}}
\tablecolumns{5}
\startdata
\cutinhead{FWHM of smoothing = $10\, \mathrm{Mpc}$}
        2 & Y & 6165 & 10 & 31\\
        4 & Y & 6165 & 4.8 & 7.8\\
\cutinhead{FWHM of smoothing = $14\, \mathrm{Mpc}$}
        2 & Y & 5401 & 11 & 37\\
        4 & Y & 5401 & 5.3 & 9.2\\
\cutinhead{FWHM of smoothing = $18\,\mathrm{Mpc}$}
        2 & N & 24,922 & 12 & 51\\
        4 & N & 24,922 & 3.9 & 5.1\\
        2 & Y & 4,836  & 9.5 & 30\\
        4 & Y & 4,836  & 2.8 & 2.6\\
\enddata
\tablecomments{The table repeats the calculations in Table~\ref{tab:summary_stats} for the Buzzard simulations, with the same 3 radial bins. $\chi^2_{\mathrm{red}}$ is only shown for the `absolute' value, as the (data-sim) value is already presented in Table~\ref{tab:summary_stats}. Because the simulation results are high-SNR, rather than calculating a PTE we simply estimate SNR by taking the square root of $\chi^2$. Generally, the detection significance is much higher for Buzzard than for ACT$\times$DES because the simulation is noiseless. $m=2$ is detected at a much higher significance than $m=4$.}
\end{deluxetable}

\begin{figure}[htbp!]
\centering
    \includegraphics[width=0.43\textwidth,trim={0.0cm 0.0cm 0.0cm 0cm},clip]{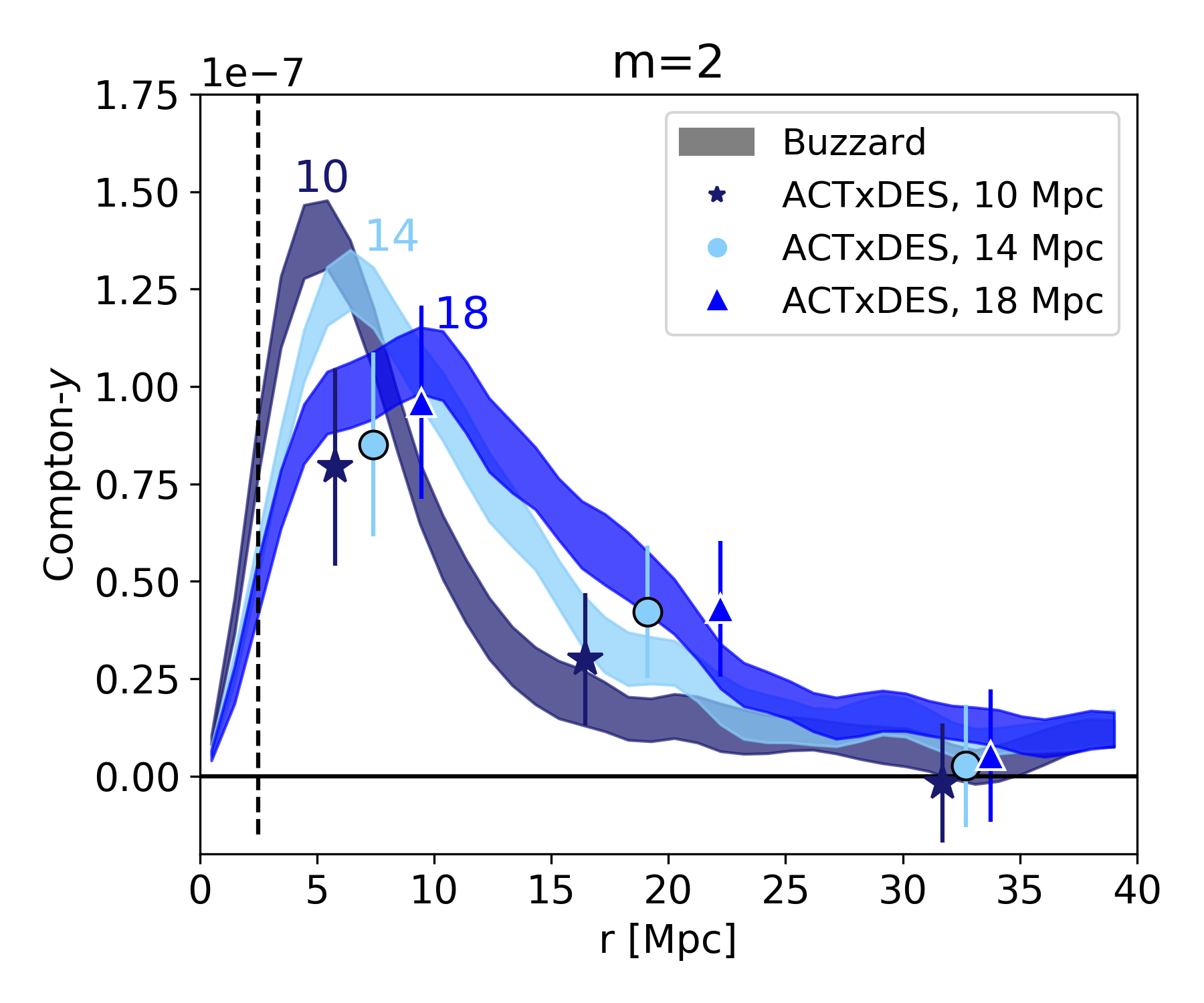}\\
    \includegraphics[width=0.43\textwidth,trim={0cm 0cm 0cm 0cm},clip]{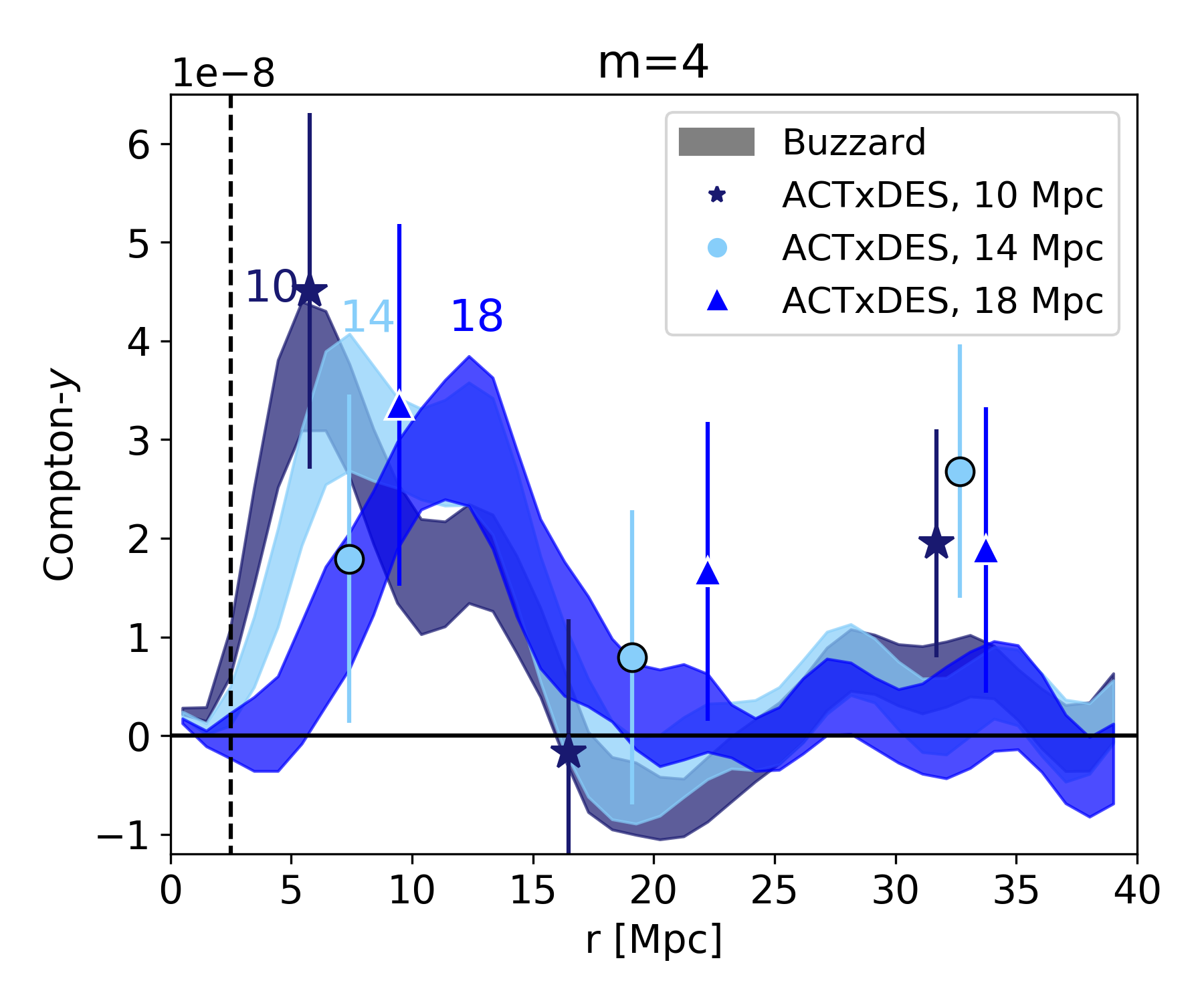}
    \caption{Comparison of the $m=2$ component (top) and $m=4$ component (bottom) of Buzzard versus ACT$\times$DES for 3 smoothing scales: FWHM=[10,14,18] Mpc. The Buzzard finely-binned data are shown for visual purposes, but is not directly comparable to the ACT$\times$DES points because of the difference in binning (see Table \ref{tab:summary_stats} for a statistical comparison with the same binning). Points are artificially spaced in $r$ for visual purposes. The cluster sample varies slightly for each scale due to the scale-dependent $\nu>2, e>0.3$ galaxy field constraints. The differences in profiles are mostly due to the dependence of orientation on smoothing, as discussed in Section~\ref{sec:parameter_dependence}. Measurements for the three scales are highly correlated with each other due to the alignment of cosmic web structure across a wide range of scales, so they should not be interpreted as independent. All data points at $r\sim33$ Mpc are significantly higher than the respective binned Buzzard point, but as this tension is at the 1.5$\sigma$ level at most, more data are necessary to determine whether this is a random fluctuation or true difference.}
    \label{fig:buzzard_act_comparison_3scale_m24}
\end{figure}

We repeat the stacking procedure for two smaller smoothing scales, FWHM=10 Mpc and FWHM=14 Mpc, with the $\nu$ and $e$ contraints enforced for the respective smoothed maps. The $m=2,4$ plots are shown in Figure \ref{fig:buzzard_act_comparison_3scale_m24}. The binning is adjusted for each scale to align the first bin with the peak location. The continuous profiles in the figures are shown for visual purposes but cannot be directly compared to the binned real data; instead, this comparison is best captured in the $\chi^2_{\mathrm{red}}$ column of Table \ref{tab:summary_stats}. 

There is evidence for $m=2$ signal at the 3.2$\sigma$ level for 14 Mpc smoothing and marginal 2.6$\sigma$ evidence for 10 Mpc. The SNR of $m=4$ is smaller in all cases, although there is marginal $2.4\sigma$ for an $m=4$ component in the smallest smoothing scale. This is consistent with the expectation that non-Gaussianity is more pronounced on smaller scales. However, because the evidence does not meet the 3$\sigma$ threshold, we leave further analysis to future work with larger data sets. Generally, the observed data are in very good agreement with the simulations, reaching at most a reduced-$\chi^2$ of 1.2.

Due to the lack of detection in $m=4$, we cannot claim to find evidence for non-Gaussian structure with the currently available data. For $m=2$, as discussed in Section~\ref{subsec:grf_comparison}, the boost from applying the $\nu$ and $e$ thresholds only occurs in realistic fields and not in a Gaussian random field. Therefore, it is likely that the strength of $m=2$ in stacks on the constrained cluster sample is indicative of non-Gaussianity. However, to robustly address this, we would need to repeat the study in Section~\ref{subsec:grf_comparison} with the exact methods applied to the final results (using thin redshift slicing, rescaling, and combining multiple slices). We leave this detailed comparison for future work.

\begin{figure}
    \includegraphics[width=.43\textwidth]{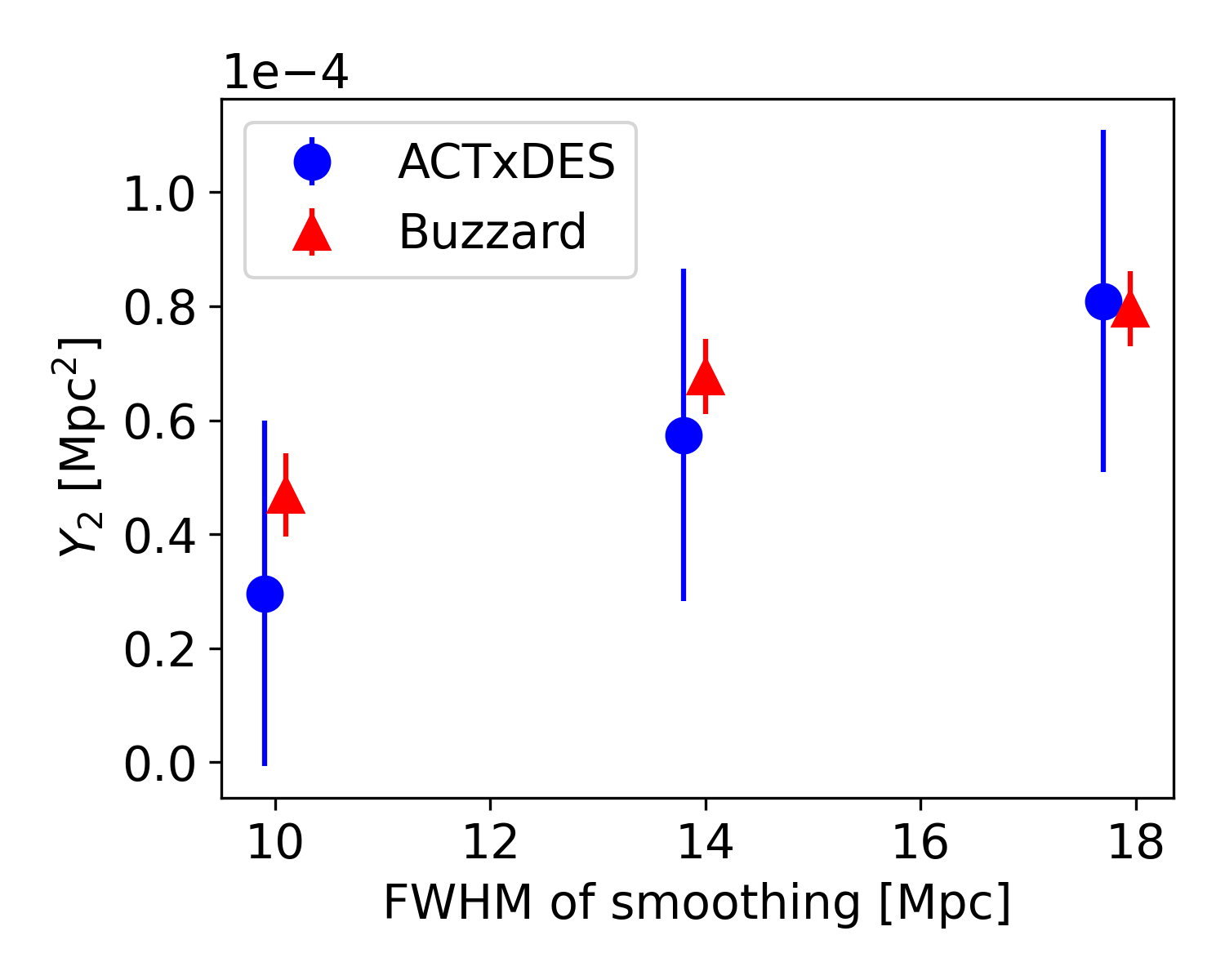}
    \caption{Comparison of the integrated $m=2$ profile to R=40 Mpc between Buzzard and ACT$\times$DES for three smoothing scales. At all scales, the simulated and observed results are in agreement within 1$\sigma$.}
    \label{fig:buzzard_act_comparison_3scale_integral}
\end{figure}

As discussed in Section \ref{sec:parameter_dependence}, the shape of each profile is highly dependent on smoothing scale and thus the integrated profile provides a useful single-value quantity for comparison between data and simulations. Figure \ref{fig:buzzard_act_comparison_3scale_integral} shows the integrated $Y$ signal (computed with a Riemann sum) over the Buzzard (red) and ACT$\times$DES (blue) coarsely-binned $m=2$ profiles for three smoothing scales. Errors are propagated with the covariance matrix for each scale. At all scales, the Buzzard and ACT integrals are within 1$\sigma$ of each other. The ACT integrals have SNR=[1.0, 2.0, 2.7] for the [10, 14, 18] Mpc smoothing scales. Therefore, we find marginal indications of integrated anisotropic clustering of thermal energy for the two coarser scales.

\section{Conclusions}\label{sec:conclusion}
\subsection{Overview}
This paper introduced a new real-space method for probing anisotropies in gas signal in the cosmic web. By combining millimeter-wavelength data from the ACT CMB survey with optical data from the DES galaxy survey, we measured the average superclustering signal from hot gas surrounding DES redMaPPer clusters. We showed that there is a significant Compton-$y$ signal from non-Gaussianity in the  late-time universe visible in the $m=4$ moment of simulated stacks. There is a marginal indication for $m=4$ signal in the observational data. Using characteristics of the galaxy field, we identified highly overdense, elongated regions. Selecting for clusters in these high-superclustering areas enhances the gas signal in the $m=2$ moment of oriented stacks, causing a distinction from a Gaussian random field. We visually demonstrated this enhancement in the observational data, leaving a more rigorous proof of non-Gaussianity to future work. Generally, with observed ACT$\times$DES data, we found marginal-to-significant evidence for extended signal in the $m=2$ moments; the significance depended on the chosen galaxy field smoothing. This evidence demonstrates that the average thermal energy distribution around clusters is not isotropic. This, of course, is expected because clusters are embedded in filamentary structures in the cosmic web spanning a wide range of scales. Comparing to the Buzzard mocks showed broad agreement in the $m=2$ and $m=4$ radial profiles. For the integrated Compton-$y$ from $m=2$, the results at all scales are in agreement.

We did not attempt to constrain the contribution of the WHIM, due to gas outside individual halos, to the anisotropic $y$ signal. In a previous filament-stacking study, \citet{deGraaff2019} found a filament signal with $y$ = $6\times10^{-9}$. The team also found that bound gas in halos only contributes $\sim20$\% of this signal. A WHIM signal at $y\sim5\times10^{-9}$, then, is $\sim5\%$ of the peak $m=2$ signal measured in our work. If the \citet{deGraaff2019} study was correct, this indicates that as expected, the dominant tSZ signal from the extended large-scale structure surrounding clusters comes from bound gas in halos. The previously-measured WHIM signal is smaller than any of the 1$\sigma$ error bars in our work and thus undetectable with the current method and data. A detection may be possible with the currently available data by stacking on all \textit{galaxies} instead of clusters, as well as by applying a more sophisticated approach to the combination of smoothing scales and field constraints. A WHIM detection may also be possible with the future, expanded version of the ACT $y$ map and greater overlap with DES cluster data. We leave a detailed study of how oriented stacking can be used to characterize the WHIM for future work.

With the Websky full-sky simulations, we demonstrated the potential of the oriented stacking method to probe the evolution of superclustering out to $z\sim2$ and the relationship between cluster mass and the surrounding environment. These theory results demonstrated that the anisotropic gas signal from superclustering is expected to grow with time from $z\sim2$ to $z\sim 0.7$, then stabilize. For fixed redshift, the strength of superclustering correlates with cluster mass, indicating that massive clusters are embedded in more massive and/or denser filaments. The limited number of available clusters in the ACT $y$ map overlapping with DES prohibited the same analyses on real data.

\subsection{Systematics and Future Outlook} \label{subsec:systematics}

We identify a few sources of systematic uncertainty which were negligible in this study but will become important as the ACT data expands and improves in the next few years. First, as noted in Section~\ref{sec:Data}, $\lambda<20$ \redmapper clusters are known to contain false detections. False clusters should be less correlated with the surrounding large-scale structure than real clusters, and therefore their inclusion should bias our results low. Nevertheless, we found evidence for superclustering in ACT$\times$DES, indicating that the impact of cluster sample impurities was not enough to drown out the signal. Estimating how much higher the signal would be with a completely clean $\lambda>10$ sample is beyond the scope of this work, as these low-richness impurities have not yet been well-characterized.

A possible source of systematic error in the comparison with simulations, as noted in Section~\ref{sec:Data}, is the failure of Buzzard to reproduce the \redmapper mass-richness relation. As the errors on ACT$\times$DES decrease with future data, this may become a source of tension between the simulation and data. However, there are reasons to suspect this has a minor effect on our measurements compared to other factors such as the imperfect gas prescription applied to Buzzard. First, because this work does not focus on the cluster interior, discrepancies in the richness distribution are only important to the extent that they correlate with the surrounding large-scale structure. We found in section \ref{sec:parameter_dependence} that cluster richness is indeed correlated with the anisotropic Compton-$y$ signal; however, this is subdominant compared to the dependence on $\nu$ evaluated at larger scales. Additionally, because the mock Compton-$y$ map is created using only halo information from the Buzzard dark-matter-only simulation, the $y$ map signal does not depend on the realistic identification of \redmapper clusters. The most important steps in our pipeline are the selection and orientation of clusters using the \redmagic mock galaxy data, where Buzzard does well in matching DES. It may be possible to entirely avoid the mass-richness systematic in future work by using the Buzzard state-of-the-art: a recent version of Buzzard improves the color-dependent clustering which heavily affects \redmapper selection \citep{DeRose2021}. In general, to fully understand the factors influencing the oriented tSZ signal, we must disentangle the effects of cosmology, galaxy and cluster modelling/selection, and gas prescription. A succeeding paper will address this by applying oriented stacking to the galaxy distribution alone and comparing the DES results to Buzzard.

Moving forward, future Compton-$y$ maps from ACT will cover a sky area with complete overlap with DES over the 5000 sq. deg. footprint \citep{Naess2020}. They will additionally include high-resolution 30, 40, and 230 GHz data collected using Advanced ACT arrays, allowing for better removal of contamination from Galactic synchrotron sources, extragalactic radio sources, the cosmic infrared background, and dust \citep{Madhavacheril2020}. Our study will be repeated with the full cluster sample from DES, which is approximately 14 times larger than the sample used in this study. The systematic error contribution is at the few percent level for the current $y$ map, so random errors will still dominate. Therefore, the error bars on the data are expected to shrink by a factor of $\sqrt{14}=3.7$. If the binned $y$ values in the $m=2$ profile for 18 Mpc smoothing remained the same with 3.7$\times$ smaller error bars, the detection of $m=2$ signal after $\nu$ and $e$ cuts will be at the 13$\sigma$ level. The $m=4$ component would be detectable at the 5$\sigma$ level for the same scale. Slight tension would emerge with Buzzard, with a $\chi^2_{\mathrm{red}}$ of 2.5. With the next iteration of ACT data, oriented stacking will be able to address questions of non-Gaussianity and assess the accuracy of the tSZ pasting prescriptions in simulations in more detail.

Additionally, the larger cluster sample will allow us to remove low-$\lambda$ clusters and repeat the study with a more pure, higher-richness sample. This is expected to boost the superclustering gas signal. In a succeeding paper, we will use this expanded data to study the impact of the WHIM on the superclustering signal from hot gas.

Looking further to the future, The Simons Observatory \citep[SO,] []{SO2019} is currently being built in Chile, expected to begin taking CMB observations in the mid-2020s. With measurements in 6 frequencies, the SO Compton-$y$ map will attain lower noise levels and higher resolution than that of ACT. In addition, upcoming telescopes like the Fred Young Sub-millimeter Telescope \citep[fomerly CCAT-p,][]{ccatp:2019} and the future CMB S4 mission \citep{cmbs4/whitepaper} will further advance tSZ measurements by mapping the millimeter sky with improved sensitivity and frequency coverage.

To optimize the science returns from oriented stacking, these higher quality CMB maps will need to be coupled with augmented cluster and galaxy data. The SO team predicts that the number of clusters detected through the tSZ effect will rise by an order of magnitude from current levels \citep{SO2019}. The sample will extend to further redshifts than the DES red-sequence-detected sample because the tSZ effect is redshift-independent, whereas received flux from optical galaxy observations (as in DES) diminishes with redshift. SO forecasts $\sim200$ detected clusters in a bin of width $\Delta z=0.1$ at $z\sim1.5$, whereas the DES RedMaPPer catalog extends only to $z\sim0.9$. Higher-redshift clusters will enable us to measure the evolution of superclustering further into the universe's past. In addition, the Dark Energy Spectroscopic Instrument \citep[DESI,][]{DESI2019} will provide many new spectroscopic redshifts for galaxies and clusters, enabling improved tomographic slicing to study redshift evolution of the LSS \citep{DESI_cluster}. The Vera C. Rubin Observatory Legacy Survey of Space and Time \citep[LSST,][]{LSST2019} will make unprecedented contributions to the optical galaxy and cluster data with deep observations of the Southern sky resulting in an estimated 20 billion galaxies. Looking to other wavelengths, eROSITA is expected to deliver hundreds of superclusters detected in the X-ray: additional objects on which to stack the tSZ signal \citep{eROSITA}. Applying the methods detailed in this paper to combinations of these varied data sets, where they overlap, will greatly increase the signal-to-noise and redshift extent of superclustering measurements.

This paper demonstrates a flexible tool for examining superclustering in the universe. We have shown its effectiveness for a combination of galaxy and gas data, but it can also be used to combine galaxy and weak lensing data or to probe the anisotropies of the galaxy field alone. By combining different probes, this method can be implemented to study the anisotropic bias of galaxies and gas with respect to dark matter. A follow-up paper will use oriented stacking on galaxy number density maps and compare the results to extended gas signal to probe information about the anisotropic bias, the baryonic content of halos, and possibly the baryons outside of halos. This may provide more insight into the census of baryonic content in filaments. In addition, future work will determine whether this method of measuring anisotropic clustering can complement two-point clustering statistics in constraining viable models of dark energy.

\section*{Acknowledgments}

We thank the anonymous referee for providing valuable comments which improved the quality of this paper.

Canadian co-authors acknowledge support from the Natural Sciences and Engineering Research Council of Canada. Computations were performed on the SciNet supercomputer at the SciNet HPC Consortium. SciNet is funded by: the Canada Foundation for Innovation; the Government of Ontario; Ontario Research Fund - Research Excellence; and the University of Toronto. 

The authors thank Bruce Partridge for useful and extensive comments on the draft.

JRB's research was funded by the Natural Sciences and Engineering Research Council of Canada Discovery Grant Program and a fellowship from the Canadian Institute for Advanced Research (CIFAR) Gravity and Extreme Universe program.

ADH acknowledges support from the Sutton Family Chair in Science, Christianity and Cultures.

R.~H. is a CIFAR Azrieli Global Scholar, Gravity \& the Extreme Universe Program, 2019, and a 2020 Alfred P. Sloan Research Fellowship. RH is supported by Natural Sciences and Engineering Research Council of Canada Discovery Grant Program and the Connaught Fund. 

JPH acknowledges funding for SZ cluster studies from NSF AAG number AST-1615657.

KM acknowledges support from the National Research Foundation of South Africa.

This work was supported by the U.S. National Science Foundation through awards AST-0408698, AST-0965625, and AST-1440226 for the ACT project, as well as awards PHY-0355328, PHY-0855887 and PHY-1214379. Funding was also provided by Princeton University, the University of Pennsylvania, and a Canada Foundation for Innovation (CFI) award to UBC. ACT operates in the Parque Astron\'omico Atacama in northern Chile under the auspices of the Comisi\'on Nacional de Investigaci\'on (CONICYT). The development of multichroic detectors and lenses was supported by NASA grants NNX13AE56G and NNX14AB58G. Detector research at NIST was supported by the NIST Innovations in Measurement Science program. 

Funding for the DES Projects has been provided by the U.S. Department of Energy, the U.S. National Science Foundation, the Ministry of Science and Education of Spain, 
the Science and Technology Facilities Council of the United Kingdom, the Higher Education Funding Council for England, the National Center for Supercomputing 
Applications at the University of Illinois at Urbana-Champaign, the Kavli Institute of Cosmological Physics at the University of Chicago, 
the Center for Cosmology and Astro-Particle Physics at the Ohio State University,
the Mitchell Institute for Fundamental Physics and Astronomy at Texas A\&M University, Financiadora de Estudos e Projetos, 
Funda{\c c}{\~a}o Carlos Chagas Filho de Amparo {\`a} Pesquisa do Estado do Rio de Janeiro, Conselho Nacional de Desenvolvimento Cient{\'i}fico e Tecnol{\'o}gico and 
the Minist{\'e}rio da Ci{\^e}ncia, Tecnologia e Inova{\c c}{\~a}o, the Deutsche Forschungsgemeinschaft and the Collaborating Institutions in the Dark Energy Survey. 

The Collaborating Institutions are Argonne National Laboratory, the University of California at Santa Cruz, the University of Cambridge, Centro de Investigaciones Energ{\'e}ticas, 
Medioambientales y Tecnol{\'o}gicas-Madrid, the University of Chicago, University College London, the DES-Brazil Consortium, the University of Edinburgh, 
the Eidgen{\"o}ssische Technische Hochschule (ETH) Z{\"u}rich, 
Fermi National Accelerator Laboratory, the University of Illinois at Urbana-Champaign, the Institut de Ci{\`e}ncies de l'Espai (IEEC/CSIC), 
the Institut de F{\'i}sica d'Altes Energies, Lawrence Berkeley National Laboratory, the Ludwig-Maximilians Universit{\"a}t M{\"u}nchen and the associated Excellence Cluster Universe, 
the University of Michigan, NFS's NOIRLab, the University of Nottingham, The Ohio State University, the University of Pennsylvania, the University of Portsmouth, 
SLAC National Accelerator Laboratory, Stanford University, the University of Sussex, Texas A\&M University, and the OzDES Membership Consortium.

Based in part on observations at Cerro Tololo Inter-American Observatory at NSF's NOIRLab (NOIRLab Prop. ID 2012B-0001; PI: J. Frieman), which is managed by the Association of Universities for Research in Astronomy (AURA) under a cooperative agreement with the National Science Foundation.

The DES data management system is supported by the National Science Foundation under Grant Numbers AST-1138766 and AST-1536171.
The DES participants from Spanish institutions are partially supported by MICINN under grants ESP2017-89838, PGC2018-094773, PGC2018-102021, SEV-2016-0588, SEV-2016-0597, and MDM-2015-0509, some of which include ERDF funds from the European Union. IFAE is partially funded by the CERCA program of the Generalitat de Catalunya.
Research leading to these results has received funding from the European Research
Council under the European Union's Seventh Framework Program (FP7/2007-2013) including ERC grant agreements 240672, 291329, and 306478.
We  acknowledge support from the Brazilian Instituto Nacional de Ci\^encia
e Tecnologia (INCT) do e-Universo (CNPq grant 465376/2014-2).

This manuscript has been authored by Fermi Research Alliance, LLC under Contract No. DE-AC02-07CH11359 with the U.S. Department of Energy, Office of Science, Office of High Energy Physics.

This work received support from the U.S. Department of Energy under contract number DE-AC02-76SF00515 at SLAC National Accelerator Laboratory. This research used computing resources at SLAC National Accelerator Laboratory and at the National Energy Research Scientific Computing Center (NERSC), a U.S. Department of Energy Office of Science User Facility located at Lawrence Berkeley National Laboratory, operated under Contract No. DE-AC02-05CH11231.

\software{Astropy \citep{astropy:2013, astropy:2018}, COOP \citep{Huang2016}, Healpy \citep{Zonca2019}, HEALPix \citep{Healpix2005}, NumPy \citep{harris2020array}}

\bibliography{superclustering}

\begin{thebibliography}{}
\expandafter\ifx\csname natexlab\endcsname\relax\def\natexlab#1{#1}\fi
\providecommand{\url}[1]{\href{#1}{#1}}
\providecommand{\dodoi}[1]{doi:~\href{http://doi.org/#1}{\nolinkurl{#1}}}
\providecommand{\doeprint}[1]{\href{http://ascl.net/#1}{\nolinkurl{http://ascl.net/#1}}}
\providecommand{\doarXiv}[1]{\href{https://arxiv.org/abs/#1}{\nolinkurl{https://arxiv.org/abs/#1}}}

\bibitem[{{Abbott} {et~al.}(2020){Abbott}, {Aguena}, {Alarcon}, {Allam},
  {Allen}, {Annis}, {Avila}, {Bacon}, {Bechtol}, {Bermeo}, {Bernstein},
  {Bertin}, {Bhargava}, {Bocquet}, {Brooks}, {Brout}, {Buckley-Geer}, {Burke},
  {Carnero Rosell}, {Carrasco Kind}, {Carretero}, {Castander}, {Cawthon},
  {Chang}, {Chen}, {Choi}, {Costanzi}, {Crocce}, {da Costa}, {Davis}, {De
  Vicente}, {DeRose}, {Desai}, {Diehl}, {Dietrich}, {Dodelson}, {Doel},
  {Drlica-Wagner}, {Eckert}, {Eifler}, {Elvin-Poole}, {Estrada}, {Everett},
  {Evrard}, {Farahi}, {Ferrero}, {Flaugher}, {Fosalba}, {Frieman},
  {Garc{\'\i}a-Bellido}, {Gatti}, {Gaztanaga}, {Gerdes}, {Giannantonio},
  {Giles}, {Grandis}, {Gruen}, {Gruendl}, {Gschwend}, {Gutierrez}, {Hartley},
  {Hinton}, {Hollowood}, {Honscheid}, {Hoyle}, {Huterer}, {James}, {Jarvis},
  {Jeltema}, {Johnson}, {Johnson}, {Kent}, {Krause}, {Kron}, {Kuehn},
  {Kuropatkin}, {Lahav}, {Li}, {Lidman}, {Lima}, {Lin}, {MacCrann}, {Maia},
  {Mantz}, {Marshall}, {Martini}, {Mayers}, {Melchior}, {Mena-Fern{\'a}ndez},
  {Menanteau}, {Miquel}, {Mohr}, {Nichol}, {Nord}, {Ogando}, {Palmese},
  {Paz-Chinch{\'o}n}, {Plazas}, {Prat}, {Rau}, {Romer}, {Roodman}, {Rooney},
  {Rozo}, {Rykoff}, {Sako}, {Samuroff}, {S{\'a}nchez}, {Sanchez}, {Saro},
  {Scarpine}, {Schubnell}, {Scolnic}, {Serrano}, {Sevilla-Noarbe}, {Sheldon},
  {Smith}, {Smith}, {Suchyta}, {Swanson}, {Tarle}, {Thomas}, {To}, {Troxel},
  {Tucker}, {Varga}, {von der Linden}, {Walker}, {Wechsler}, {Weller},
  {Wilkinson}, {Wu}, {Yanny}, {Zhang}, {Zhang}, {Zuntz}, \& {DES
  Collaboration}}]{Abbott2020}
{Abbott}, T.~M.~C., {Aguena}, M., {Alarcon}, A., {et~al.} 2020, \prd, 102,
  023509, \dodoi{10.1103/PhysRevD.102.023509}

\bibitem[{{Ade} {et~al.}(2019){Ade}, {Aguirre}, {Ahmed}, {Aiola}, {Ali},
  {Alonso}, {Alvarez}, {Arnold}, {Ashton}, {Austermann}, {Awan}, {Baccigalupi},
  {Baildon}, {Barron}, {Battaglia}, {Battye}, {Baxter}, {Bazarko}, {Beall},
  {Bean}, {Beck}, {Beckman}, {Beringue}, {Bianchini}, {Boada}, {Boettger},
  {Bond}, {Borrill}, {Brown}, {Bruno}, {Bryan}, {Calabrese}, {Calafut},
  {Calisse}, {Carron}, {Challinor}, {Chesmore}, {Chinone}, {Chluba}, {Cho},
  {Choi}, {Coppi}, {Cothard}, {Coughlin}, {Crichton}, {Crowley}, {Crowley},
  {Cukierman}, {D'Ewart}, {D{\"u}nner}, {de Haan}, {Devlin}, {Dicker},
  {Didier}, {Dobbs}, {Dober}, {Duell}, {Duff}, {Duivenvoorden}, {Dunkley},
  {Dusatko}, {Errard}, {Fabbian}, {Feeney}, {Ferraro}, {Flux{\`a}}, {Freese},
  {Frisch}, {Frolov}, {Fuller}, {Fuzia}, {Galitzki}, {Gallardo}, {Tomas Galvez
  Ghersi}, {Gao}, {Gawiser}, {Gerbino}, {Gluscevic}, {Goeckner-Wald}, {Golec},
  {Gordon}, {Gralla}, {Green}, {Grigorian}, {Groh}, {Groppi}, {Guan},
  {Gudmundsson}, {Han}, {Hargrave}, {Hasegawa}, {Hasselfield}, {Hattori},
  {Haynes}, {Hazumi}, {He}, {Healy}, {Henderson}, {Hervias-Caimapo}, {Hill},
  {Hill}, {Hilton}, {Hilton}, {Hincks}, {Hinshaw}, {Hlo{\v{z}}ek}, {Ho}, {Ho},
  {Howe}, {Huang}, {Hubmayr}, {Huffenberger}, {Hughes}, {Ijjas}, {Ikape},
  {Irwin}, {Jaffe}, {Jain}, {Jeong}, {Kaneko}, {Karpel}, {Katayama}, {Keating},
  {Kernasovskiy}, {Keskitalo}, {Kisner}, {Kiuchi}, {Klein}, {Knowles},
  {Koopman}, {Kosowsky}, {Krachmalnicoff}, {Kuenstner}, {Kuo}, {Kusaka},
  {Lashner}, {Lee}, {Lee}, {Leon}, {Leung}, {Lewis}, {Li}, {Li}, {Limon},
  {Linder}, {Lopez-Caraballo}, {Louis}, {Lowry}, {Lungu}, {Madhavacheril},
  {Mak}, {Maldonado}, {Mani}, {Mates}, {Matsuda}, {Maurin}, {Mauskopf}, {May},
  {McCallum}, {McKenney}, {McMahon}, {Meerburg}, {Meyers}, {Miller},
  {Mirmelstein}, {Moodley}, {Munchmeyer}, {Munson}, {Naess}, {Nati},
  {Navaroli}, {Newburgh}, {Nguyen}, {Niemack}, {Nishino}, {Orlowski-Scherer},
  {Page}, {Partridge}, {Peloton}, {Perrotta}, {Piccirillo}, {Pisano},
  {Poletti}, {Puddu}, {Puglisi}, {Raum}, {Reichardt}, {Remazeilles},
  {Rephaeli}, {Riechers}, {Rojas}, {Roy}, {Sadeh}, {Sakurai}, {Salatino},
  {Sathyanarayana Rao}, {Schaan}, {Schmittfull}, {Sehgal}, {Seibert}, {Seljak},
  {Sherwin}, {Shimon}, {Sierra}, {Sievers}, {Sikhosana}, {Silva-Feaver},
  {Simon}, {Sinclair}, {Siritanasak}, {Smith}, {Smith}, {Spergel}, {Staggs},
  {Stein}, {Stevens}, {Stompor}, {Suzuki}, {Tajima}, {Takakura}, {Teply},
  {Thomas}, {Thorne}, {Thornton}, {Trac}, {Tsai}, {Tucker}, {Ullom},
  {Vagnozzi}, {van Engelen}, {Van Lanen}, {Van Winkle}, {Vavagiakis},
  {Verg{\`e}s}, {Vissers}, {Wagoner}, {Walker}, {Ward}, {Westbrook},
  {Whitehorn}, {Williams}, {Williams}, {Wollack}, {Xu}, {Yu}, {Yu}, {Zago},
  {Zhang}, {Zhu}, \& {Simons Observatory Collaboration}}]{SO2019}
{Ade}, P., {Aguirre}, J., {Ahmed}, Z., {et~al.} 2019, \jcap, 2019, 056,
  \dodoi{10.1088/1475-7516/2019/02/056}

\bibitem[{{Aiola} {et~al.}(2020){Aiola}, {Calabrese}, {Maurin}, {Naess},
  {Schmitt}, {Abitbol}, {Addison}, {Ade}, {Alonso}, {Amiri}, {Amodeo},
  {Angile}, {Austermann}, {Baildon}, {Battaglia}, {Beall}, {Bean}, {Becker},
  {Bond}, {Bruno}, {Calafut}, {Campusano}, {Carrero}, {Chesmore}, {Cho},
  {Choi}, {Clark}, {Cothard}, {Crichton}, {Crowley}, {Darwish}, {Datta},
  {Denison}, {Devlin}, {Duell}, {Duff}, {Duivenvoorden}, {Dunkley},
  {D{\"u}nner}, {Essinger-Hileman}, {Fankhanel}, {Ferraro}, {Fox}, {Fuzia},
  {Gallardo}, {Gluscevic}, {Golec}, {Grace}, {Gralla}, {Guan}, {Hall},
  {Halpern}, {Han}, {Hargrave}, {Hasselfield}, {Helton}, {Henderson},
  {Hensley}, {Hill}, {Hilton}, {Hilton}, {Hincks}, {Hlo{\v{z}}ek}, {Ho},
  {Hubmayr}, {Huffenberger}, {Hughes}, {Infante}, {Irwin}, {Jackson}, {Klein},
  {Knowles}, {Koopman}, {Kosowsky}, {Lakey}, {Li}, {Li}, {Li}, {Lokken},
  {Louis}, {Lungu}, {MacInnis}, {Madhavacheril}, {Maldonado}, {Mallaby-Kay},
  {Marsden}, {McMahon}, {Menanteau}, {Moodley}, {Morton}, {Namikawa}, {Nati},
  {Newburgh}, {Nibarger}, {Nicola}, {Niemack}, {Nolta}, {Orlowski-Sherer},
  {Page}, {Pappas}, {Partridge}, {Phakathi}, {Pisano}, {Prince}, {Puddu}, {Qu},
  {Rivera}, {Robertson}, {Rojas}, {Salatino}, {Schaan}, {Schillaci}, {Sehgal},
  {Sherwin}, {Sierra}, {Sievers}, {Sifon}, {Sikhosana}, {Simon}, {Spergel},
  {Staggs}, {Stevens}, {Storer}, {Sunder}, {Switzer}, {Thorne}, {Thornton},
  {Trac}, {Treu}, {Tucker}, {Vale}, {Van Engelen}, {Van Lanen}, {Vavagiakis},
  {Wagoner}, {Wang}, {Ward}, {Wollack}, {Xu}, {Zago}, \& {Zhu}}]{Aiola2020}
{Aiola}, S., {Calabrese}, E., {Maurin}, L., {et~al.} 2020, \jcap, 2020, 047,
  \dodoi{10.1088/1475-7516/2020/12/047}

\bibitem[{{Alam} {et~al.}(2017){Alam}, {Ata}, {Bailey}, {Beutler}, {Bizyaev},
  {Blazek}, {Bolton}, {Brownstein}, {Burden}, {Chuang}, {Comparat}, {Cuesta},
  {Dawson}, {Eisenstein}, {Escoffier}, {Gil-Mar{\'\i}n}, {Grieb}, {Hand}, {Ho},
  {Kinemuchi}, {Kirkby}, {Kitaura}, {Malanushenko}, {Malanushenko}, {Maraston},
  {McBride}, {Nichol}, {Olmstead}, {Oravetz}, {Padmanabhan},
  {Palanque-Delabrouille}, {Pan}, {Pellejero-Ibanez}, {Percival}, {Petitjean},
  {Prada}, {Price-Whelan}, {Reid}, {Rodr{\'\i}guez-Torres}, {Roe}, {Ross},
  {Ross}, {Rossi}, {Rubi{\~n}o-Mart{\'\i}n}, {Saito}, {Salazar-Albornoz},
  {Samushia}, {S{\'a}nchez}, {Satpathy}, {Schlegel}, {Schneider},
  {Sc{\'o}ccola}, {Seo}, {Sheldon}, {Simmons}, {Slosar}, {Strauss}, {Swanson},
  {Thomas}, {Tinker}, {Tojeiro}, {Maga{\~n}a}, {Vazquez}, {Verde}, {Wake},
  {Wang}, {Weinberg}, {White}, {Wood-Vasey}, {Y{\`e}che}, {Zehavi}, {Zhai}, \&
  {Zhao}}]{boss_dr12}
{Alam}, S., {Ata}, M., {Bailey}, S., {et~al.} 2017, \mnras, 470, 2617,
  \dodoi{10.1093/mnras/stx721}

\bibitem[{{Amodeo} {et~al.}(2021){Amodeo}, {Battaglia}, {Schaan}, {Ferraro},
  {Moser}, {Aiola}, {Austermann}, {Beall}, {Bean}, {Becker}, {Bond},
  {Calabrese}, {Calafut}, {Choi}, {Denison}, {Devlin}, {Duff}, {Duivenvoorden},
  {Dunkley}, {D{\"u}nner}, {Gallardo}, {Hall}, {Han}, {Hill}, {Hilton},
  {Hilton}, {Hlo{\v{z}}ek}, {Hubmayr}, {Huffenberger}, {Hughes}, {Koopman},
  {MacInnis}, {McMahon}, {Madhavacheril}, {Moodley}, {Mroczkowski}, {Naess},
  {Nati}, {Newburgh}, {Niemack}, {Page}, {Partridge}, {Schillaci}, {Sehgal},
  {Sif{\'o}n}, {Spergel}, {Staggs}, {Storer}, {Ullom}, {Vale}, {van Engelen},
  {Van Lanen}, {Vavagiakis}, {Wollack}, \& {Xu}}]{Amodeo2021}
{Amodeo}, S., {Battaglia}, N., {Schaan}, E., {et~al.} 2021, \prd, 103, 063514,
  \dodoi{10.1103/PhysRevD.103.063514}

\bibitem[{{Arag{\'o}n-Calvo} {et~al.}(2010){Arag{\'o}n-Calvo}, {van de
  Weygaert}, \& {Jones}}]{AragonCalvo2010}
{Arag{\'o}n-Calvo}, M.~A., {van de Weygaert}, R., \& {Jones}, B. J.~T. 2010,
  \mnras, 408, 2163, \dodoi{10.1111/j.1365-2966.2010.17263.x}

\bibitem[{{Astropy Collaboration} {et~al.}(2013){Astropy Collaboration},
  {Robitaille}, {Tollerud}, {Greenfield}, {Droettboom}, {Bray}, {Aldcroft},
  {Davis}, {Ginsburg}, {Price-Whelan}, {Kerzendorf}, {Conley}, {Crighton},
  {Barbary}, {Muna}, {Ferguson}, {Grollier}, {Parikh}, {Nair}, {Unther},
  {Deil}, {Woillez}, {Conseil}, {Kramer}, {Turner}, {Singer}, {Fox}, {Weaver},
  {Zabalza}, {Edwards}, {Azalee Bostroem}, {Burke}, {Casey}, {Crawford},
  {Dencheva}, {Ely}, {Jenness}, {Labrie}, {Lim}, {Pierfederici}, {Pontzen},
  {Ptak}, {Refsdal}, {Servillat}, \& {Streicher}}]{astropy:2013}
{Astropy Collaboration}, {Robitaille}, T.~P., {Tollerud}, E.~J., {et~al.} 2013,
  \aap, 558, A33, \dodoi{10.1051/0004-6361/201322068}

\bibitem[{{Astropy Collaboration} {et~al.}(2018){Astropy Collaboration},
  {Price-Whelan}, {Sip{\H{o}}cz}, {G{\"u}nther}, {Lim}, {Crawford}, {Conseil},
  {Shupe}, {Craig}, {Dencheva}, {Ginsburg}, {Vand erPlas}, {Bradley},
  {P{\'e}rez-Su{\'a}rez}, {de Val-Borro}, {Aldcroft}, {Cruz}, {Robitaille},
  {Tollerud}, {Ardelean}, {Babej}, {Bach}, {Bachetti}, {Bakanov}, {Bamford},
  {Barentsen}, {Barmby}, {Baumbach}, {Berry}, {Biscani}, {Boquien}, {Bostroem},
  {Bouma}, {Brammer}, {Bray}, {Breytenbach}, {Buddelmeijer}, {Burke},
  {Calderone}, {Cano Rodr{\'\i}guez}, {Cara}, {Cardoso}, {Cheedella}, {Copin},
  {Corrales}, {Crichton}, {D'Avella}, {Deil}, {Depagne}, {Dietrich}, {Donath},
  {Droettboom}, {Earl}, {Erben}, {Fabbro}, {Ferreira}, {Finethy}, {Fox},
  {Garrison}, {Gibbons}, {Goldstein}, {Gommers}, {Greco}, {Greenfield},
  {Groener}, {Grollier}, {Hagen}, {Hirst}, {Homeier}, {Horton}, {Hosseinzadeh},
  {Hu}, {Hunkeler}, {Ivezi{\'c}}, {Jain}, {Jenness}, {Kanarek}, {Kendrew},
  {Kern}, {Kerzendorf}, {Khvalko}, {King}, {Kirkby}, {Kulkarni}, {Kumar},
  {Lee}, {Lenz}, {Littlefair}, {Ma}, {Macleod}, {Mastropietro}, {McCully},
  {Montagnac}, {Morris}, {Mueller}, {Mumford}, {Muna}, {Murphy}, {Nelson},
  {Nguyen}, {Ninan}, {N{\"o}the}, {Ogaz}, {Oh}, {Parejko}, {Parley}, {Pascual},
  {Patil}, {Patil}, {Plunkett}, {Prochaska}, {Rastogi}, {Reddy Janga},
  {Sabater}, {Sakurikar}, {Seifert}, {Sherbert}, {Sherwood-Taylor}, {Shih},
  {Sick}, {Silbiger}, {Singanamalla}, {Singer}, {Sladen}, {Sooley},
  {Sornarajah}, {Streicher}, {Teuben}, {Thomas}, {Tremblay}, {Turner},
  {Terr{\'o}n}, {van Kerkwijk}, {de la Vega}, {Watkins}, {Weaver}, {Whitmore},
  {Woillez}, {Zabalza}, \& {Astropy Contributors}}]{astropy:2018}
{Astropy Collaboration}, {Price-Whelan}, A.~M., {Sip{\H{o}}cz}, B.~M., {et~al.}
  2018, \aj, 156, 123, \dodoi{10.3847/1538-3881/aabc4f}

\bibitem[{{Bagchi} {et~al.}(2017){Bagchi}, {Sankhyayan}, {Sarkar},
  {Raychaudhury}, {Jacob}, \& {Dabhade}}]{Bagchi2017}
{Bagchi}, J., {Sankhyayan}, S., {Sarkar}, P., {et~al.} 2017, \apj, 844, 25,
  \dodoi{10.3847/1538-4357/aa7949}

\bibitem[{{Bardeen} {et~al.}(1986){Bardeen}, {Bond}, {Kaiser}, \&
  {Szalay}}]{BBKS1986}
{Bardeen}, J.~M., {Bond}, J.~R., {Kaiser}, N., \& {Szalay}, A.~S. 1986, \apj,
  304, 15, \dodoi{10.1086/164143}

\bibitem[{{Basilakos} {et~al.}(2001){Basilakos}, {Plionis}, \&
  {Rowan-Robinson}}]{Basilakos2001}
{Basilakos}, S., {Plionis}, M., \& {Rowan-Robinson}, M. 2001, \mnras, 323, 47,
  \dodoi{10.1046/j.1365-8711.2001.04226.x}

\bibitem[{{Battaglia} {et~al.}(2012{\natexlab{a}}){Battaglia}, {Bond},
  {Pfrommer}, \& {Sievers}}]{Battaglia2012a}
{Battaglia}, N., {Bond}, J.~R., {Pfrommer}, C., \& {Sievers}, J.~L.
  2012{\natexlab{a}}, \apj, 758, 74, \dodoi{10.1088/0004-637X/758/2/74}

\bibitem[{{Battaglia} {et~al.}(2012{\natexlab{b}}){Battaglia}, {Bond},
  {Pfrommer}, \& {Sievers}}]{Battaglia2012b}
---. 2012{\natexlab{b}}, \apj, 758, 75, \dodoi{10.1088/0004-637X/758/2/75}

\bibitem[{{Battaglia} {et~al.}(2010){Battaglia}, {Bond}, {Pfrommer}, {Sievers},
  \& {Sijacki}}]{Battaglia2010}
{Battaglia}, N., {Bond}, J.~R., {Pfrommer}, C., {Sievers}, J.~L., \& {Sijacki},
  D. 2010, \apj, 725, 91, \dodoi{10.1088/0004-637X/725/1/91}

\bibitem[{{Bharadwaj} \& {Pandey}(2004)}]{Bharadwaj2004}
{Bharadwaj}, S., \& {Pandey}, B. 2004, \apj, 615, 1, \dodoi{10.1086/424476}

\bibitem[{{Bond} \& {Efstathiou}(1987)}]{BE1987}
{Bond}, J.~R., \& {Efstathiou}, G. 1987, \mnras, 226, 655,
  \dodoi{10.1093/mnras/226.3.655}

\bibitem[{{Bond} {et~al.}(1996){Bond}, {Kofman}, \& {Pogosyan}}]{BKP}
{Bond}, J.~R., {Kofman}, L., \& {Pogosyan}, D. 1996, \nat, 380, 603,
  \dodoi{10.1038/380603a0}

\bibitem[{{Bond} \& {Myers}(1996)}]{BondMyers1996}
{Bond}, J.~R., \& {Myers}, S.~T. 1996, \apjs, 103, 1, \dodoi{10.1086/192267}

\bibitem[{{Borgani}(1995)}]{Borgani1995}
{Borgani}, S. 1995, \physrep, 251, 1, \dodoi{10.1016/0370-1573(94)00073-C}

\bibitem[{{Bouma} {et~al.}(2021){Bouma}, {Richter}, \& {Wendt}}]{Bouma2021}
{Bouma}, S.~J.~D., {Richter}, P., \& {Wendt}, M. 2021, \aap, 647, A166,
  \dodoi{10.1051/0004-6361/202039786}

\bibitem[{{Carlstrom} {et~al.}(2019){Carlstrom}, {Abazajian}, {Addison},
  {Adshead}, {Ahmed}, {Allen}, {Alonso}, {Alvarez}, {Anderson}, {Arnold},
  {Baccigalupi}, {Bailey}, {Barkats}, {Barron}, {Barry}, {Bartlett}, {Basu
  Thakur}, {Battaglia}, {Baxter}, {Bean}, {Bebek}, {Bender}, {Benson},
  {Berger}, {Bhimani}, {Bischoff}, {Bleem}, {Bocquet}, {Boddy}, {Bonato},
  {Bond}, {Borrill}, {Bouchet}, {Brown}, {Bryan}, {Burkhart}, {Buza}, {Byrum},
  {Calabrese}, {Calafut}, {Caldwell}, {Carlstrom}, {Carron}, {Cecil},
  {Challinor}, {Chang}, {Chinone}, {Cho}, {Cooray}, {Crawford}, {Crites},
  {Cukierman}, {Cyr-Racine}, {de Haan}, {de Zotti}, {Delabrouille},
  {Demarteau}, {Devlin}, {Di Valentino}, {Dobbs}, {Duff}, {Duivenvoorden},
  {Dvorkin}, {Edwards}, {Eimer}, {Errard}, {Essinger-Hileman}, {Fabbian},
  {Feng}, {Ferraro}, {Filippini}, {Flauger}, {Flaugher}, {Fraisse}, {Frolov},
  {Galitzki}, {Galli}, {Ganga}, {Gerbino}, {Gilchriese}, {Gluscevic}, {Green},
  {Grin}, {Grohs}, {Gualtieri}, {Guarino}, {Gudmundsson}, {Habib}, {Haller},
  {Halpern}, {Halverson}, {Hanany}, {Harrington}, {Hasegawa}, {Hasselfield},
  {Hazumi}, {Heitmann}, {Henderson}, {Henning}, {Hill}, {Hlo{\v{z}}ek},
  {Holder}, {Holzapfel}, {Hubmayr}, {Huffenberger}, {Huffer}, {Hui}, {Irwin},
  {Johnson}, {Johnstone}, {Jones}, {Karkare}, {Katayama}, {Kerby}, {Kernovsky},
  {Keskitalo}, {Kisner}, {Knox}, {Kosowsky}, {Kovac}, {Kovetz}, {Kuhlmann},
  {Kuo}, {Kurita}, {Kusaka}, {Lahteenmaki}, {Lawrence}, {Lee}, {Lewis}, {Li},
  {Linder}, {Loverde}, {Lowitz}, {Madhavacheril}, {Mantz}, {Matsuda},
  {Mauskopf}, {McMahon}, {Meerburg}, {Melin}, {Meyers}, {Millea}, {Mohr},
  {Moncelsi}, {Mroczkowski}, {Mukherjee}, {Munchmeyer}, {Nagai}, {Nagy},
  {Namikawa}, {Nati}, {Natoli}, {Negrello}, {Newburgh}, {Niemack}, {Nishino},
  {Nordby}, {Novosad}, {O'Connor}, {Obied}, {Padin}, {Pandey}, {Partridge},
  {Pierpaoli}, {Pogosian}, {Pryke}, {Puglisi}, {Racine}, {Raghunathan},
  {Rahlin}, {Rajagopalan}, {Raveri}, {Reichanadter}, {Reichardt},
  {Remazeilles}, {Rocha}, {Roe}, {Roy}, {Ruhl}, {Salatino}, {Saliwanchik},
  {Schaan}, {Schillaci}, {Schmittfull}, {Scott}, {Sehgal}, {Shandera},
  {Sheehy}, {Sherwin}, {Shirokoff}, {Simon}, {Slosar}, {Somerville}, {Staggs},
  {Stark}, {Stompor}, {Story}, {Stoughton}, {Suzuki}, {Tajima}, {Teply},
  {Thompson}, {Timbie}, {Tomasi}, {Treu}, {Tristram}, {Tucker}, {Umilta}, {van
  Engelen}, {Vieira}, {Vieregg}, {Vogelsberger}, {Wang}, {Watson}, {White},
  {Whitehorn}, {Wollack}, {Wu}, {Xu}, {Yasini}, {Yeck}, {Yoon}, {Young}, \&
  {Zonca}}]{cmbs4/whitepaper}
{Carlstrom}, J., {Abazajian}, K., {Addison}, G., {et~al.} 2019, in Bulletin of
  the American Astronomical Society, Vol.~51, 209.
\newblock \doarXiv{1908.01062}

\bibitem[{{Carlstrom} {et~al.}(2002){Carlstrom}, {Holder}, \&
  {Reese}}]{carlstrom/etal:2002}
{Carlstrom}, J.~E., {Holder}, G.~P., \& {Reese}, E.~D. 2002, \araa, 40, 643,
  \dodoi{10.1146/annurev.astro.40.060401.093803}

\bibitem[{{Cen}(1994)}]{Cen1994}
{Cen}, R. 1994, \apj, 424, 22, \dodoi{10.1086/173868}

\bibitem[{{Cen} \& {Ostriker}(1999)}]{CenOstriker1999}
{Cen}, R., \& {Ostriker}, J.~P. 1999, \apj, 514, 1, \dodoi{10.1086/306949}

\bibitem[{{Cen} \& {Ostriker}(2006)}]{CenOstriker2006}
---. 2006, \apj, 650, 560, \dodoi{10.1086/506505}

\bibitem[{{Choi} {et~al.}(2018){Choi}, {Austermann}, {Beall}, {Crowley},
  {Datta}, {Duff}, {Gallardo}, {Ho}, {Hubmayr}, {Koopman}, {Li}, {Nati},
  {Niemack}, {Page}, {Salatino}, {Simon}, {Staggs}, {Stevens}, {Ullom}, \&
  {Wollack}}]{Choi2018}
{Choi}, S.~K., {Austermann}, J., {Beall}, J.~A., {et~al.} 2018, Journal of Low
  Temperature Physics, 193, 267, \dodoi{10.1007/s10909-018-1982-4}

\bibitem[{{Clampitt} {et~al.}(2017){Clampitt}, {S{\'a}nchez}, {Kwan}, {Krause},
  {MacCrann}, {Park}, {Troxel}, {Jain}, {Rozo}, {Rykoff}, {Wechsler}, {Blazek},
  {Bonnett}, {Crocce}, {Fang}, {Gaztanaga}, {Gruen}, {Jarvis}, {Miquel},
  {Prat}, {Ross}, {Sheldon}, {Zuntz}, {Abbott}, {Abdalla}, {Armstrong},
  {Becker}, {Benoit-L{\'e}vy}, {Bernstein}, {Bertin}, {Brooks}, {Burke},
  {Carnero Rosell}, {Carrasco Kind}, {Cunha}, {D'Andrea}, {da Costa}, {Desai},
  {Diehl}, {Dietrich}, {Doel}, {Estrada}, {Evrard}, {Fausti Neto}, {Flaugher},
  {Fosalba}, {Frieman}, {Gruendl}, {Honscheid}, {James}, {Kuehn}, {Kuropatkin},
  {Lahav}, {Lima}, {March}, {Marshall}, {Martini}, {Melchior}, {Mohr},
  {Nichol}, {Nord}, {Plazas}, {Romer}, {Sanchez}, {Scarpine}, {Schubnell},
  {Sevilla-Noarbe}, {Smith}, {Soares-Santos}, {Sobreira}, {Suchyta}, {Swanson},
  {Tarle}, {Thomas}, {Vikram}, \& {Walker}}]{Clampitt2017}
{Clampitt}, J., {S{\'a}nchez}, C., {Kwan}, J., {et~al.} 2017, \mnras, 465,
  4204, \dodoi{10.1093/mnras/stw2988}

\bibitem[{{Codis} {et~al.}(2018{\natexlab{a}}){Codis}, {Jindal}, {Chisari},
  {Vibert}, {Dubois}, {Pichon}, \& {Devriendt}}]{CodisJindal2018}
{Codis}, S., {Jindal}, A., {Chisari}, N.~E., {et~al.} 2018{\natexlab{a}},
  \mnras, 481, 4753, \dodoi{10.1093/mnras/sty2567}

\bibitem[{{Codis} {et~al.}(2018{\natexlab{b}}){Codis}, {Pogosyan}, \&
  {Pichon}}]{Codis2018}
{Codis}, S., {Pogosyan}, D., \& {Pichon}, C. 2018{\natexlab{b}}, \mnras, 479,
  973, \dodoi{10.1093/mnras/sty1643}

\bibitem[{{Coil}(2013)}]{Coil2013}
{Coil}, A.~L. 2013, in Planets, Stars and Stellar Systems. Volume 6:
  Extragalactic Astronomy and Cosmology, ed. T.~D. {Oswalt} \& W.~C. {Keel},
  Vol.~6 (Dordrecht:Springer), 387, \dodoi{10.1007/978-94-007-5609-0_8}

\bibitem[{{Costanzi} {et~al.}(2021){Costanzi}, {Saro}, {Bocquet}, {Abbott},
  {Aguena}, {Allam}, {Amara}, {Annis}, {Avila}, {Bacon}, {Benson}, {Bhargava},
  {Brooks}, {Buckley-Geer}, {Burke}, {Carnero Rosell}, {Carrasco Kind},
  {Carretero}, {Choi}, {da Costa}, {Pereira}, {De Vicente}, {Desai}, {Diehl},
  {Dietrich}, {Doel}, {Eifler}, {Everett}, {Ferrero}, {Fert{\'e}}, {Flaugher},
  {Fosalba}, {Frieman}, {Garc{\'\i}a-Bellido}, {Gaztanaga}, {Gerdes},
  {Giannantonio}, {Giles}, {Grandis}, {Gruen}, {Gruendl}, {Gupta}, {Gutierrez},
  {Hartley}, {Hinton}, {Hollowood}, {Honscheid}, {James}, {Jeltema}, {Krause},
  {Kuehn}, {Kuropatkin}, {Lahav}, {Lima}, {MacCrann}, {Maia}, {Marshall},
  {Menanteau}, {Miquel}, {Mohr}, {Morgan}, {Myles}, {Ogando}, {Palmese},
  {Paz-Chinch{\'o}n}, {Plazas}, {Rapetti}, {Reichardt}, {Romer}, {Roodman},
  {Ruppin}, {Salvati}, {Samuroff}, {Sanchez}, {Scarpine}, {Serrano},
  {Sevilla-Noarbe}, {Singh}, {Smith}, {Soares-Santos}, {Stark}, {Suchyta},
  {Swanson}, {Tarle}, {Thomas}, {To}, {Tucker}, {Varga}, {Wechsler}, {Zhang},
  {DES}, \& {SPT Collaborations}}]{Costanzi2021}
{Costanzi}, M., {Saro}, A., {Bocquet}, S., {et~al.} 2021, \prd, 103, 043522,
  \dodoi{10.1103/PhysRevD.103.043522}

\bibitem[{{Darragh Ford} {et~al.}(2019){Darragh Ford}, {Laigle}, {Gozaliasl},
  {Pichon}, {Devriendt}, {Slyz}, {Arnouts}, {Dubois}, {Finoguenov},
  {Griffiths}, {Kraljic}, {Pan}, {Peirani}, \& {Sarron}}]{DarraghFord2019}
{Darragh Ford}, E., {Laigle}, C., {Gozaliasl}, G., {et~al.} 2019, \mnras, 489,
  5695, \dodoi{10.1093/mnras/stz2490}

\bibitem[{{Dawson} {et~al.}(2013){Dawson}, {Schlegel}, {Ahn}, {Anderson},
  {Aubourg}, {Bailey}, {Barkhouser}, {Bautista}, {Beifiori}, {Berlind},
  {Bhardwaj}, {Bizyaev}, {Blake}, {Blanton}, {Blomqvist}, {Bolton}, {Borde},
  {Bovy}, {Brandt}, {Brewington}, {Brinkmann}, {Brown}, {Brownstein}, {Bundy},
  {Busca}, {Carithers}, {Carnero}, {Carr}, {Chen}, {Comparat}, {Connolly},
  {Cope}, {Croft}, {Cuesta}, {da Costa}, {Davenport}, {Delubac}, {de Putter},
  {Dhital}, {Ealet}, {Ebelke}, {Eisenstein}, {Escoffier}, {Fan}, {Filiz Ak},
  {Finley}, {Font-Ribera}, {G{\'e}nova-Santos}, {Gunn}, {Guo}, {Haggard},
  {Hall}, {Hamilton}, {Harris}, {Harris}, {Ho}, {Hogg}, {Holder}, {Honscheid},
  {Huehnerhoff}, {Jordan}, {Jordan}, {Kauffmann}, {Kazin}, {Kirkby}, {Klaene},
  {Kneib}, {Le Goff}, {Lee}, {Long}, {Loomis}, {Lundgren}, {Lupton}, {Maia},
  {Makler}, {Malanushenko}, {Malanushenko}, {Mandelbaum}, {Manera}, {Maraston},
  {Margala}, {Masters}, {McBride}, {McDonald}, {McGreer}, {McMahon}, {Mena},
  {Miralda-Escud{\'e}}, {Montero-Dorta}, {Montesano}, {Muna}, {Myers},
  {Naugle}, {Nichol}, {Noterdaeme}, {Nuza}, {Olmstead}, {Oravetz}, {Oravetz},
  {Owen}, {Padmanabhan}, {Palanque-Delabrouille}, {Pan}, {Parejko},
  {P{\^a}ris}, {Percival}, {P{\'e}rez-Fournon}, {P{\'e}rez-R{\`a}fols},
  {Petitjean}, {Pfaffenberger}, {Pforr}, {Pieri}, {Prada}, {Price-Whelan},
  {Raddick}, {Rebolo}, {Rich}, {Richards}, {Rockosi}, {Roe}, {Ross}, {Ross},
  {Rossi}, {Rubi{\~n}o-Martin}, {Samushia}, {S{\'a}nchez}, {Sayres}, {Schmidt},
  {Schneider}, {Sc{\'o}ccola}, {Seo}, {Shelden}, {Sheldon}, {Shen}, {Shu},
  {Slosar}, {Smee}, {Snedden}, {Stauffer}, {Steele}, {Strauss}, {Streblyanska},
  {Suzuki}, {Swanson}, {Tal}, {Tanaka}, {Thomas}, {Tinker}, {Tojeiro},
  {Tremonti}, {Vargas Maga{\~n}a}, {Verde}, {Viel}, {Wake}, {Watson}, {Weaver},
  {Weinberg}, {Weiner}, {West}, {White}, {Wood-Vasey}, {Yeche}, {Zehavi},
  {Zhao}, \& {Zheng}}]{BOSS2013}
{Dawson}, K.~S., {Schlegel}, D.~J., {Ahn}, C.~P., {et~al.} 2013, \aj, 145, 10,
  \dodoi{10.1088/0004-6256/145/1/10}

\bibitem[{{de Graaff} {et~al.}(2019){de Graaff}, {Cai}, {Heymans}, \&
  {Peacock}}]{deGraaff2019}
{de Graaff}, A., {Cai}, Y.-C., {Heymans}, C., \& {Peacock}, J.~A. 2019, \aap,
  624, A48, \dodoi{10.1051/0004-6361/201935159}

\bibitem[{{DeRose} {et~al.}(2019){DeRose}, {Wechsler}, {Becker}, {Busha},
  {Rykoff}, {MacCrann}, {Erickson}, {Evrard}, {Kravtsov}, {Gruen}, {Allam},
  {Avila}, {Bridle}, {Brooks}, {Buckley-Geer}, {Carnero Rosell}, {Carrasco
  Kind}, {Carretero}, {Castander}, {Cawthon}, {Crocce}, {da Costa}, {Davis},
  {De Vicente}, {Dietrich}, {Doel}, {Drlica-Wagner}, {Fosalba}, {Frieman},
  {Garcia-Bellido}, {Gutierrez}, {Hartley}, {Hollowood}, {Hoyle}, {James},
  {Krause}, {Kuehn}, {Kuropatkin}, {Lima}, {Maia}, {Menanteau}, {Miller},
  {Miquel}, {Ogand o}, {Plazas Malag{\'o}n}, {Romer}, {Sanchez}, {Schindler},
  {Serrano}, {Sevilla-Noarbe}, {Smith}, {Suchyta}, {Swanson}, {Tarle}, \&
  {Vikram}}]{DeRose2019}
{DeRose}, J., {Wechsler}, R.~H., {Becker}, M.~R., {et~al.} 2019, arXiv
  e-prints, arXiv:1901.02401.
\newblock \doarXiv{1901.02401}

\bibitem[{{DeRose} {et~al.}(2022){DeRose}, {Wechsler}, {Becker}, {Rykoff},
  {Pandey}, {MacCrann}, {Amon}, {Myles}, {Krause}, {Gruen}, {Jain}, {Troxel},
  {Prat}, {Alarcon}, {S{\'a}nchez}, {Blazek}, {Crocce}, {Giannini}, {Gatti},
  {Bernstein}, {Zuntz}, {Dodelson}, {Fang}, {Friedrich}, {Secco},
  {Elvin-Poole}, {Porredon}, {Everett}, {Choi}, {Harrison}, {Cordero},
  {Rodriguez-Monroy}, {McCullough}, {Cawthon}, {Chen}, {Alves},
  {Andrade-Oliveira}, {Bechtol}, {Camacho}, {Campos}, {Rosell}, {Kind},
  {Diehl}, {Drlica-Wagner}, {Eckert}, {Eifler}, {Gruendl}, {Hartley}, {Huang},
  {Huff}, {Kuropatkin}, {Raveri}, {Rosenfeld}, {Ross}, {Sanchez},
  {Sevilla-Noarbe}, {Sheldon}, {Yanny}, {Yin}, {Zhang}, {Fosalba}, {Aguena},
  {Allam}, {Annis}, {Avila}, {Bacon}, {Bhargava}, {Brooks}, {Buckley-Geer},
  {Burke}, {Carretero}, {Castander}, {Chang}, {Costanzi}, {da Costa},
  {Pereira}, {De Vicente}, {Desai}, {Dietrich}, {Doel}, {Evrard}, {Ferrero},
  {Fert{\'e}}, {Flaugher}, {Frieman}, {Garc{\'\i}a-Bellido}, {Gaztanaga},
  {Giannantonio}, {Gschwend}, {Gutierrez}, {Hinton}, {Hollowood}, {Honscheid},
  {Huterer}, {James}, {Kuehn}, {Lahav}, {Lima}, {Maia}, {Marshall}, {Melchior},
  {Menanteau}, {Miquel}, {Mohr}, {Morgan}, {Palmese}, {Paz-Chinch{\'o}n},
  {Petravick}, {Pieres}, {Malag{\'o}n}, {Sanchez}, {Scarpine}, {Serrano},
  {Smith}, {Soares-Santos}, {Suchyta}, {Tarle}, {Thomas}, {To}, {Varga}, \&
  {DES Collaboration}}]{DeRose2021}
---. 2022, \prd, 105, 123520, \dodoi{10.1103/PhysRevD.105.123520}

\bibitem[{{Desjacques} {et~al.}(2018){Desjacques}, {Jeong}, \&
  {Schmidt}}]{Desjacques2018}
{Desjacques}, V., {Jeong}, D., \& {Schmidt}, F. 2018, \physrep, 733, 1,
  \dodoi{10.1016/j.physrep.2017.12.002}

\bibitem[{{Dey} {et~al.}(2019){Dey}, {Schlegel}, {Lang}, {Blum}, {Burleigh},
  {Fan}, {Findlay}, {Finkbeiner}, {Herrera}, {Juneau}, {Landriau}, {Levi},
  {McGreer}, {Meisner}, {Myers}, {Moustakas}, {Nugent}, {Patej}, {Schlafly},
  {Walker}, {Valdes}, {Weaver}, {Y{\`e}che}, {Zou}, {Zhou}, {Abareshi},
  {Abbott}, {Abolfathi}, {Aguilera}, {Alam}, {Allen}, {Alvarez}, {Annis},
  {Ansarinejad}, {Aubert}, {Beechert}, {Bell}, {BenZvi}, {Beutler}, {Bielby},
  {Bolton}, {Brice{\~n}o}, {Buckley-Geer}, {Butler}, {Calamida}, {Carlberg},
  {Carter}, {Casas}, {Castander}, {Choi}, {Comparat}, {Cukanovaite}, {Delubac},
  {DeVries}, {Dey}, {Dhungana}, {Dickinson}, {Ding}, {Donaldson}, {Duan},
  {Duckworth}, {Eftekharzadeh}, {Eisenstein}, {Etourneau}, {Fagrelius},
  {Farihi}, {Fitzpatrick}, {Font-Ribera}, {Fulmer}, {G{\"a}nsicke},
  {Gaztanaga}, {George}, {Gerdes}, {Gontcho}, {Gorgoni}, {Green}, {Guy},
  {Harmer}, {Hernandez}, {Honscheid}, {Huang}, {James}, {Jannuzi}, {Jiang},
  {Joyce}, {Karcher}, {Karkar}, {Kehoe}, {Kneib}, {Kueter-Young}, {Lan},
  {Lauer}, {Le Guillou}, {Le Van Suu}, {Lee}, {Lesser}, {Perreault Levasseur},
  {Li}, {Mann}, {Marshall}, {Mart{\'\i}nez-V{\'a}zquez}, {Martini}, {du Mas des
  Bourboux}, {McManus}, {Meier}, {M{\'e}nard}, {Metcalfe},
  {Mu{\~n}oz-Guti{\'e}rrez}, {Najita}, {Napier}, {Narayan}, {Newman}, {Nie},
  {Nord}, {Norman}, {Olsen}, {Paat}, {Palanque-Delabrouille}, {Peng},
  {Poppett}, {Poremba}, {Prakash}, {Rabinowitz}, {Raichoor}, {Rezaie},
  {Robertson}, {Roe}, {Ross}, {Ross}, {Rudnick}, {Safonova}, {Saha},
  {S{\'a}nchez}, {Savary}, {Schweiker}, {Scott}, {Seo}, {Shan}, {Silva},
  {Slepian}, {Soto}, {Sprayberry}, {Staten}, {Stillman}, {Stupak}, {Summers},
  {Sien Tie}, {Tirado}, {Vargas-Maga{\~n}a}, {Vivas}, {Wechsler}, {Williams},
  {Yang}, {Yang}, {Yapici}, {Zaritsky}, {Zenteno}, {Zhang}, {Zhang}, {Zhou}, \&
  {Zhou}}]{DESI2019}
{Dey}, A., {Schlegel}, D.~J., {Lang}, D., {et~al.} 2019, \aj, 157, 168,
  \dodoi{10.3847/1538-3881/ab089d}

\bibitem[{{Einasto} {et~al.}(1997){Einasto}, {Tago}, {Jaaniste}, {Einasto}, \&
  {Andernach}}]{Einasto1997}
{Einasto}, M., {Tago}, E., {Jaaniste}, J., {Einasto}, J., \& {Andernach}, H.
  1997, \aaps, 123, 119, \dodoi{10.1051/aas:1997340}

\bibitem[{{Flaugher} {et~al.}(2015){Flaugher}, {Diehl}, {Honscheid}, {Abbott},
  {Alvarez}, {Angstadt}, {Annis}, {Antonik}, {Ballester}, {Beaufore},
  {Bernstein}, {Bernstein}, {Bigelow}, {Bonati}, {Boprie}, {Brooks},
  {Buckley-Geer}, {Campa}, {Cardiel-Sas}, {Castander}, {Castilla}, {Cease},
  {Cela-Ruiz}, {Chappa}, {Chi}, {Cooper}, {da Costa}, {Dede}, {Derylo},
  {DePoy}, {de Vicente}, {Doel}, {Drlica-Wagner}, {Eiting}, {Elliott}, {Emes},
  {Estrada}, {Fausti Neto}, {Finley}, {Flores}, {Frieman}, {Gerdes},
  {Gladders}, {Gregory}, {Gutierrez}, {Hao}, {Holland}, {Holm}, {Huffman},
  {Jackson}, {James}, {Jonas}, {Karcher}, {Karliner}, {Kent}, {Kessler},
  {Kozlovsky}, {Kron}, {Kubik}, {Kuehn}, {Kuhlmann}, {Kuk}, {Lahav}, {Lathrop},
  {Lee}, {Levi}, {Lewis}, {Li}, {Mandrichenko}, {Marshall}, {Martinez},
  {Merritt}, {Miquel}, {Mu{\~n}oz}, {Neilsen}, {Nichol}, {Nord}, {Ogando},
  {Olsen}, {Palaio}, {Patton}, {Peoples}, {Plazas}, {Rauch}, {Reil}, {Rheault},
  {Roe}, {Rogers}, {Roodman}, {Sanchez}, {Scarpine}, {Schindler}, {Schmidt},
  {Schmitt}, {Schubnell}, {Schultz}, {Schurter}, {Scott}, {Serrano}, {Shaw},
  {Smith}, {Soares-Santos}, {Stefanik}, {Stuermer}, {Suchyta}, {Sypniewski},
  {Tarle}, {Thaler}, {Tighe}, {Tran}, {Tucker}, {Walker}, {Wang}, {Watson},
  {Weaverdyck}, {Wester}, {Woods}, {Yanny}, \& {DES
  Collaboration}}]{Flaugher2015}
{Flaugher}, B., {Diehl}, H.~T., {Honscheid}, K., {et~al.} 2015, \aj, 150, 150,
  \dodoi{10.1088/0004-6256/150/5/150}

\bibitem[{{Fowler} {et~al.}(2007){Fowler}, {Niemack}, {Dicker}, {Aboobaker},
  {Ade}, {Battistelli}, {Devlin}, {Fisher}, {Halpern}, {Hargrave}, {Hincks},
  {Kaul}, {Klein}, {Lau}, {Limon}, {Marriage}, {Mauskopf}, {Page}, {Staggs},
  {Swetz}, {Switzer}, {Thornton}, \& {Tucker}}]{Fowler2007}
{Fowler}, J.~W., {Niemack}, M.~D., {Dicker}, S.~R., {et~al.} 2007, \ao, 46,
  3444, \dodoi{10.1364/AO.46.003444}

\bibitem[{{Frisch} {et~al.}(1995){Frisch}, {Einasto}, {Einasto}, {Freudling},
  {Fricke}, {Gramann}, {Saar}, \& {Toomet}}]{Frisch1995}
{Frisch}, P., {Einasto}, J., {Einasto}, M., {et~al.} 1995, \aap, 296, 611.
\newblock \doarXiv{astro-ph/9503037}

\bibitem[{{Ghirardini} {et~al.}(2021){Ghirardini}, {Bulbul}, {Hoang}, {Klein},
  {Okabe}, {Biffi}, {Br{\"u}ggen}, {Ramos-Ceja}, {Comparat}, {Oguri},
  {Shimwell}, {Basu}, {Bonafede}, {Botteon}, {Brunetti}, {Cassano}, {de
  Gasperin}, {Dennerl}, {Gatuzz}, {Gastaldello}, {Intema}, {Merloni}, {Nandra},
  {Pacaud}, {Predehl}, {Reiprich}, {Robrade}, {R{\"o}ttgering}, {Sanders}, {van
  Weeren}, \& {Williams}}]{eROSITA}
{Ghirardini}, V., {Bulbul}, E., {Hoang}, D.~N., {et~al.} 2021, \aap, 647, A4,
  \dodoi{10.1051/0004-6361/202039554}

\bibitem[{{Giannantonio} {et~al.}(2012){Giannantonio}, {Porciani}, {Carron},
  {Amara}, \& {Pillepich}}]{Giannantonio2012}
{Giannantonio}, T., {Porciani}, C., {Carron}, J., {Amara}, A., \& {Pillepich},
  A. 2012, \mnras, 422, 2854, \dodoi{10.1111/j.1365-2966.2012.20604.x}

\bibitem[{{Giodini} {et~al.}(2013){Giodini}, {Lovisari}, {Pointecouteau},
  {Ettori}, {Reiprich}, \& {Hoekstra}}]{giodini/etal:2013}
{Giodini}, S., {Lovisari}, L., {Pointecouteau}, E., {et~al.} 2013, \ssr, 177,
  247, \dodoi{10.1007/s11214-013-9994-5}

\bibitem[{{Gitti} {et~al.}(2012){Gitti}, {Brighenti}, \&
  {McNamara}}]{Gitti2012}
{Gitti}, M., {Brighenti}, F., \& {McNamara}, B.~R. 2012, Advances in Astronomy,
  2012, 950641, \dodoi{10.1155/2012/950641}

\bibitem[{{G{\'o}rski} {et~al.}(2005){G{\'o}rski}, {Hivon}, {Banday},
  {Wandelt}, {Hansen}, {Reinecke}, \& {Bartelmann}}]{Healpix2005}
{G{\'o}rski}, K.~M., {Hivon}, E., {Banday}, A.~J., {et~al.} 2005, \apj, 622,
  759, \dodoi{10.1086/427976}

\bibitem[{{Hand} {et~al.}(2011){Hand}, {Appel}, {Battaglia}, {Bond}, {Das},
  {Devlin}, {Dunkley}, {D{\"u}nner}, {Essinger-Hileman}, {Fowler}, {Hajian},
  {Halpern}, {Hasselfield}, {Hilton}, {Hincks}, {Hlozek}, {Hughes}, {Irwin},
  {Klein}, {Kosowsky}, {Lin}, {Marriage}, {Marsden}, {McLaren}, {Menanteau},
  {Moodley}, {Niemack}, {Nolta}, {Page}, {Parker}, {Partridge}, {Plimpton},
  {Reese}, {Rojas}, {Sehgal}, {Sherwin}, {Sievers}, {Spergel}, {Staggs},
  {Swetz}, {Switzer}, {Thornton}, {Trac}, {Visnjic}, \& {Wollack}}]{Hand2011}
{Hand}, N., {Appel}, J.~W., {Battaglia}, N., {et~al.} 2011, \apj, 736, 39,
  \dodoi{10.1088/0004-637X/736/1/39}

\bibitem[{Harris {et~al.}(2020)Harris, Millman, van~der Walt, Gommers,
  Virtanen, Cournapeau, Wieser, Taylor, Berg, Smith, Kern, Picus, Hoyer, van
  Kerkwijk, Brett, Haldane, del R{\'{i}}o, Wiebe, Peterson,
  G{\'{e}}rard-Marchant, Sheppard, Reddy, Weckesser, Abbasi, Gohlke, \&
  Oliphant}]{harris2020array}
Harris, C.~R., Millman, K.~J., van~der Walt, S.~J., {et~al.} 2020, Nature, 585,
  357, \dodoi{10.1038/s41586-020-2649-2}

\bibitem[{{Henderson} {et~al.}(2016){Henderson}, {Allison}, {Austermann},
  {Baildon}, {Battaglia}, {Beall}, {Becker}, {De Bernardis}, {Bond},
  {Calabrese}, {Choi}, {Coughlin}, {Crowley}, {Datta}, {Devlin}, {Duff},
  {Dunkley}, {D{\"u}nner}, {van Engelen}, {Gallardo}, {Grace}, {Hasselfield},
  {Hills}, {Hilton}, {Hincks}, {Hloẑek}, {Ho}, {Hubmayr}, {Huffenberger},
  {Hughes}, {Irwin}, {Koopman}, {Kosowsky}, {Li}, {McMahon}, {Munson}, {Nati},
  {Newburgh}, {Niemack}, {Niraula}, {Page}, {Pappas}, {Salatino}, {Schillaci},
  {Schmitt}, {Sehgal}, {Sherwin}, {Sievers}, {Simon}, {Spergel}, {Staggs},
  {Stevens}, {Thornton}, {Van Lanen}, {Vavagiakis}, {Ward}, \&
  {Wollack}}]{Henderson2016}
{Henderson}, S.~W., {Allison}, R., {Austermann}, J., {et~al.} 2016, Journal of
  Low Temperature Physics, 184, 772, \dodoi{10.1007/s10909-016-1575-z}

\bibitem[{{Hill} {et~al.}(2018){Hill}, {Baxter}, {Lidz}, {Greco}, \&
  {Jain}}]{Hill2018}
{Hill}, J.~C., {Baxter}, E.~J., {Lidz}, A., {Greco}, J.~P., \& {Jain}, B. 2018,
  \prd, 97, 083501, \dodoi{10.1103/PhysRevD.97.083501}

\bibitem[{{Ho} {et~al.}(2018){Ho}, {Gronke}, {Falck}, \& {Mota}}]{Ho2018}
{Ho}, A., {Gronke}, M., {Falck}, B., \& {Mota}, D.~F. 2018, \aap, 619, A122,
  \dodoi{10.1051/0004-6361/201833899}

\bibitem[{Ho {et~al.}(2017)Ho, Austermann, Beall, Choi, Cothard, Crowley,
  Datta, Devlin, Duff, Gallardo, Hasselfield, Henderson, Hilton, Hubmayr,
  Koopman, Li, McMahon, Niemack, Salatino, Simon, Staggs, Ward, Ullom,
  Vavagiakis, \& Wollack}]{Shuay2017}
Ho, S.-P.~P., Austermann, J., Beall, J.~A., {et~al.} 2017, in Millimeter,
  Submillimeter, and Far-Infrared Detectors and Instrumentation for Astronomy
  VIII, ed. W.~S. Holland \& J.~Zmuidzinas, Vol. 9914, International Society
  for Optics and Photonics (SPIE), 301 -- 315.
\newblock \url{https://doi.org/10.1117/12.2233113}

\bibitem[{{Hopkins} {et~al.}(2005){Hopkins}, {Bahcall}, \&
  {Bode}}]{Hopkins2005}
{Hopkins}, P.~F., {Bahcall}, N.~A., \& {Bode}, P. 2005, \apj, 618, 1,
  \dodoi{10.1086/425993}

\bibitem[{{Huang}(2016)}]{Huang2016}
{Huang}, Z. 2016, \prd, 93, 043538, \dodoi{10.1103/PhysRevD.93.043538}

\bibitem[{{Ivezi{\'c}} {et~al.}(2019){Ivezi{\'c}}, {Kahn}, {Tyson}, {Abel},
  {Acosta}, {Allsman}, {Alonso}, {AlSayyad}, {Anderson}, {Andrew}, {Angel},
  {Angeli}, {Ansari}, {Antilogus}, {Araujo}, {Armstrong}, {Arndt}, {Astier},
  {Aubourg}, {Auza}, {Axelrod}, {Bard}, {Barr}, {Barrau}, {Bartlett}, {Bauer},
  {Bauman}, {Baumont}, {Bechtol}, {Bechtol}, {Becker}, {Becla}, {Beldica},
  {Bellavia}, {Bianco}, {Biswas}, {Blanc}, {Blazek}, {Bland ford}, {Bloom},
  {Bogart}, {Bond}, {Booth}, {Borgland}, {Borne}, {Bosch}, {Boutigny},
  {Brackett}, {Bradshaw}, {Brand t}, {Brown}, {Bullock}, {Burchat}, {Burke},
  {Cagnoli}, {Calabrese}, {Callahan}, {Callen}, {Carlin}, {Carlson}, {Chand
  rasekharan}, {Charles-Emerson}, {Chesley}, {Cheu}, {Chiang}, {Chiang},
  {Chirino}, {Chow}, {Ciardi}, {Claver}, {Cohen-Tanugi}, {Cockrum}, {Coles},
  {Connolly}, {Cook}, {Cooray}, {Covey}, {Cribbs}, {Cui}, {Cutri}, {Daly},
  {Daniel}, {Daruich}, {Daubard}, {Daues}, {Dawson}, {Delgado}, {Dellapenna},
  {de Peyster}, {de Val-Borro}, {Digel}, {Doherty}, {Dubois},
  {Dubois-Felsmann}, {Durech}, {Economou}, {Eifler}, {Eracleous}, {Emmons},
  {Fausti Neto}, {Ferguson}, {Figueroa}, {Fisher-Levine}, {Focke}, {Foss},
  {Frank}, {Freemon}, {Gangler}, {Gawiser}, {Geary}, {Gee}, {Geha}, {Gessner},
  {Gibson}, {Gilmore}, {Glanzman}, {Glick}, {Goldina}, {Goldstein}, {Goodenow},
  {Graham}, {Gressler}, {Gris}, {Guy}, {Guyonnet}, {Haller}, {Harris},
  {Hascall}, {Haupt}, {Hernand ez}, {Herrmann}, {Hileman}, {Hoblitt},
  {Hodgson}, {Hogan}, {Howard}, {Huang}, {Huffer}, {Ingraham}, {Innes},
  {Jacoby}, {Jain}, {Jammes}, {Jee}, {Jenness}, {Jernigan}, {Jevremovi{\'c}},
  {Johns}, {Johnson}, {Johnson}, {Jones}, {Juramy-Gilles}, {Juri{\'c}},
  {Kalirai}, {Kallivayalil}, {Kalmbach}, {Kantor}, {Karst}, {Kasliwal},
  {Kelly}, {Kessler}, {Kinnison}, {Kirkby}, {Knox}, {Kotov}, {Krabbendam},
  {Krughoff}, {Kub{\'a}nek}, {Kuczewski}, {Kulkarni}, {Ku}, {Kurita}, {Lage},
  {Lambert}, {Lange}, {Langton}, {Le Guillou}, {Levine}, {Liang}, {Lim},
  {Lintott}, {Long}, {Lopez}, {Lotz}, {Lupton}, {Lust}, {MacArthur}, {Mahabal},
  {Mand elbaum}, {Markiewicz}, {Marsh}, {Marshall}, {Marshall}, {May},
  {McKercher}, {McQueen}, {Meyers}, {Migliore}, {Miller}, {Mills}, {Miraval},
  {Moeyens}, {Moolekamp}, {Monet}, {Moniez}, {Monkewitz}, {Montgomery},
  {Morrison}, {Mueller}, {Muller}, {Mu{\~n}oz Arancibia}, {Neill}, {Newbry},
  {Nief}, {Nomerotski}, {Nordby}, {O'Connor}, {Oliver}, {Olivier}, {Olsen},
  {O'Mullane}, {Ortiz}, {Osier}, {Owen}, {Pain}, {Palecek}, {Parejko},
  {Parsons}, {Pease}, {Peterson}, {Peterson}, {Petravick}, {Libby Petrick},
  {Petry}, {Pierfederici}, {Pietrowicz}, {Pike}, {Pinto}, {Plante}, {Plate},
  {Plutchak}, {Price}, {Prouza}, {Radeka}, {Rajagopal}, {Rasmussen},
  {Regnault}, {Reil}, {Reiss}, {Reuter}, {Ridgway}, {Riot}, {Ritz}, {Robinson},
  {Roby}, {Roodman}, {Rosing}, {Roucelle}, {Rumore}, {Russo}, {Saha},
  {Sassolas}, {Schalk}, {Schellart}, {Schindler}, {Schmidt}, {Schneider},
  {Schneider}, {Schoening}, {Schumacher}, {Schwamb}, {Sebag}, {Selvy},
  {Sembroski}, {Seppala}, {Serio}, {Serrano}, {Shaw}, {Shipsey}, {Sick},
  {Silvestri}, {Slater}, {Smith}, {Smith}, {Sobhani}, {Soldahl},
  {Storrie-Lombardi}, {Stover}, {Strauss}, {Street}, {Stubbs}, {Sullivan},
  {Sweeney}, {Swinbank}, {Szalay}, {Takacs}, {Tether}, {Thaler}, {Thayer},
  {Thomas}, {Thornton}, {Thukral}, {Tice}, {Trilling}, {Turri}, {Van Berg},
  {Vanden Berk}, {Vetter}, {Virieux}, {Vucina}, {Wahl}, {Walkowicz}, {Walsh},
  {Walter}, {Wang}, {Wang}, {Warner}, {Wiecha}, {Willman}, {Winters},
  {Wittman}, {Wolff}, {Wood-Vasey}, {Wu}, {Xin}, {Yoachim}, \&
  {Zhan}}]{LSST2019}
{Ivezi{\'c}}, {\v{Z}}., {Kahn}, S.~M., {Tyson}, J.~A., {et~al.} 2019, \apj,
  873, 111, \dodoi{10.3847/1538-4357/ab042c}

\bibitem[{{Kolokotronis} {et~al.}(2002){Kolokotronis}, {Basilakos}, \&
  {Plionis}}]{Kolokotronis2002}
{Kolokotronis}, V., {Basilakos}, S., \& {Plionis}, M. 2002, \mnras, 331, 1020,
  \dodoi{10.1046/j.1365-8711.2002.05263.x}

\bibitem[{{Kotecha} {et~al.}(2022){Kotecha}, {Welker}, {Zhou}, {Wadsley},
  {Kraljic}, {Sorce}, {Rasia}, {Roberts}, {Gray}, {Yepes}, \&
  {Cui}}]{Kotecha2022}
{Kotecha}, S., {Welker}, C., {Zhou}, Z., {et~al.} 2022, \mnras, 512, 926,
  \dodoi{10.1093/mnras/stac300}

\bibitem[{{Kraljic} {et~al.}(2020){Kraljic}, {Pichon}, {Codis}, {Laigle},
  {Dav{\'e}}, {Dubois}, {Hwang}, {Pogosyan}, {Arnouts}, {Devriendt}, {Musso},
  {Peirani}, {Slyz}, \& {Treyer}}]{Kraljic2020}
{Kraljic}, K., {Pichon}, C., {Codis}, S., {et~al.} 2020, \mnras, 491, 4294,
  \dodoi{10.1093/mnras/stz3319}

\bibitem[{{Kravtsov} {et~al.}(2004){Kravtsov}, {Berlind}, {Wechsler}, {Klypin},
  {Gottl{\"o}ber}, {Allgood}, \& {Primack}}]{Kravtsov2004}
{Kravtsov}, A.~V., {Berlind}, A.~A., {Wechsler}, R.~H., {et~al.} 2004, \apj,
  609, 35, \dodoi{10.1086/420959}

\bibitem[{{Krolewski} {et~al.}(2019){Krolewski}, {Ho}, {Chen}, {Chan},
  {Tenneti}, {Bizyaev}, \& {Kraljic}}]{Krolewski2019}
{Krolewski}, A., {Ho}, S., {Chen}, Y.-C., {et~al.} 2019, \apj, 876, 52,
  \dodoi{10.3847/1538-4357/ab1010}

\bibitem[{{Kuchner} {et~al.}(2022){Kuchner}, {Haggar}, {Arag{\'o}n-Salamanca},
  {Pearce}, {Gray}, {Rost}, {Cui}, {Knebe}, \& {Yepes}}]{Kuchner2022}
{Kuchner}, U., {Haggar}, R., {Arag{\'o}n-Salamanca}, A., {et~al.} 2022, \mnras,
  510, 581, \dodoi{10.1093/mnras/stab3419}

\bibitem[{{Li} {et~al.}(2018){Li}, {Austermann}, {Beall}, {Bruno}, {Choi},
  {Cothard}, {Crowley}, {Duff}, {Gallardo}, {Henderson}, {Ho}, {Hubmayr},
  {Koopman}, {McMahon}, {Niemack}, {Salatino}, {Simon}, {Staggs}, {Stevens},
  {Ullom}, {Ward}, \& {Wollack}}]{Li2018}
{Li}, Y., {Austermann}, J.~E., {Beall}, J.~A., {et~al.} 2018, in Society of
  Photo-Optical Instrumentation Engineers (SPIE) Conference Series, Vol. 10708,
  Millimeter, Submillimeter, and Far-Infrared Detectors and Instrumentation for
  Astronomy IX, ed. J.~{Zmuidzinas} \& J.-R. {Gao}, 107080A,
  \dodoi{10.1117/12.2313942}

\bibitem[{{Libeskind} {et~al.}(2018){Libeskind}, {van de Weygaert}, {Cautun},
  {Falck}, {Tempel}, {Abel}, {Alpaslan}, {Arag{\'o}n-Calvo}, {Forero-Romero},
  {Gonzalez}, {Gottl{\"o}ber}, {Hahn}, {Hellwing}, {Hoffman}, {Jones},
  {Kitaura}, {Knebe}, {Manti}, {Neyrinck}, {Nuza}, {Padilla}, {Platen},
  {Ramachandra}, {Robotham}, {Saar}, {Shand arin}, {Steinmetz}, {Stoica},
  {Sousbie}, \& {Yepes}}]{Libeskind2018}
{Libeskind}, N.~I., {van de Weygaert}, R., {Cautun}, M., {et~al.} 2018, \mnras,
  473, 1195, \dodoi{10.1093/mnras/stx1976}

\bibitem[{{Lim} {et~al.}(2018){Lim}, {Mo}, {Li}, {Liu}, {Ma}, {Wang}, \&
  {Yang}}]{Lim2018}
{Lim}, S.~H., {Mo}, H.~J., {Li}, R., {et~al.} 2018, \apj, 854, 181,
  \dodoi{10.3847/1538-4357/aaaa21}

\bibitem[{{Louis} {et~al.}(2017){Louis}, {Grace}, {Hasselfield}, {Lungu},
  {Maurin}, {Addison}, {Ade}, {Aiola}, {Allison}, {Amiri}, {Angile},
  {Battaglia}, {Beall}, {de Bernardis}, {Bond}, {Britton}, {Calabrese}, {Cho},
  {Choi}, {Coughlin}, {Crichton}, {Crowley}, {Datta}, {Devlin}, {Dicker},
  {Dunkley}, {D{\"u}nner}, {Ferraro}, {Fox}, {Gallardo}, {Gralla}, {Halpern},
  {Henderson}, {Hill}, {Hilton}, {Hilton}, {Hincks}, {Hlozek}, {Ho}, {Huang},
  {Hubmayr}, {Huffenberger}, {Hughes}, {Infante}, {Irwin}, {Muya Kasanda},
  {Klein}, {Koopman}, {Kosowsky}, {Li}, {Madhavacheril}, {Marriage}, {McMahon},
  {Menanteau}, {Moodley}, {Munson}, {Naess}, {Nati}, {Newburgh}, {Nibarger},
  {Niemack}, {Nolta}, {Nu{\~n}ez}, {Page}, {Pappas}, {Partridge}, {Rojas},
  {Schaan}, {Schmitt}, {Sehgal}, {Sherwin}, {Sievers}, {Simon}, {Spergel},
  {Staggs}, {Switzer}, {Thornton}, {Trac}, {Treu}, {Tucker}, {Van Engelen},
  {Ward}, \& {Wollack}}]{Louis2017}
{Louis}, T., {Grace}, E., {Hasselfield}, M., {et~al.} 2017, \jcap, 2017, 031,
  \dodoi{10.1088/1475-7516/2017/06/031}

\bibitem[{{Madhavacheril} {et~al.}(2020){Madhavacheril}, {Hill}, {N{\ae}ss},
  {Addison}, {Aiola}, {Baildon}, {Battaglia}, {Bean}, {Bond}, {Calabrese},
  {Calafut}, {Choi}, {Darwish}, {Datta}, {Devlin}, {Dunkley}, {D{\"u}nner},
  {Ferraro}, {Gallardo}, {Gluscevic}, {Halpern}, {Han}, {Hasselfield},
  {Hilton}, {Hincks}, {Hlo{\v{z}}ek}, {Ho}, {Huffenberger}, {Hughes},
  {Koopman}, {Kosowsky}, {Lokken}, {Louis}, {Lungu}, {MacInnis}, {Maurin},
  {McMahon}, {Moodley}, {Nati}, {Niemack}, {Page}, {Partridge}, {Robertson},
  {Sehgal}, {Schaan}, {Schillaci}, {Sherwin}, {Sif{\'o}n}, {Simon}, {Spergel},
  {Staggs}, {Storer}, {van Engelen}, {Vavagiakis}, {Wollack}, \&
  {Xu}}]{Madhavacheril2020}
{Madhavacheril}, M.~S., {Hill}, J.~C., {N{\ae}ss}, S., {et~al.} 2020, \prd,
  102, 023534, \dodoi{10.1103/PhysRevD.102.023534}

\bibitem[{{Maraston} {et~al.}(2013){Maraston}, {Pforr}, {Henriques}, {Thomas},
  {Wake}, {Brownstein}, {Capozzi}, {Tinker}, {Bundy}, {Skibba}, {Beifiori},
  {Nichol}, {Edmondson}, {Schneider}, {Chen}, {Masters}, {Steele}, {Bolton},
  {York}, {Weaver}, {Higgs}, {Bizyaev}, {Brewington}, {Malanushenko},
  {Malanushenko}, {Snedden}, {Oravetz}, {Pan}, {Shelden}, \&
  {Simmons}}]{Maraston2013}
{Maraston}, C., {Pforr}, J., {Henriques}, B.~M., {et~al.} 2013, \mnras, 435,
  2764, \dodoi{10.1093/mnras/stt1424}

\bibitem[{{McCarthy} {et~al.}(2003){McCarthy}, {Babul}, {Holder}, \&
  {Balogh}}]{mccarthy/etal:2003}
{McCarthy}, I.~G., {Babul}, A., {Holder}, G.~P., \& {Balogh}, M.~L. 2003, \apj,
  591, 515, \dodoi{10.1086/375486}

\bibitem[{{McClintock} {et~al.}(2019){McClintock}, {Varga}, {Gruen}, {Rozo},
  {Rykoff}, {Shin}, {Melchior}, {DeRose}, {Seitz}, {Dietrich}, {Sheldon},
  {Zhang}, {von der Linden}, {Jeltema}, {Mantz}, {Romer}, {Allen}, {Becker},
  {Bermeo}, {Bhargava}, {Costanzi}, {Everett}, {Farahi}, {Hamaus}, {Hartley},
  {Hollowood}, {Hoyle}, {Israel}, {Li}, {MacCrann}, {Morris}, {Palmese},
  {Plazas}, {Pollina}, {Rau}, {Simet}, {Soares-Santos}, {Troxel}, {Vergara
  Cervantes}, {Wechsler}, {Zuntz}, {Abbott}, {Abdalla}, {Allam}, {Annis},
  {Avila}, {Bridle}, {Brooks}, {Burke}, {Carnero Rosell}, {Carrasco Kind},
  {Carretero}, {Castander}, {Crocce}, {Cunha}, {D'Andrea}, {da Costa}, {Davis},
  {De Vicente}, {Diehl}, {Doel}, {Drlica-Wagner}, {Evrard}, {Flaugher},
  {Fosalba}, {Frieman}, {Garc{\'\i}a-Bellido}, {Gaztanaga}, {Gerdes},
  {Giannantonio}, {Gruendl}, {Gutierrez}, {Honscheid}, {James}, {Kirk},
  {Krause}, {Kuehn}, {Lahav}, {Li}, {Lima}, {March}, {Marshall}, {Menanteau},
  {Miquel}, {Mohr}, {Nord}, {Ogando}, {Roodman}, {Sanchez}, {Scarpine},
  {Schindler}, {Sevilla-Noarbe}, {Smith}, {Smith}, {Sobreira}, {Suchyta},
  {Swanson}, {Tarle}, {Tucker}, {Vikram}, {Walker}, \&
  {Weller}}]{McClintock2019}
{McClintock}, T., {Varga}, T.~N., {Gruen}, D., {et~al.} 2019, \mnras, 482,
  1352, \dodoi{10.1093/mnras/sty2711}

\bibitem[{{Miller} \& {LAMBDA group}(2018)}]{lambda}
{Miller}, N., \& {LAMBDA group}. 2018, in American Astronomical Society Meeting
  Abstracts, Vol. 231, American Astronomical Society Meeting Abstracts \#231,
  430.05

\bibitem[{{Mroczkowski} {et~al.}(2019){Mroczkowski}, {Nagai}, {Basu}, {Chluba},
  {Sayers}, {Adam}, {Churazov}, {Crites}, {Di Mascolo}, {Eckert},
  {Macias-Perez}, {Mayet}, {Perotto}, {Pointecouteau}, {Romero}, {Ruppin},
  {Scannapieco}, \& {ZuHone}}]{Mroczkowski/etal:2019}
{Mroczkowski}, T., {Nagai}, D., {Basu}, K., {et~al.} 2019, \ssr, 215, 17,
  \dodoi{10.1007/s11214-019-0581-2}

\bibitem[{{Naess} {et~al.}(2020){Naess}, {Aiola}, {Austermann}, {Battaglia},
  {Beall}, {Becker}, {Bond}, {Calabrese}, {Choi}, {Cothard}, {Crowley},
  {Darwish}, {Datta}, {Denison}, {Devlin}, {Duell}, {Duff}, {Duivenvoorden},
  {Dunkley}, {D{\"u}nner}, {Fox}, {Gallardo}, {Halpern}, {Han}, {Hasselfield},
  {Hill}, {Hilton}, {Hilton}, {Hincks}, {Hlo{\v{z}}ek}, {Ho}, {Hubmayr},
  {Huffenberger}, {Hughes}, {Kosowsky}, {Louis}, {Madhavacheril}, {McMahon},
  {Moodley}, {Nati}, {Nibarger}, {Niemack}, {Page}, {Partridge}, {Salatino},
  {Schaan}, {Schillaci}, {Schmitt}, {Sherwin}, {Sehgal}, {Sif{\'o}n},
  {Spergel}, {Staggs}, {Stevens}, {Storer}, {Ullom}, {Vale}, {Van Engelen},
  {Van Lanen}, {Vavagiakis}, {Wollack}, \& {Xu}}]{Naess2020}
{Naess}, S., {Aiola}, S., {Austermann}, J.~E., {et~al.} 2020, \jcap, 2020, 046,
  \dodoi{10.1088/1475-7516/2020/12/046}

\bibitem[{{Oort}(1983)}]{Oort1983}
{Oort}, J.~H. 1983, \araa, 21, 373, \dodoi{10.1146/annurev.aa.21.090183.002105}

\bibitem[{{Pandey} {et~al.}(2021){Pandey}, {Krause}, {DeRose}, {MacCrann},
  {Jain}, {Crocce}, {Blazek}, {Choi}, {Huang}, {To}, {Fang}, {Elvin-Poole},
  {Prat}, {Porredon}, {Secco}, {Rodriguez-Monroy}, {Weaverdyck}, {Park},
  {Raveri}, {Rozo}, {Rykoff}, {Bernstein}, {S{\'a}nchez}, {Jarvis}, {Troxel},
  {Zacharegkas}, {Chang}, {Alarcon}, {Alves}, {Amon}, {Andrade-Oliveira},
  {Baxter}, {Bechtol}, {Becker}, {Camacho}, {Campos}, {Carnero Rosell},
  {Carrasco Kind}, {Cawthon}, {Chen}, {Chintalapati}, {Davis}, {Di Valentino},
  {Diehl}, {Dodelson}, {Doux}, {Drlica-Wagner}, {Eckert}, {Eifler}, {Elsner},
  {Everett}, {Farahi}, {Fert{\'e}}, {Fosalba}, {Friedrich}, {Gatti},
  {Giannini}, {Gruen}, {Gruendl}, {Harrison}, {Hartley}, {Huff}, {Huterer},
  {Leget}, {McCullough}, {Muir}, {Myles}, {Navarro-Alsina}, {Omori}, {Rollins},
  {Roodman}, {Rosenfeld}, {Sevilla-Noarbe}, {Sheldon}, {Shin}, {Troja},
  {Tutusaus}, {Varga}, {Wechsler}, {Yanny}, {Yin}, {Zhang}, {Zuntz}, {Abbott},
  {Aguena}, {Allam}, {Annis}, {Bacon}, {Bertin}, {Brooks}, {Burke},
  {Carretero}, {Conselice}, {Costanzi}, {da Costa}, {Pereira}, {De Vicente},
  {Dietrich}, {Doel}, {Evrard}, {Ferrero}, {Flaugher}, {Frieman},
  {Garc{\'\i}a-Bellido}, {Gaztanaga}, {Gerdes}, {Giannantonio}, {Gschwend},
  {Gutierrez}, {Hinton}, {Hollowood}, {Honscheid}, {James}, {Jeltema}, {Kuehn},
  {Kuropatkin}, {Lahav}, {Lima}, {Lin}, {Maia}, {Marshall}, {Melchior},
  {Menanteau}, {Miller}, {Miquel}, {Mohr}, {Morgan}, {Palmese},
  {Paz-Chinch{\'o}n}, {Petravick}, {Pieres}, {Plazas Malag{\'o}n}, {Sanchez},
  {Scarpine}, {Serrano}, {Smith}, {Soares-Santos}, {Suchyta}, {Tarle},
  {Thomas}, \& {Weller}}]{Pandey2021}
{Pandey}, S., {Krause}, E., {DeRose}, J., {et~al.} 2021, arXiv e-prints,
  arXiv:2105.13545.
\newblock \doarXiv{2105.13545}

\bibitem[{{Peebles}(1980)}]{Peebles1980}
{Peebles}, P.~J.~E. 1980, {The large-scale structure of the universe}
  (Princeton University Press)

\bibitem[{{Pessa} {et~al.}(2018){Pessa}, {Tejos}, {Barrientos}, {Werk},
  {Bielby}, {Padilla}, {Morris}, {Prochaska}, {Lopez}, \&
  {Hummels}}]{Pessa2018}
{Pessa}, I., {Tejos}, N., {Barrientos}, L.~F., {et~al.} 2018, \mnras, 477,
  2991, \dodoi{10.1093/mnras/sty723}

\bibitem[{{Plagge} {et~al.}(2010){Plagge}, {Benson}, {Ade}, {Aird}, {Bleem},
  {Carlstrom}, {Chang}, {Cho}, {Crawford}, {Crites}, {de Haan}, {Dobbs},
  {George}, {Hall}, {Halverson}, {Holder}, {Holzapfel}, {Hrubes}, {Joy},
  {Keisler}, {Knox}, {Lee}, {Leitch}, {Lueker}, {Marrone}, {McMahon}, {Mehl},
  {Meyer}, {Mohr}, {Montroy}, {Padin}, {Pryke}, {Reichardt}, {Ruhl},
  {Schaffer}, {Shaw}, {Shirokoff}, {Spieler}, {Stalder}, {Staniszewski},
  {Stark}, {Vanderlinde}, {Vieira}, {Williamson}, \& {Zahn}}]{Plagge2010}
{Plagge}, T., {Benson}, B.~A., {Ade}, P.~A.~R., {et~al.} 2010, \apj, 716, 1118,
  \dodoi{10.1088/0004-637X/716/2/1118}

\bibitem[{{Planck Collaboration} {et~al.}(2013){Planck Collaboration}, {Ade},
  {Aghanim}, {Arnaud}, {Ashdown}, {Atrio-Barandela}, {Aumont}, {Baccigalupi},
  {Balbi}, {Banday}, \& et~al.}]{PlanckStackedSZ2013}
{Planck Collaboration}, {Ade}, P.~A.~R., {Aghanim}, N., {et~al.} 2013, \aap,
  550, A131, \dodoi{10.1051/0004-6361/201220040}

\bibitem[{{Planck Collaboration} {et~al.}(2016{\natexlab{a}}){Planck
  Collaboration}, {Aghanim}, {Arnaud}, {Ashdown}, {Aumont}, {Baccigalupi},
  {Band ay}, {Barreiro}, {Bartlett}, {Bartolo}, {Battaner}, {Battye},
  {Benabed}, {Beno{\^\i}t}, {Benoit-L{\'e}vy}, {Bernard}, {Bersanelli},
  {Bielewicz}, {Bock}, {Bonaldi}, {Bonavera}, {Bond}, {Borrill}, {Bouchet},
  {Burigana}, {Butler}, {Calabrese}, {Cardoso}, {Catalano}, {Challinor},
  {Chiang}, {Christensen}, {Churazov}, {Clements}, {Colombo}, {Combet},
  {Comis}, {Coulais}, {Crill}, {Curto}, {Cuttaia}, {Danese}, {Davies}, {Davis},
  {de Bernardis}, {de Rosa}, {de Zotti}, {Delabrouille}, {D{\'e}sert},
  {Dickinson}, {Diego}, {Dolag}, {Dole}, {Donzelli}, {Dor{\'e}}, {Douspis},
  {Ducout}, {Dupac}, {Efstathiou}, {Elsner}, {En{\ss}lin}, {Eriksen},
  {Fergusson}, {Finelli}, {Forni}, {Frailis}, {Fraisse}, {Franceschi},
  {Frejsel}, {Galeotta}, {Galli}, {Ganga}, {G{\'e}nova-Santos}, {Giard},
  {Gonz{\'a}lez-Nuevo}, {G{\'o}rski}, {Gregorio}, {Gruppuso}, {Gudmundsson},
  {Hansen}, {Harrison}, {Henrot-Versill{\'e}}, {Hern{\'a}ndez-Monteagudo},
  {Herranz}, {Hildebrand t}, {Hivon}, {Holmes}, {Hornstrup}, {Huffenberger},
  {Hurier}, {Jaffe}, {Jones}, {Juvela}, {Keih{\"a}nen}, {Keskitalo}, {Kneissl},
  {Knoche}, {Kunz}, {Kurki-Suonio}, {Lacasa}, {Lagache}, {L{\"a}hteenm{\"a}ki},
  {Lamarre}, {Lasenby}, {Lattanzi}, {Leonardi}, {Lesgourgues}, {Levrier},
  {Liguori}, {Lilje}, {Linden-V{\o}rnle}, {L{\'o}pez-Caniego},
  {Mac{\'\i}as-P{\'e}rez}, {Maffei}, {Maggio}, {Maino}, {Mandolesi},
  {Mangilli}, {Maris}, {Martin}, {Mart{\'\i}nez-Gonz{\'a}lez}, {Masi},
  {Matarrese}, {Melchiorri}, {Melin}, {Migliaccio}, {Miville-Desch{\^e}nes},
  {Moneti}, {Montier}, {Morgante}, {Mortlock}, {Munshi}, {Murphy}, {Naselsky},
  {Nati}, {Natoli}, {Noviello}, {Novikov}, {Novikov}, {Paci}, {Pagano},
  {Pajot}, {Paoletti}, {Pasian}, {Patanchon}, {Perdereau}, {Perotto},
  {Pettorino}, {Piacentini}, {Piat}, {Pierpaoli}, {Pietrobon}, {Plaszczynski},
  {Pointecouteau}, {Polenta}, {Ponthieu}, {Pratt}, {Prunet}, {Puget}, {Rachen},
  {Reinecke}, {Remazeilles}, {Renault}, {Renzi}, {Ristorcelli}, {Rocha},
  {Rossetti}, {Roudier}, {Rubi{\~n}o-Mart{\'\i}n}, {Rusholme}, {Sandri},
  {Santos}, {Sauv{\'e}}, {Savelainen}, {Savini}, {Scott}, {Spencer},
  {Stolyarov}, {Stompor}, {Sunyaev}, {Sutton}, {Suur-Uski}, {Sygnet}, {Tauber},
  {Terenzi}, {Toffolatti}, {Tomasi}, {Tramonte}, {Tristram}, {Tucci},
  {Tuovinen}, {Valenziano}, {Valiviita}, {Van Tent}, {Vielva}, {Villa}, {Wade},
  {Wandelt}, {Wehus}, {Yvon}, {Zacchei}, \& {Zonca}}]{Planck2016tSZ}
{Planck Collaboration}, {Aghanim}, N., {Arnaud}, M., {et~al.}
  2016{\natexlab{a}}, \aap, 594, A22, \dodoi{10.1051/0004-6361/201525826}

\bibitem[{{Planck Collaboration} {et~al.}(2016{\natexlab{b}}){Planck
  Collaboration}, {Ade}, {Aghanim}, {Akrami}, {Aluri}, {Arnaud}, {Ashdown},
  {Aumont}, {Baccigalupi}, {Banday}, {Barreiro}, {Bartolo}, {Basak},
  {Battaner}, {Benabed}, {Beno{\^\i}t}, {Benoit-L{\'e}vy}, {Bernard},
  {Bersanelli}, {Bielewicz}, {Bock}, {Bonaldi}, {Bonavera}, {Bond}, {Borrill},
  {Bouchet}, {Boulanger}, {Bucher}, {Burigana}, {Butler}, {Calabrese},
  {Cardoso}, {Casaponsa}, {Catalano}, {Challinor}, {Chamballu}, {Chiang},
  {Christensen}, {Church}, {Clements}, {Colombi}, {Colombo}, {Combet},
  {Contreras}, {Couchot}, {Coulais}, {Crill}, {Cruz}, {Curto}, {Cuttaia},
  {Danese}, {Davies}, {Davis}, {de Bernardis}, {de Rosa}, {de Zotti},
  {Delabrouille}, {D{\'e}sert}, {Diego}, {Dole}, {Donzelli}, {Dor{\'e}},
  {Douspis}, {Ducout}, {Dupac}, {Efstathiou}, {Elsner}, {En{\ss}lin},
  {Eriksen}, {Fantaye}, {Fergusson}, {Fernandez-Cobos}, {Finelli}, {Forni},
  {Frailis}, {Fraisse}, {Franceschi}, {Frejsel}, {Frolov}, {Galeotta}, {Galli},
  {Ganga}, {Gauthier}, {Ghosh}, {Giard}, {Giraud-H{\'e}raud}, {Gjerl{\o}w},
  {Gonz{\'a}lez-Nuevo}, {G{\'o}rski}, {Gratton}, {Gregorio}, {Gruppuso},
  {Gudmundsson}, {Hansen}, {Hanson}, {Harrison}, {Henrot-Versill{\'e}},
  {Hern{\'a}ndez-Monteagudo}, {Herranz}, {Hildebrandt}, {Hivon}, {Hobson},
  {Holmes}, {Hornstrup}, {Hovest}, {Huang}, {Huffenberger}, {Hurier}, {Jaffe},
  {Jaffe}, {Jones}, {Juvela}, {Keih{\"a}nen}, {Keskitalo}, {Kim}, {Kisner},
  {Knoche}, {Kunz}, {Kurki-Suonio}, {Lagache}, {L{\"a}hteenm{\"a}ki},
  {Lamarre}, {Lasenby}, {Lattanzi}, {Lawrence}, {Leonardi}, {Lesgourgues},
  {Levrier}, {Liguori}, {Lilje}, {Linden-V{\o}rnle}, {Liu},
  {L{\'o}pez-Caniego}, {Lubin}, {Mac{\'\i}as-P{\'e}rez}, {Maggio}, {Maino},
  {Mandolesi}, {Mangilli}, {Marinucci}, {Maris}, {Martin},
  {Mart{\'\i}nez-Gonz{\'a}lez}, {Masi}, {Matarrese}, {McGehee}, {Meinhold},
  {Melchiorri}, {Mendes}, {Mennella}, {Migliaccio}, {Mikkelsen}, {Mitra},
  {Miville-Desch{\^e}nes}, {Molinari}, {Moneti}, {Montier}, {Morgante},
  {Mortlock}, {Moss}, {Munshi}, {Murphy}, {Naselsky}, {Nati}, {Natoli},
  {Netterfield}, {N{\o}rgaard-Nielsen}, {Noviello}, {Novikov}, {Novikov},
  {Oxborrow}, {Paci}, {Pagano}, {Pajot}, {Pant}, {Paoletti}, {Pasian},
  {Patanchon}, {Pearson}, {Perdereau}, {Perotto}, {Perrotta}, {Pettorino},
  {Piacentini}, {Piat}, {Pierpaoli}, {Pietrobon}, {Plaszczynski},
  {Pointecouteau}, {Polenta}, {Popa}, {Pratt}, {Pr{\'e}zeau}, {Prunet},
  {Puget}, {Rachen}, {Rebolo}, {Reinecke}, {Remazeilles}, {Renault}, {Renzi},
  {Ristorcelli}, {Rocha}, {Rosset}, {Rossetti}, {Rotti}, {Roudier},
  {Rubi{\~n}o-Mart{\'\i}n}, {Rusholme}, {Sandri}, {Santos}, {Savelainen},
  {Savini}, {Scott}, {Seiffert}, {Shellard}, {Souradeep}, {Spencer},
  {Stolyarov}, {Stompor}, {Sudiwala}, {Sunyaev}, {Sutton}, {Suur-Uski},
  {Sygnet}, {Tauber}, {Terenzi}, {Toffolatti}, {Tomasi}, {Tristram},
  {Trombetti}, {Tucci}, {Tuovinen}, {Valenziano}, {Valiviita}, {Van Tent},
  {Vielva}, {Villa}, {Wade}, {Wandelt}, {Wehus}, {Yvon}, {Zacchei}, {Zibin}, \&
  {Zonca}}]{Planck2016_Isotropy_Statistics}
{Planck Collaboration}, {Ade}, P.~A.~R., {Aghanim}, N., {et~al.}
  2016{\natexlab{b}}, \aap, 594, A16, \dodoi{10.1051/0004-6361/201526681}

\bibitem[{{Planck Collaboration} {et~al.}(2016{\natexlab{c}}){Planck
  Collaboration}, {Ade}, {Aghanim}, {Arnaud}, {Ashdown}, {Aumont},
  {Baccigalupi}, {Banday}, {Barreiro}, {Bartlett}, {Bartolo}, {Battaner},
  {Battye}, {Benabed}, {Beno{\^\i}t}, {Benoit-L{\'e}vy}, {Bernard},
  {Bersanelli}, {Bielewicz}, {Bock}, {Bonaldi}, {Bonavera}, {Bond}, {Borrill},
  {Bouchet}, {Boulanger}, {Bucher}, {Burigana}, {Butler}, {Calabrese},
  {Cardoso}, {Catalano}, {Challinor}, {Chamballu}, {Chary}, {Chiang}, {Chluba},
  {Christensen}, {Church}, {Clements}, {Colombi}, {Colombo}, {Combet},
  {Coulais}, {Crill}, {Curto}, {Cuttaia}, {Danese}, {Davies}, {Davis}, {de
  Bernardis}, {de Rosa}, {de Zotti}, {Delabrouille}, {D{\'e}sert}, {Di
  Valentino}, {Dickinson}, {Diego}, {Dolag}, {Dole}, {Donzelli}, {Dor{\'e}},
  {Douspis}, {Ducout}, {Dunkley}, {Dupac}, {Efstathiou}, {Elsner},
  {En{\ss}lin}, {Eriksen}, {Farhang}, {Fergusson}, {Finelli}, {Forni},
  {Frailis}, {Fraisse}, {Franceschi}, {Frejsel}, {Galeotta}, {Galli}, {Ganga},
  {Gauthier}, {Gerbino}, {Ghosh}, {Giard}, {Giraud-H{\'e}raud}, {Giusarma},
  {Gjerl{\o}w}, {Gonz{\'a}lez-Nuevo}, {G{\'o}rski}, {Gratton}, {Gregorio},
  {Gruppuso}, {Gudmundsson}, {Hamann}, {Hansen}, {Hanson}, {Harrison}, {Helou},
  {Henrot-Versill{\'e}}, {Hern{\'a}ndez-Monteagudo}, {Herranz}, {Hildebrandt},
  {Hivon}, {Hobson}, {Holmes}, {Hornstrup}, {Hovest}, {Huang}, {Huffenberger},
  {Hurier}, {Jaffe}, {Jaffe}, {Jones}, {Juvela}, {Keih{\"a}nen}, {Keskitalo},
  {Kisner}, {Kneissl}, {Knoche}, {Knox}, {Kunz}, {Kurki-Suonio}, {Lagache},
  {L{\"a}hteenm{\"a}ki}, {Lamarre}, {Lasenby}, {Lattanzi}, {Lawrence}, {Leahy},
  {Leonardi}, {Lesgourgues}, {Levrier}, {Lewis}, {Liguori}, {Lilje},
  {Linden-V{\o}rnle}, {L{\'o}pez-Caniego}, {Lubin}, {Mac{\'\i}as-P{\'e}rez},
  {Maggio}, {Maino}, {Mandolesi}, {Mangilli}, {Marchini}, {Maris}, {Martin},
  {Martinelli}, {Mart{\'\i}nez-Gonz{\'a}lez}, {Masi}, {Matarrese}, {McGehee},
  {Meinhold}, {Melchiorri}, {Melin}, {Mendes}, {Mennella}, {Migliaccio},
  {Millea}, {Mitra}, {Miville-Desch{\^e}nes}, {Moneti}, {Montier}, {Morgante},
  {Mortlock}, {Moss}, {Munshi}, {Murphy}, {Naselsky}, {Nati}, {Natoli},
  {Netterfield}, {N{\o}rgaard-Nielsen}, {Noviello}, {Novikov}, {Novikov},
  {Oxborrow}, {Paci}, {Pagano}, {Pajot}, {Paladini}, {Paoletti}, {Partridge},
  {Pasian}, {Patanchon}, {Pearson}, {Perdereau}, {Perotto}, {Perrotta},
  {Pettorino}, {Piacentini}, {Piat}, {Pierpaoli}, {Pietrobon}, {Plaszczynski},
  {Pointecouteau}, {Polenta}, {Popa}, {Pratt}, {Pr{\'e}zeau}, {Prunet},
  {Puget}, {Rachen}, {Reach}, {Rebolo}, {Reinecke}, {Remazeilles}, {Renault},
  {Renzi}, {Ristorcelli}, {Rocha}, {Rosset}, {Rossetti}, {Roudier},
  {Rouill{\'e} d'Orfeuil}, {Rowan-Robinson}, {Rubi{\~n}o-Mart{\'\i}n},
  {Rusholme}, {Said}, {Salvatelli}, {Salvati}, {Sandri}, {Santos},
  {Savelainen}, {Savini}, {Scott}, {Seiffert}, {Serra}, {Shellard}, {Spencer},
  {Spinelli}, {Stolyarov}, {Stompor}, {Sudiwala}, {Sunyaev}, {Sutton},
  {Suur-Uski}, {Sygnet}, {Tauber}, {Terenzi}, {Toffolatti}, {Tomasi},
  {Tristram}, {Trombetti}, {Tucci}, {Tuovinen}, {T{\"u}rler}, {Umana},
  {Valenziano}, {Valiviita}, {Van Tent}, {Vielva}, {Villa}, {Wade}, {Wandelt},
  {Wehus}, {White}, {White}, {Wilkinson}, {Yvon}, {Zacchei}, \&
  {Zonca}}]{Planck2016}
---. 2016{\natexlab{c}}, \aap, 594, A13, \dodoi{10.1051/0004-6361/201525830}

\bibitem[{{Puchwein} {et~al.}(2008){Puchwein}, {Sijacki}, \&
  {Springel}}]{Puchwein2008}
{Puchwein}, E., {Sijacki}, D., \& {Springel}, V. 2008, \apjl, 687, L53,
  \dodoi{10.1086/593352}

\bibitem[{{Regaldo-Saint Blancard} {et~al.}(2021){Regaldo-Saint Blancard},
  {Codis}, {Bond}, \& {Stein}}]{Regaldo2021}
{Regaldo-Saint Blancard}, B., {Codis}, S., {Bond}, J.~R., \& {Stein}, G. 2021,
  \mnras, 504, 1694, \dodoi{10.1093/mnras/stab927}

\bibitem[{{Remazeilles} {et~al.}(2011){Remazeilles}, {Delabrouille}, \&
  {Cardoso}}]{Remazeilles2011}
{Remazeilles}, M., {Delabrouille}, J., \& {Cardoso}, J.-F. 2011, \mnras, 410,
  2481, \dodoi{10.1111/j.1365-2966.2010.17624.x}

\bibitem[{{Rozo} {et~al.}(2016){Rozo}, {Rykoff}, {Abate}, {Bonnett}, {Crocce},
  {Davis}, {Hoyle}, {Leistedt}, {Peiris}, {Wechsler}, {Abbott}, {Abdalla},
  {Banerji}, {Bauer}, {Benoit-L{\'e}vy}, {Bernstein}, {Bertin}, {Brooks},
  {Buckley-Geer}, {Burke}, {Capozzi}, {Rosell}, {Carollo}, {Kind}, {Carretero},
  {Castander}, {Childress}, {Cunha}, {D'Andrea}, {Davis}, {DePoy}, {Desai},
  {Diehl}, {Dietrich}, {Doel}, {Eifler}, {Evrard}, {Neto}, {Flaugher},
  {Fosalba}, {Frieman}, {Gaztanaga}, {Gerdes}, {Glazebrook}, {Gruen},
  {Gruendl}, {Honscheid}, {James}, {Jarvis}, {Kim}, {Kuehn}, {Kuropatkin},
  {Lahav}, {Lidman}, {Lima}, {Maia}, {March}, {Martini}, {Melchior}, {Miller},
  {Miquel}, {Mohr}, {Nichol}, {Nord}, {O'Neill}, {Ogando}, {Plazas}, {Romer},
  {Roodman}, {Sako}, {Sanchez}, {Santiago}, {Schubnell}, {Sevilla-Noarbe},
  {Smith}, {Soares-Santos}, {Sobreira}, {Suchyta}, {Swanson}, {Thaler},
  {Thomas}, {Uddin}, {Vikram}, {Walker}, {Wester}, {Zhang}, \& {da
  Costa}}]{Rozo2016}
{Rozo}, E., {Rykoff}, E.~S., {Abate}, A., {et~al.} 2016, \mnras, 461, 1431,
  \dodoi{10.1093/mnras/stw1281}

\bibitem[{{Rykoff} {et~al.}(2014){Rykoff}, {Rozo}, {Busha}, {Cunha},
  {Finoguenov}, {Evrard}, {Hao}, {Koester}, {Leauthaud}, {Nord}, {Pierre},
  {Reddick}, {Sadibekova}, {Sheldon}, \& {Wechsler}}]{Rykoff2014}
{Rykoff}, E.~S., {Rozo}, E., {Busha}, M.~T., {et~al.} 2014, \apj, 785, 104,
  \dodoi{10.1088/0004-637X/785/2/104}

\bibitem[{{Rykoff} {et~al.}(2016){Rykoff}, {Rozo}, {Hollowood}, {Bermeo-Hernand
  ez}, {Jeltema}, {Mayers}, {Romer}, {Rooney}, {Saro}, {Vergara Cervantes},
  {Wechsler}, {Wilcox}, {Abbott}, {Abdalla}, {Allam}, {Annis},
  {Benoit-L{\'e}vy}, {Bernstein}, {Bertin}, {Brooks}, {Burke}, {Capozzi},
  {Carnero Rosell}, {Carrasco Kind}, {Castander}, {Childress}, {Collins},
  {Cunha}, {D'Andrea}, {da Costa}, {Davis}, {Desai}, {Diehl}, {Dietrich},
  {Doel}, {Evrard}, {Finley}, {Flaugher}, {Fosalba}, {Frieman}, {Glazebrook},
  {Goldstein}, {Gruen}, {Gruendl}, {Gutierrez}, {Hilton}, {Honscheid}, {Hoyle},
  {James}, {Kay}, {Kuehn}, {Kuropatkin}, {Lahav}, {Lewis}, {Lidman}, {Lima},
  {Maia}, {Mann}, {Marshall}, {Martini}, {Melchior}, {Miller}, {Miquel},
  {Mohr}, {Nichol}, {Nord}, {Ogando}, {Plazas}, {Reil}, {Sahl{\'e}n},
  {Sanchez}, {Santiago}, {Scarpine}, {Schubnell}, {Sevilla-Noarbe}, {Smith},
  {Soares-Santos}, {Sobreira}, {Stott}, {Suchyta}, {Swanson}, {Tarle},
  {Thomas}, {Tucker}, {Uddin}, {Viana}, {Vikram}, {Walker}, {Zhang}, \& {DES
  Collaboration}}]{Rykoff2016}
{Rykoff}, E.~S., {Rozo}, E., {Hollowood}, D., {et~al.} 2016, \apjs, 224, 1,
  \dodoi{10.3847/0067-0049/224/1/1}

\bibitem[{{Santiago-Bautista} {et~al.}(2020){Santiago-Bautista}, {Caretta},
  {Bravo-Alfaro}, {Pointecouteau}, \& {Andernach}}]{SantiagoBautista2020}
{Santiago-Bautista}, I., {Caretta}, C.~A., {Bravo-Alfaro}, H., {Pointecouteau},
  E., \& {Andernach}, H. 2020, \aap, 637, A31,
  \dodoi{10.1051/0004-6361/201936397}

\bibitem[{{Schaan} {et~al.}(2021){Schaan}, {Ferraro}, {Amodeo}, {Battaglia},
  {Aiola}, {Austermann}, {Beall}, {Bean}, {Becker}, {Bond}, {Calabrese},
  {Calafut}, {Choi}, {Denison}, {Devlin}, {Duff}, {Duivenvoorden}, {Dunkley},
  {D{\"u}nner}, {Gallardo}, {Guan}, {Han}, {Hill}, {Hilton}, {Hilton},
  {Hlo{\v{z}}ek}, {Hubmayr}, {Huffenberger}, {Hughes}, {Koopman}, {MacInnis},
  {McMahon}, {Madhavacheril}, {Moodley}, {Mroczkowski}, {Naess}, {Nati},
  {Newburgh}, {Niemack}, {Page}, {Partridge}, {Salatino}, {Sehgal},
  {Schillaci}, {Sif{\'o}n}, {Smith}, {Spergel}, {Staggs}, {Storer}, {Trac},
  {Ullom}, {Van Lanen}, {Vale}, {van Engelen}, {Maga{\~n}a}, {Vavagiakis},
  {Wollack}, {Xu}, \& {Atacama Cosmology Telescope Collaboration}}]{Schaan2021}
{Schaan}, E., {Ferraro}, S., {Amodeo}, S., {et~al.} 2021, \prd, 103, 063513,
  \dodoi{10.1103/PhysRevD.103.063513}

\bibitem[{{Sefusatti} {et~al.}(2006){Sefusatti}, {Crocce}, {Pueblas}, \&
  {Scoccimarro}}]{Sefusatti2006}
{Sefusatti}, E., {Crocce}, M., {Pueblas}, S., \& {Scoccimarro}, R. 2006, \prd,
  74, 023522, \dodoi{10.1103/PhysRevD.74.023522}

\bibitem[{{Sehgal} {et~al.}(2010){Sehgal}, {Bode}, {Das},
  {Hernandez-Monteagudo}, {Huffenberger}, {Lin}, {Ostriker}, \&
  {Trac}}]{Sehgal2010}
{Sehgal}, N., {Bode}, P., {Das}, S., {et~al.} 2010, \apj, 709, 920,
  \dodoi{10.1088/0004-637X/709/2/920}

\bibitem[{{Sehgal} {et~al.}(2011){Sehgal}, {Trac}, {Acquaviva}, {Ade},
  {Aguirre}, {Amiri}, {Appel}, {Barrientos}, {Battistelli}, {Bond}, {Brown},
  {Burger}, {Chervenak}, {Das}, {Devlin}, {Dicker}, {Bertrand Doriese},
  {Dunkley}, {D{\"u}nner}, {Essinger-Hileman}, {Fisher}, {Fowler}, {Hajian},
  {Halpern}, {Hasselfield}, {Hern{\'a}ndez-Monteagudo}, {Hilton}, {Hilton},
  {Hincks}, {Hlozek}, {Holtz}, {Huffenberger}, {Hughes}, {Hughes}, {Infante},
  {Irwin}, {Jones}, {Baptiste Juin}, {Klein}, {Kosowsky}, {Lau}, {Limon},
  {Lin}, {Lupton}, {Marriage}, {Marsden}, {Martocci}, {Mauskopf}, {Menanteau},
  {Moodley}, {Moseley}, {Netterfield}, {Niemack}, {Nolta}, {Page}, {Parker},
  {Partridge}, {Reid}, {Sherwin}, {Sievers}, {Spergel}, {Staggs}, {Swetz},
  {Switzer}, {Thornton}, {Tucker}, {Warne}, {Wollack}, \& {Zhao}}]{Sehgal2011}
{Sehgal}, N., {Trac}, H., {Acquaviva}, V., {et~al.} 2011, \apj, 732, 44,
  \dodoi{10.1088/0004-637X/732/1/44}

\bibitem[{{Sehgal} {et~al.}(2013){Sehgal}, {Addison}, {Battaglia},
  {Battistelli}, {Bond}, {Das}, {Devlin}, {Dunkley}, {D{\"u}nner}, {Gralla},
  {Hajian}, {Halpern}, {Hasselfield}, {Hilton}, {Hincks}, {Hlozek}, {Hughes},
  {Kosowsky}, {Lin}, {Louis}, {Marriage}, {Marsden}, {Menanteau}, {Moodley},
  {Niemack}, {Page}, {Partridge}, {Reese}, {Sherwin}, {Sievers}, {Sif{\'o}n},
  {Spergel}, {Staggs}, {Swetz}, {Switzer}, \& {Wollack}}]{Sehgal2013}
{Sehgal}, N., {Addison}, G., {Battaglia}, N., {et~al.} 2013, \apj, 767, 38,
  \dodoi{10.1088/0004-637X/767/1/38}

\bibitem[{{Sevilla-Noarbe} {et~al.}(2021){Sevilla-Noarbe}, {Bechtol}, {Carrasco
  Kind}, {Carnero Rosell}, {Becker}, {Drlica-Wagner}, {Gruendl}, {Rykoff},
  {Sheldon}, {Yanny}, {Alarcon}, {Allam}, {Amon}, {Benoit-L{\'e}vy},
  {Bernstein}, {Bertin}, {Burke}, {Carretero}, {Choi}, {Diehl}, {Everett},
  {Flaugher}, {Gaztanaga}, {Gschwend}, {Harrison}, {Hartley}, {Hoyle},
  {Jarvis}, {Johnson}, {Kessler}, {Kron}, {Kuropatkin}, {Leistedt}, {Li},
  {Menanteau}, {Morganson}, {Ogando}, {Palmese}, {Paz-Chinch{\'o}n}, {Pieres},
  {Pond}, {Rodriguez-Monroy}, {Smith}, {Stringer}, {Troxel}, {Tucker}, {de
  Vicente}, {Wester}, {Zhang}, {Abbott}, {Aguena}, {Annis}, {Avila},
  {Bhargava}, {Bridle}, {Brooks}, {Brout}, {Castander}, {Cawthon}, {Chang},
  {Conselice}, {Costanzi}, {Crocce}, {da Costa}, {Pereira}, {Davis}, {Desai},
  {Dietrich}, {Doel}, {Eckert}, {Evrard}, {Ferrero}, {Fosalba},
  {Garc{\'\i}a-Bellido}, {Gerdes}, {Giannantonio}, {Gruen}, {Gutierrez},
  {Hinton}, {Hollowood}, {Honscheid}, {Huff}, {Huterer}, {James}, {Jeltema},
  {Kuehn}, {Lahav}, {Lidman}, {Lima}, {Lin}, {Maia}, {Marshall}, {Martini},
  {Melchior}, {Miquel}, {Mohr}, {Morgan}, {Neilsen}, {Plazas}, {Romer},
  {Roodman}, {Sanchez}, {Scarpine}, {Schubnell}, {Serrano}, {Smith}, {Suchyta},
  {Tarle}, {Thomas}, {To}, {Varga}, {Wechsler}, {Weller}, {Wilkinson}, \& {DES
  Collaboration}}]{DESY32021}
{Sevilla-Noarbe}, I., {Bechtol}, K., {Carrasco Kind}, M., {et~al.} 2021, \apjs,
  254, 24, \dodoi{10.3847/1538-4365/abeb66}

\bibitem[{{Shull} {et~al.}(2012){Shull}, {Smith}, \& {Danforth}}]{Shull2012}
{Shull}, J.~M., {Smith}, B.~D., \& {Danforth}, C.~W. 2012, \apj, 759, 23,
  \dodoi{10.1088/0004-637X/759/1/23}

\bibitem[{{Sijacki} {et~al.}(2008){Sijacki}, {Pfrommer}, {Springel}, \&
  {En{\ss}lin}}]{Sijacki2008}
{Sijacki}, D., {Pfrommer}, C., {Springel}, V., \& {En{\ss}lin}, T.~A. 2008,
  \mnras, 387, 1403, \dodoi{10.1111/j.1365-2966.2008.13310.x}

\bibitem[{{Sijacki} {et~al.}(2007){Sijacki}, {Springel}, {Di Matteo}, \&
  {Hernquist}}]{Sijacki2007}
{Sijacki}, D., {Springel}, V., {Di Matteo}, T., \& {Hernquist}, L. 2007,
  \mnras, 380, 877, \dodoi{10.1111/j.1365-2966.2007.12153.x}

\bibitem[{{Sonnenfeld} {et~al.}(2019){Sonnenfeld}, {Wang}, \&
  {Bahcall}}]{Sonnenfeld2019}
{Sonnenfeld}, A., {Wang}, W., \& {Bahcall}, N. 2019, \aap, 622, A30,
  \dodoi{10.1051/0004-6361/201834260}

\bibitem[{{Springel} {et~al.}(2018){Springel}, {Pakmor}, {Pillepich},
  {Weinberger}, {Nelson}, {Hernquist}, {Vogelsberger}, {Genel}, {Torrey},
  {Marinacci}, \& {Naiman}}]{Illustris2018}
{Springel}, V., {Pakmor}, R., {Pillepich}, A., {et~al.} 2018, \mnras, 475, 676,
  \dodoi{10.1093/mnras/stx3304}

\bibitem[{{Stein} {et~al.}(2019){Stein}, {Alvarez}, \& {Bond}}]{Stein2019}
{Stein}, G., {Alvarez}, M.~A., \& {Bond}, J.~R. 2019, \mnras, 483, 2236,
  \dodoi{10.1093/mnras/sty3226}

\bibitem[{{Stein} {et~al.}(2020){Stein}, {Alvarez}, {Bond}, {van Engelen}, \&
  {Battaglia}}]{Stein2020}
{Stein}, G., {Alvarez}, M.~A., {Bond}, J.~R., {van Engelen}, A., \&
  {Battaglia}, N. 2020, \jcap, 2020, 012, \dodoi{10.1088/1475-7516/2020/10/012}

\bibitem[{{Sunyaev} \& {Zeldovich}(1970)}]{SZ1970}
{Sunyaev}, R.~A., \& {Zeldovich}, Y.~B. 1970, Comments on Astrophysics and
  Space Physics, 2, 66

\bibitem[{{Sunyaev} \& {Zeldovich}(1972)}]{SZ1972}
---. 1972, Comments on Astrophysics and Space Physics, 4, 173

\bibitem[{{Swetz} {et~al.}(2011){Swetz}, {Ade}, {Amiri}, {Appel},
  {Battistelli}, {Burger}, {Chervenak}, {Devlin}, {Dicker}, {Doriese},
  {D{\"u}nner}, {Essinger-Hileman}, {Fisher}, {Fowler}, {Halpern},
  {Hasselfield}, {Hilton}, {Hincks}, {Irwin}, {Jarosik}, {Kaul}, {Klein},
  {Lau}, {Limon}, {Marriage}, {Marsden}, {Martocci}, {Mauskopf}, {Moseley},
  {Netterfield}, {Niemack}, {Nolta}, {Page}, {Parker}, {Staggs}, {Stryzak},
  {Switzer}, {Thornton}, {Tucker}, {Wollack}, \& {Zhao}}]{Swetz2011}
{Swetz}, D.~S., {Ade}, P.~A.~R., {Amiri}, M., {et~al.} 2011, \apjs, 194, 41,
  \dodoi{10.1088/0067-0049/194/2/41}

\bibitem[{{Takada} \& {Jain}(2004)}]{Takada2004}
{Takada}, M., \& {Jain}, B. 2004, \mnras, 348, 897,
  \dodoi{10.1111/j.1365-2966.2004.07410.x}

\bibitem[{{Tanimura} {et~al.}(2020){Tanimura}, {Aghanim}, {Bonjean},
  {Malavasi}, \& {Douspis}}]{Tanimura2020}
{Tanimura}, H., {Aghanim}, N., {Bonjean}, V., {Malavasi}, N., \& {Douspis}, M.
  2020, \aap, 637, A41, \dodoi{10.1051/0004-6361/201937158}

\bibitem[{{Tanimura} {et~al.}(2019){Tanimura}, {Hinshaw}, {McCarthy}, {Van
  Waerbeke}, {Aghanim}, {Ma}, {Mead}, {Hojjati}, \&
  {Tr{\"o}ster}}]{Tanimura2019}
{Tanimura}, H., {Hinshaw}, G., {McCarthy}, I.~G., {et~al.} 2019, \mnras, 483,
  223, \dodoi{10.1093/mnras/sty3118}

\bibitem[{{Tejos} {et~al.}(2016){Tejos}, {Prochaska}, {Crighton}, {Morris},
  {Werk}, {Theuns}, {Padilla}, {Bielby}, \& {Finn}}]{Tejos2016}
{Tejos}, N., {Prochaska}, J.~X., {Crighton}, N. H.~M., {et~al.} 2016, \mnras,
  455, 2662, \dodoi{10.1093/mnras/stv2376}

\bibitem[{{Terry} {et~al.}(2019){Terry}, {Battaglia}, {Basu}, {Beringue},
  {Bertoldi}, {Chapman}, {Choi}, {Cothard}, {Chung}, {Erler}, {Fich},
  {Foreman}, {Gallardo}, {Gao}, {Graf}, {Haynes}, {Herter}, {Hilton},
  {Hubmayr}, {Johnstone}, {Komatsu}, {Magnelli}, {Mauskopf}, {McMahon},
  {Meerburg}, {Meyers}, {Mittal}, {Niemack}, {Nikola}, {Parshley}, {Riechers},
  {Stacey}, {Stutzki}, {Vavagiakis}, {Viero}, \& {Vissers}}]{ccatp:2019}
{Terry}, H., {Battaglia}, N., {Basu}, K., {et~al.} 2019, in Bulletin of the
  American Astronomical Society, Vol.~51, 213.
\newblock \doarXiv{1909.02587}

\bibitem[{{The Dark Energy Survey Collaboration}(2005)}]{DES2005}
{The Dark Energy Survey Collaboration}. 2005, arXiv e-prints, astro.
\newblock \doarXiv{astro-ph/0510346}

\bibitem[{{Thornton} {et~al.}(2016){Thornton}, {Ade}, {Aiola}, {Angil{\`e}},
  {Amiri}, {Beall}, {Becker}, {Cho}, {Choi}, {Corlies}, {Coughlin}, {Datta},
  {Devlin}, {Dicker}, {D{\"u}nner}, {Fowler}, {Fox}, {Gallardo}, {Gao},
  {Grace}, {Halpern}, {Hasselfield}, {Henderson}, {Hilton}, {Hincks}, {Ho},
  {Hubmayr}, {Irwin}, {Klein}, {Koopman}, {Li}, {Louis}, {Lungu}, {Maurin},
  {McMahon}, {Munson}, {Naess}, {Nati}, {Newburgh}, {Nibarger}, {Niemack},
  {Niraula}, {Nolta}, {Page}, {Pappas}, {Schillaci}, {Schmitt}, {Sehgal},
  {Sievers}, {Simon}, {Staggs}, {Tucker}, {Uehara}, {van Lanen}, {Ward}, \&
  {Wollack}}]{thornton/etal:2016}
{Thornton}, R.~J., {Ade}, P.~A.~R., {Aiola}, S., {et~al.} 2016, \apjs, 227, 21,
  \dodoi{10.3847/1538-4365/227/2/21}

\bibitem[{{To} {et~al.}(2021){To}, {Krause}, {Rozo}, {Wu}, {Gruen}, {Wechsler},
  {Eifler}, {Rykoff}, {Costanzi}, {Becker}, {Bernstein}, {Blazek}, {Bocquet},
  {Bridle}, {Cawthon}, {Choi}, {Crocce}, {Davis}, {DeRose}, {Drlica-Wagner},
  {Elvin-Poole}, {Fang}, {Farahi}, {Friedrich}, {Gatti}, {Gaztanaga},
  {Giannantonio}, {Hartley}, {Hoyle}, {Jarvis}, {MacCrann}, {McClintock},
  {Miranda}, {Pereira}, {Park}, {Porredon}, {Prat}, {Rau}, {Ross}, {Samuroff},
  {S{\'a}nchez}, {Sevilla-Noarbe}, {Sheldon}, {Troxel}, {Varga}, {Vielzeuf},
  {Zhang}, {Zuntz}, {Abbott}, {Aguena}, {Amon}, {Annis}, {Avila}, {Bertin},
  {Bhargava}, {Brooks}, {Burke}, {Carnero Rosell}, {Carrasco Kind},
  {Carretero}, {Chang}, {Conselice}, {da Costa}, {Davis}, {Desai}, {Diehl},
  {Dietrich}, {Everett}, {Evrard}, {Ferrero}, {Flaugher}, {Fosalba}, {Frieman},
  {Garc{\'\i}a-Bellido}, {Gruendl}, {Gutierrez}, {Hinton}, {Hollowood},
  {Honscheid}, {Huterer}, {James}, {Jeltema}, {Kron}, {Kuehn}, {Kuropatkin},
  {Lima}, {Maia}, {Marshall}, {Menanteau}, {Miquel}, {Morgan}, {Muir}, {Myles},
  {Palmese}, {Paz-Chinch{\'o}n}, {Plazas}, {Romer}, {Roodman}, {Sanchez},
  {Santiago}, {Scarpine}, {Serrano}, {Smith}, {Suchyta}, {Swanson}, {Tarle},
  {Thomas}, {Tucker}, {Weller}, {Wester}, {Wilkinson}, \& {DES
  Collaboration}}]{To2021}
{To}, C., {Krause}, E., {Rozo}, E., {et~al.} 2021, \prl, 126, 141301,
  \dodoi{10.1103/PhysRevLett.126.141301}

\bibitem[{{Tumlinson} {et~al.}(2017){Tumlinson}, {Peeples}, \&
  {Werk}}]{Tumlinson2017}
{Tumlinson}, J., {Peeples}, M.~S., \& {Werk}, J.~K. 2017, \araa, 55, 389,
  \dodoi{10.1146/annurev-astro-091916-055240}

\bibitem[{{van de Weygaert} \& {Bond}(2008{\natexlab{a}})}]{vdWetal:2008a}
{van de Weygaert}, R., \& {Bond}, J.~R. 2008{\natexlab{a}}, in A Pan-Chromatic
  View of Clusters of Galaxies and the Large-Scale Structure, ed. M.~{Plionis},
  O.~{L{\'o}pez-Cruz}, \& D.~{Hughes}, Vol. 740 (Dordrecht: Springer), 24,
  \dodoi{10.1007/978-1-4020-6941-3_11}

\bibitem[{{van de Weygaert} \& {Bond}(2008{\natexlab{b}})}]{vdWetal:2008b}
---. 2008{\natexlab{b}}, in A Pan-Chromatic View of Clusters of Galaxies and
  the Large-Scale Structure, ed. M.~{Plionis}, O.~{L{\'o}pez-Cruz}, \&
  D.~{Hughes}, Vol. 740 (Dordrecht: Springer), 335,
  \dodoi{10.1007/978-1-4020-6941-3_10}

\bibitem[{{Wechsler} {et~al.}(2022){Wechsler}, {DeRose}, {Busha}, {Becker},
  {Rykoff}, \& {Evrard}}]{Weschler2021}
{Wechsler}, R.~H., {DeRose}, J., {Busha}, M.~T., {et~al.} 2022, \apj, 931, 145,
  \dodoi{10.3847/1538-4357/ac5b0a}

\bibitem[{{Welker} {et~al.}(2020){Welker}, {Bland-Hawthorn}, {Van de Sande},
  {Lagos}, {Elahi}, {Obreschkow}, {Bryant}, {Pichon}, {Cortese}, {Richards},
  {Croom}, {Goodwin}, {Lawrence}, {Sweet}, {Lopez-Sanchez}, {Medling}, {Owers},
  {Dubois}, \& {Devriendt}}]{Welker2020}
{Welker}, C., {Bland-Hawthorn}, J., {Van de Sande}, J., {et~al.} 2020, \mnras,
  491, 2864, \dodoi{10.1093/mnras/stz2860}

\bibitem[{{Yang} {et~al.}(2020){Yang}, {Hudson}, \& {Afshordi}}]{Yang2020}
{Yang}, T., {Hudson}, M.~J., \& {Afshordi}, N. 2020, \mnras, 498, 3158,
  \dodoi{10.1093/mnras/staa2547}

\bibitem[{{Zel'Dovich}(1970)}]{zeldovich:1970}
{Zel'Dovich}, Y.~B. 1970, \aap, 500, 13

\bibitem[{{Zeldovich} \& {Sunyaev}(1969)}]{SZ1969}
{Zeldovich}, Y.~B., \& {Sunyaev}, R.~A. 1969, \apss, 4, 301,
  \dodoi{10.1007/BF00661821}

\bibitem[{{Zhang} {et~al.}(2019){Zhang}, {Jeltema}, {Hollowood}, {Everett},
  {Rozo}, {Farahi}, {Bermeo}, {Bhargava}, {Giles}, {Romer}, {Wilkinson},
  {Rykoff}, {Mantz}, {Diehl}, {Evrard}, {Stern}, {Gruen}, {von der Linden},
  {Splettstoesser}, {Chen}, {Costanzi}, {Allen}, {Collins}, {Hilton}, {Klein},
  {Mann}, {Manolopoulou}, {Morris}, {Mayers}, {Sahlen}, {Stott}, {Vergara
  Cervantes}, {Viana}, {Wechsler}, {Allam}, {Avila}, {Bechtol}, {Bertin},
  {Brooks}, {Burke}, {Carnero Rosell}, {Carrasco Kind}, {Carretero},
  {Castander}, {da Costa}, {De Vicente}, {Desai}, {Dietrich}, {Doel},
  {Flaugher}, {Fosalba}, {Frieman}, {Garc{\'\i}a-Bellido}, {Gaztanaga},
  {Gruendl}, {Gschwend}, {Gutierrez}, {Hartley}, {Honscheid}, {Hoyle},
  {Krause}, {Kuehn}, {Kuropatkin}, {Lima}, {Maia}, {Marshall}, {Melchior},
  {Menanteau}, {Miller}, {Miquel}, {Ogando}, {Plazas}, {Sanchez}, {Scarpine},
  {Schindler}, {Serrano}, {Sevilla-Noarbe}, {Smith}, {Soares-Santos},
  {Suchyta}, {Swanson}, {Tarle}, {Thomas}, {Tucker}, {Vikram}, {Wester}, \&
  {DES Collaboration}}]{Zhang2019}
{Zhang}, Y., {Jeltema}, T., {Hollowood}, D.~L., {et~al.} 2019, \mnras, 487,
  2578, \dodoi{10.1093/mnras/stz1361}

\bibitem[{Zonca {et~al.}(2019)Zonca, Singer, Lenz, Reinecke, Rosset, Hivon, \&
  Gorski}]{Zonca2019}
Zonca, A., Singer, L., Lenz, D., {et~al.} 2019, Journal of Open Source
  Software, 4, 1298, \dodoi{10.21105/joss.01298}

\bibitem[{{Zou} {et~al.}(2021){Zou}, {Gao}, {Xu}, {Zhou}, {Ma}, {Zhou},
  {Zhang}, {Nie}, {Wang}, \& {Xue}}]{DESI_cluster}
{Zou}, H., {Gao}, J., {Xu}, X., {et~al.} 2021, \apjs, 253, 56,
  \dodoi{10.3847/1538-4365/abe5b0}

\end{thebibliography}

% \section*{Affiliations}
% \input{affiliations}
\end{document}